\documentclass[aoas,preprint]{imsart}

\usepackage[OT1]{fontenc}
\usepackage{amsthm,amsmath}
\usepackage{natbib}
\usepackage{graphicx, amssymb, mathrsfs,subfig,amsfonts,url}
\usepackage{soul}
\usepackage{tikz}
\usepackage[pdfversion=1.0]{hyperref}
\usepackage{cleveref, autonum}
\usepackage[subnum]{cases}

\usepackage{color} 
\usepackage{pstricks,pst-node,pst-plot}
\usepackage{rotating}
\usepackage{pdfpages}

  \newcommand{\bR}{\mathbb{\bar{R}}}

\newcommand{\bmu}{\mu} \newcommand{\bSigma}{\Sigma} \newcommand{\tmu}{\widetilde{\mu}}
\newcommand{\tSigma}{\widetilde{\Sigma}} \newcommand{\ts}{\widetilde{s}}

\newcommand{\Ind}{1\!\mathrm{l}}
\newcommand{\bs}{\boldsymbol}
\newcommand{\E}{{\rm E}}

\newcommand{\plotUnivariate}{plot.dir/output_10_12} %
\newcommand{\plotHist}{plot.dir/output_univariate_histgram}
\newcommand{\plotBivariate}{plot.dir/bivariate_11_7}
\newcommand{\plotBivariateTail}{plot.dir/bivariate_tail}

\newtheorem{theorem}{Theorem}[section] 
\newtheorem{lemma}[theorem]{Lemma}

\theoremstyle{remark} 
\newtheorem{remark}[theorem]{Remark}

\arxiv{arXiv:1710.08508}

\startlocaldefs
\numberwithin{equation}{section}
\theoremstyle{plain}

\endlocaldefs

\begin{document}

\begin{frontmatter}
\title{Standardization of multivariate Gaussian mixture models and background adjustment of PET images in brain oncology\thanksref{T1}} 
\runtitle{Background Adjustment via Robust Spatial GMM}
\thankstext{T1}{This work was partially supported by NIH grant R21EB13795. }

\begin{aug}
	\author{\fnms{Meng} \snm{Li}\thanksref{m1}\ead[label=e1]{meng@rice.edu}}
	\and
	\author{\fnms{Armin} \snm{Schwartzman}\thanksref{m2}\ead[label=e2]{armins@ucsd.edu}}

	\runauthor{Li and Schwartzman}
	
	\affiliation{Rice University\thanksmark{m1} and University of California, San Diego\thanksmark{m2}}
	
	\address{Rice University \\ Department of Statistics \\ 6100 Main St, MS 138\\
		Houston, TX 77005, USA\\
		\printead{e1}}

	\address{University of California, San Diego\\
		Division of Biostatistics\\
		9500 Gilman Drive \#0725\\
		La Jolla, CA 92093, USA\\
		\printead{e2}}
\end{aug}

\begin{abstract}
	In brain oncology, it is routine to evaluate the progress or remission of the disease based on the differences between a pre-treatment and a post-treatment Positron Emission Tomography (PET) scan. Background adjustment is necessary to reduce confounding by tissue-dependent changes not related to the disease. When modeling the voxel intensities for the two scans as a bivariate Gaussian mixture, background adjustment translates into standardizing the mixture at each voxel, while tumor lesions present themselves as outliers to be detected. In this paper, we address the question of how to standardize the mixture to a standard multivariate normal distribution, so that the outliers (i.e., tumor lesions) can be detected using a statistical test. We show theoretically and numerically that the tail distribution of the standardized scores is favorably close to standard normal in a wide range of scenarios while being conservative at the tails, validating voxelwise hypothesis testing based on standardized scores. To address standardization in spatially heterogeneous image data, we propose a spatial and robust multivariate expectation-maximization (EM) algorithm, where prior class membership probabilities are provided by transformation of spatial probability template maps and the estimation of the class mean and covariances are robust to outliers. Simulations in both univariate and bivariate cases suggest that standardized scores with soft assignment have tail probabilities that are either very close to or more conservative than standard normal. The proposed methods are applied to a real data set from a PET phantom experiment, yet they are generic and can be used in other contexts.
\end{abstract}


\end{frontmatter}

\section{Introduction} 

In brain oncology, it is routine to evaluate the progression or remission of the disease based on differences between a pre-treatment and a post-treatment Positron Emission Tomography (PET) three-dimensional scan~\citep{Valk2003}. Using markers such as injected F-18-fluorodeoxyglucose (FDG), these images measure the activity of glucose metabolism in human brains, which is normalized by dose and patient weight to standard uptake value (SUV) units. 

A standard practice in this setting is to conduct analyses based on differences between {\it scalar summaries} of the two PET scans~\citep{Young+:99, Wah+:09}, such as the maximum SUV, calculated within user-defined regions of interest (ROIs)~\citep{Zas+:93,Lee+:09,Tak+:11}. However, due to variations in scanner settings, neurological activity or pharmacological drug effects, the scans may exhibit background differences even without significant progression or remission of the disease itself~\citep{Soret2007a,Bai2013, Soffientini2016}. The background effect may even be spatially heterogeneous, varying according to tissue type~\citep{Guo+:14,Qin+:}. Moreover, lesion changes may be missed if those lesions are not included in the user-defined ROIs. For these reasons, \citet{Guo+:14} proposed a new approach comparing the two PET scans {\it voxelwise}.

\begin{figure}[ht!]
	\caption{Effects of background adjustment in detecting changes in tumor lesions in a phantom study. The 1st row shows a transverse slice of the original PET scans. Their direct difference in the last column exhibits a large positive background change between the scans. The 2nd row shows the respective standardized images via background adjustment and their difference based on our proposed method detailed later.} 
	\label{fig:motivation}
	\begin{tabular}{@{}{c}*{3}{c@{\hskip -0pt}}@{}}
		& {Scan 1} \rule{0.75cm}{0pt} & {Scan 2} \rule{0.75cm}{0pt} & Difference
		\rule{0.75cm}{0pt}\\
		\begin{sideways} \rule[0pt]{0.7in}{0pt} Original \end{sideways}
		&\includegraphics[trim = 150 30 100 28,
		clip,width=0.32\linewidth]{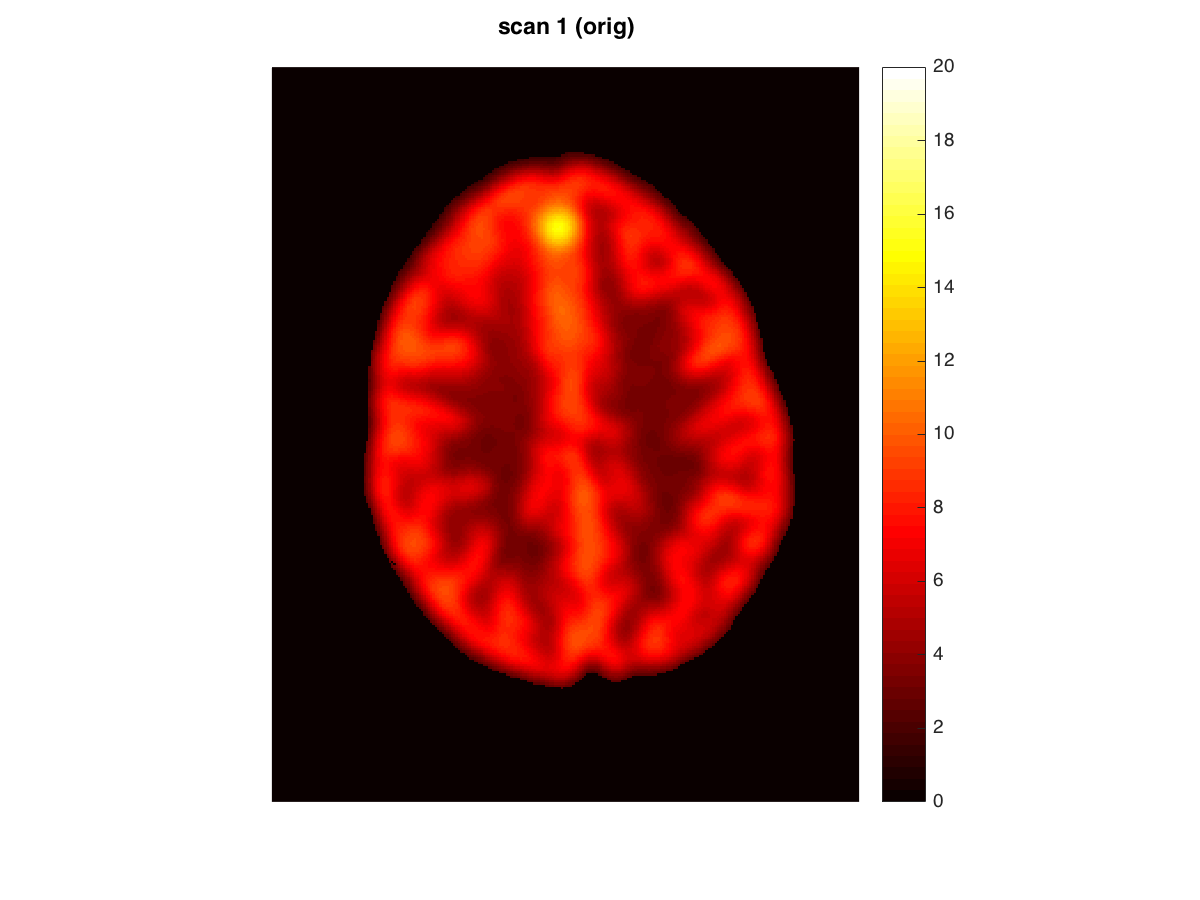} & \includegraphics[trim = 150 30 100 28,
		clip,width=0.32\linewidth]{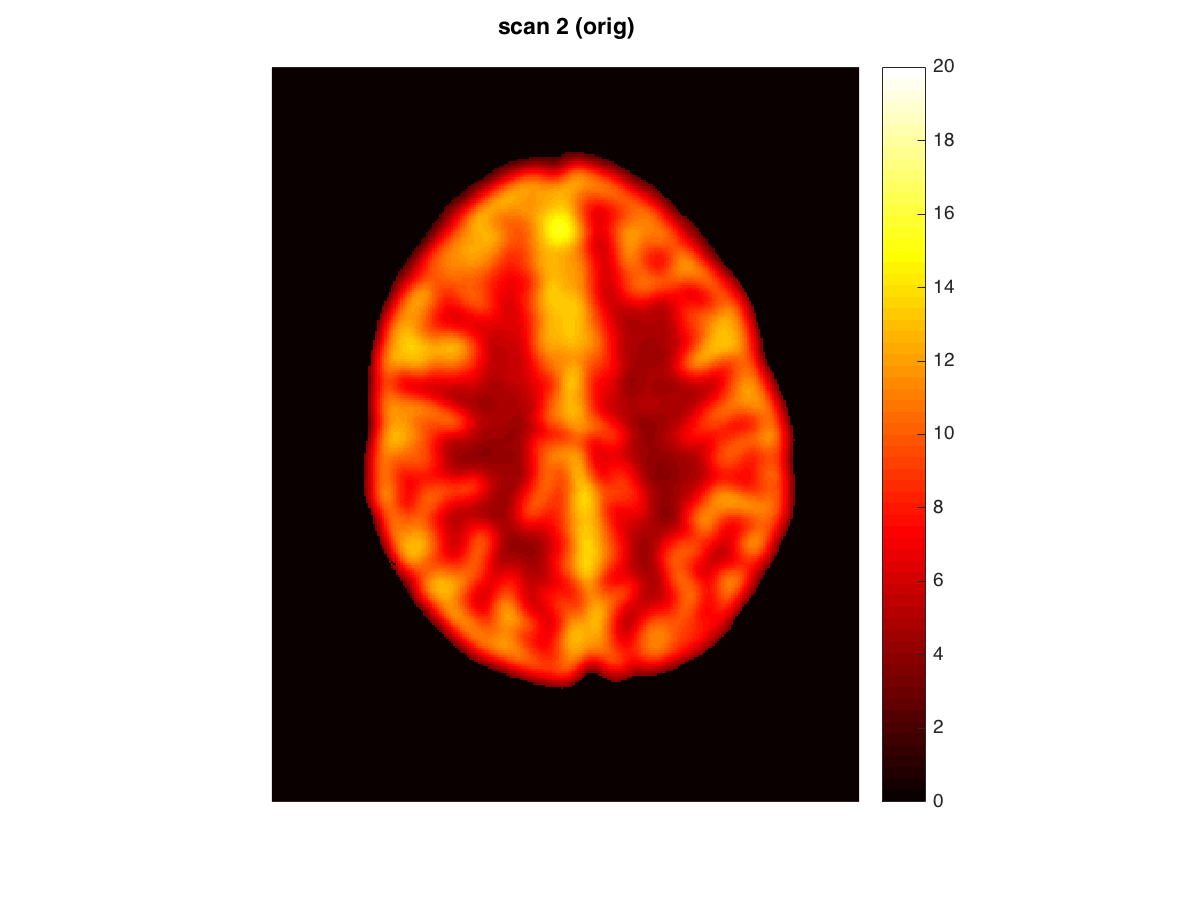} & \includegraphics[trim = 150 30 100 28,
		clip,width=0.32\linewidth]{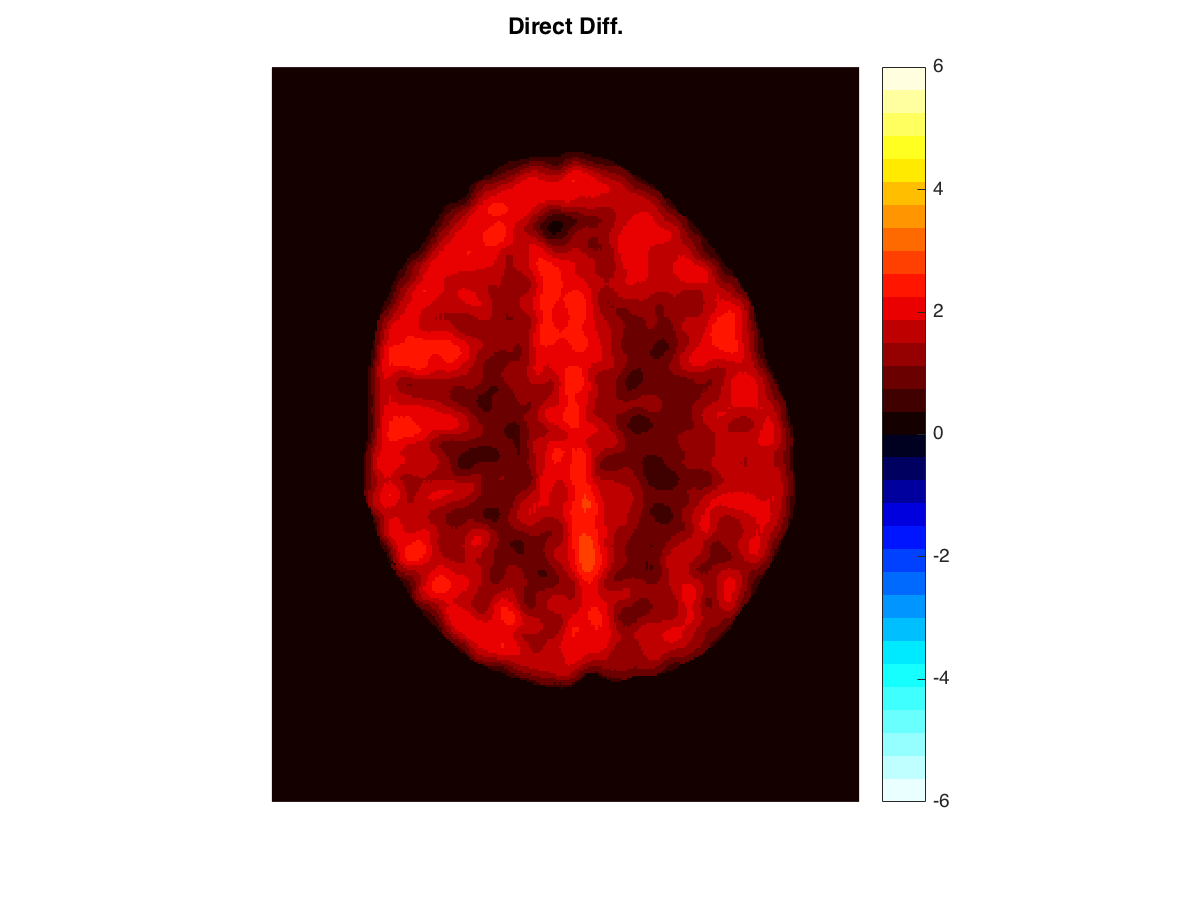} \\ \begin{sideways} \rule[0pt]{0.25in}{0pt}
			Background Adjustment \end{sideways} &\includegraphics[trim = 150 30 100 28, clip,width=0.32\linewidth]{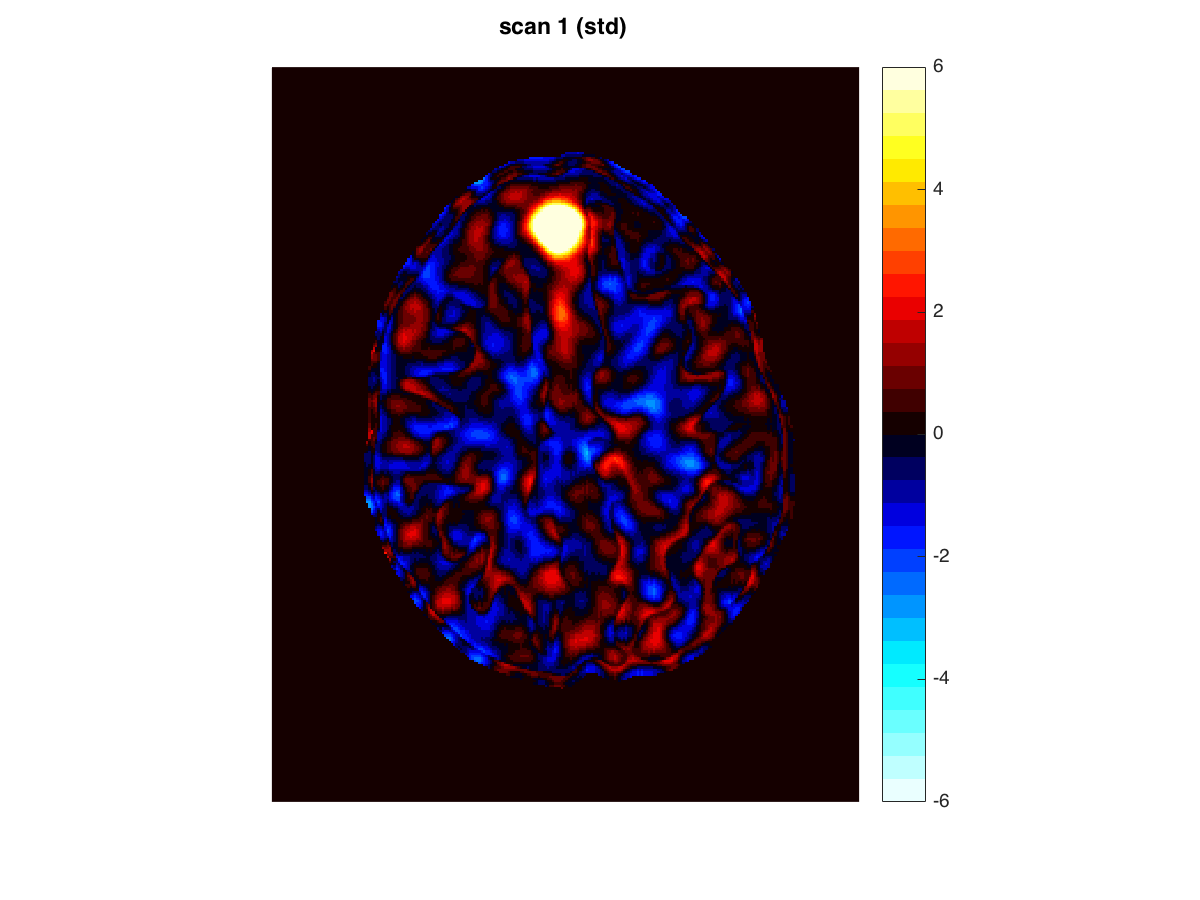} & \includegraphics[trim = 150 30 100 28, clip,width=0.32\linewidth]{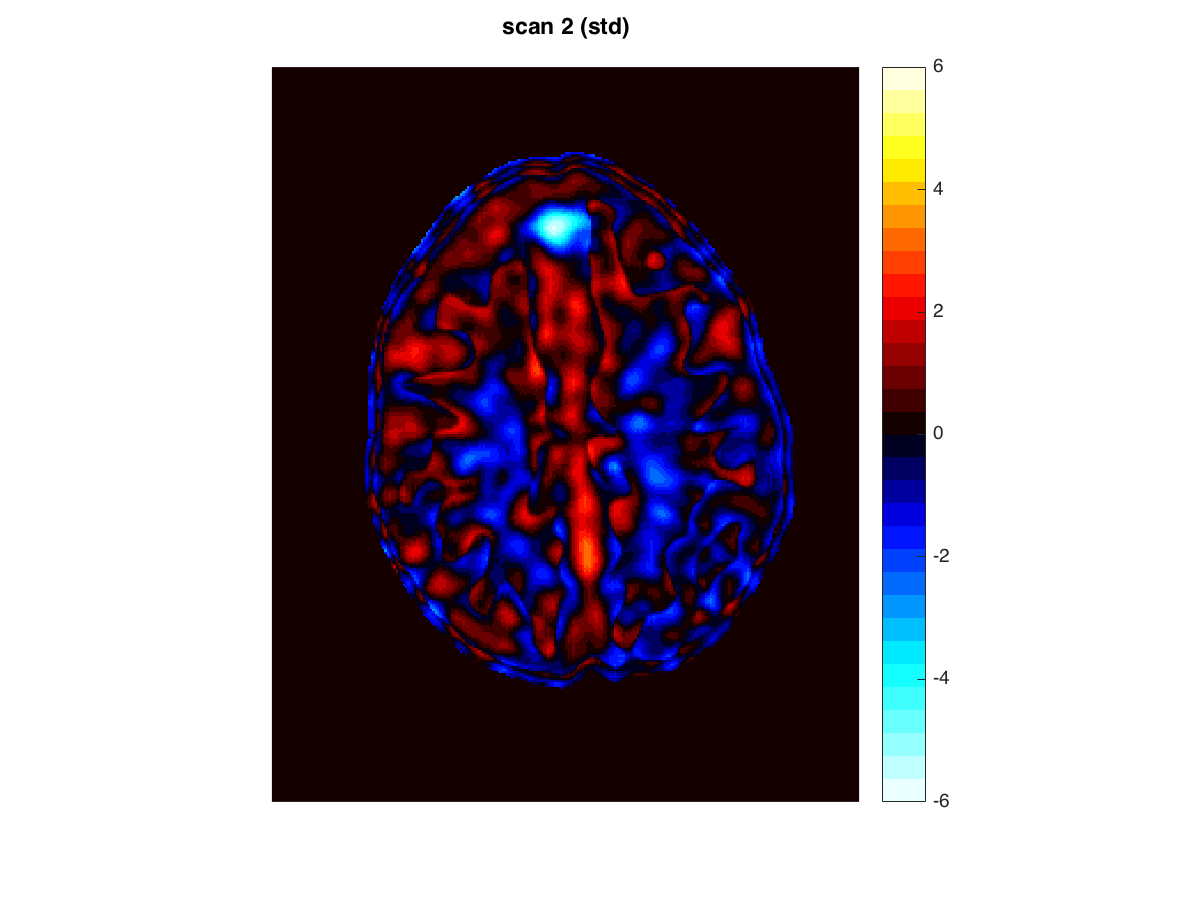} & \includegraphics[trim = 150 30 100 28, clip,width=0.32\linewidth]{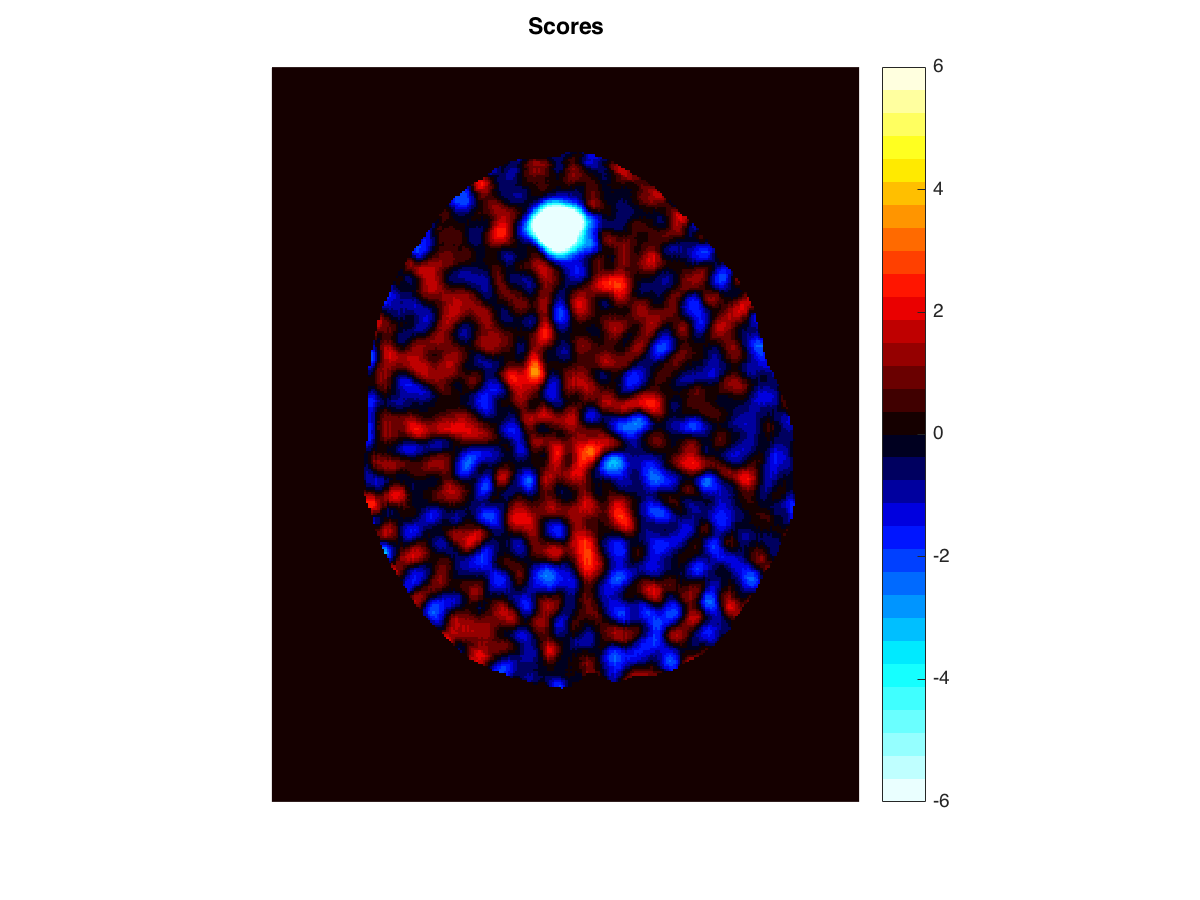} 
	\end{tabular} 
\end{figure}

As a motivating example that will be detailed in Section~\ref{section:data.application}, Figure~\ref{fig:motivation} shows data from the phantom experiment in \cite{Qin+:} simulating pre- and post-treatment scans with a tumor lesion. A direct voxelwise difference between the two scans shows a global non-homogeneous background change while failing to detect changes in the lesion (Figure~\ref{fig:motivation}, Row 1 and Column 3). This observation suggests that \textit{background adjustment} is necessary in voxelwise comparisons to reduce confounding by tissue-dependent changes not related to the disease, in order to isolate localized differences that are relevant to assess the disease status. 

Statistical models have been intensively used in quantitative analysis for PET imaging to provide automated and objective assessment~\citep{leahy2000statistical, borghammer2009data, o2014voxel}. Gaussian mixture models (GMM) are one of most popular approaches~\citep{zhang1994, Guo+:14,Soffientini2016}, commonly assuming that voxel intensities belong to a mixture model with three components representing gray matter (GM), white matter (WM) and cerebro-spinal fluid (CSF). This is standard, for example, in the widely used software Statistical Parametric Mapping or SPM~\citep{SPM:05, Ashburner2012}.

Modeling the voxel intensities in the two scans as a bivariate mixture model with GM, WM and CSF components, the background adjustment procedure in \citet{Guo+:14} consisted of standardizing a multivariate Gaussian mixture model at each voxel. Large localized changes, presumably representing tumors, were then detected by performing a statistical test at each voxel with respect to an empirical null distribution. Following up on that work,~\citet{Qin+:} made a heuristic observation that the distribution of the standardized difference scores {\it across voxels} was close to standard normal, supporting the use of the standard normal as the null distribution for the voxelwise tests.
From a pure frequentist perspective, however, the p-value for {\it each voxel} should be computed from the distribution of the standardized score at {\it that voxel}. \citet{Qin+:} offered no investigation of whether and why the distribution of the standardized scores {\it at each voxel}, whose standardization is imperfect and different, should be standard normal.

Motivated by the problem above, this paper addresses the question of whether the distribution of the standardized score at each voxel is close to standard normal, considering a range of challenges from realistic PET image settings with lesions present, spatial heterogeneity, and tissue dependent background changes. This comparison is most important at the tails of the distribution (one-sided or two-sided), where inference usually occurs.

\begin{figure}[ht!]
	\centering 
	\includegraphics[trim=100 290 100 300, clip, width=\linewidth]{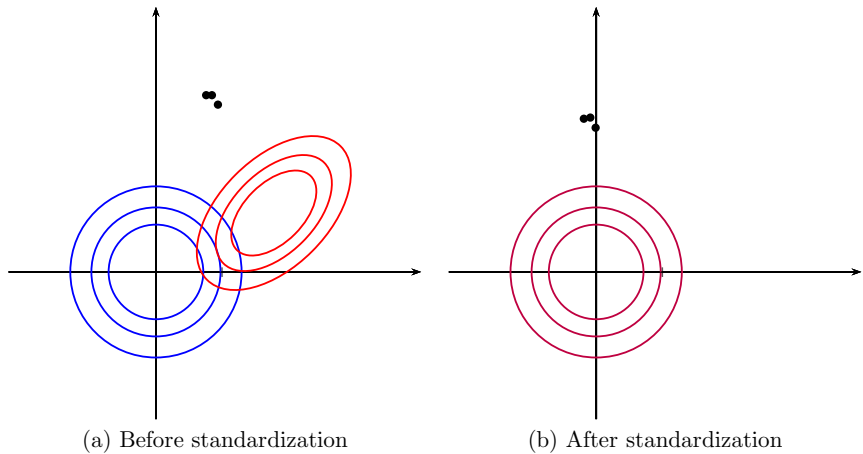} 
	\caption{Standardization of a two-component bivariate Gaussian mixture model with outliers. (a) density contours at (40\%, 50\%, 60\%)th quantiles (blue and red) of each of the two classes in the mixture (same parameters as in Figure~\ref{figure:setting} (b)). (b) density contour at (40\%, 50\%, 60\%)th quantiles (purple) of the standardized mixture, obtained using equation~\eqref{eq:Ti} with equal membership probabilities. The black dots in (a) and (b) represent the outliers before and after standardization, respectively.}  
	\label{figure:demo} 
\end{figure}

In its simplest form, the question can be formulated more abstractly as follows. Suppose we have observations from a multivariate Gaussian mixture model, among which there are some outliers we wish to detect. As an illustration, Figure \ref{figure:demo}(a) shows a bivariate mixture model with some outliers that do not belong to any of the classes but also do not constitute enough data to fit a third class. Is it possible to standardize the mixture so that the distribution will be centered around the origin with identity covariance? If so, the standardized mixture represents a null distribution against which the outlying observations can be detected using a statistical test, as illustrated in Figure \ref{figure:demo}(b). But will the tails of the distribution be close enough to standard normal so that p-values for detection are valid?

There is a rich literature on the estimation and application of finite mixture models~\citep{mclachlan2004finite}, leading to numerous research areas such as the expectation-maximization (EM) algorithm~\citep{Dempster+Laird+Rubin:77, redner1984mixture,gupta2011theory}, the estimation of unknown components~\citep{richardson1997bayesian,vlassis1999kurtosis,stephens2000bayesian, lo2001testing, figueiredo2002unsupervised}, and mixture of general distributions other than Gaussian~\citep{peel2000robust, dasgupta2005learning, Hanson2006, lin2007finite, venturini2008gamma}, only to name a few. However, there is limited work on inference problems such as hypothesis testing in a mixture model context.

In the first half of this paper, we study the standardization of a Gaussian mixture model systematically in various but simple ways. We show that, surprisingly, the tail distribution of the standardized scores is favorably close to standard normal in a wide range of scenarios while being conservative at the tails, making it suitable for statistical inference. Compared to the standardization method for background adjustment in \cite{Guo+:14} and \cite{Qin+:}, we consider several variations using both soft and hard assignment of the observations to latent classes. In the data application in Figure~\ref{fig:motivation}, the analysis based on the model-based standardized differences proposed in this paper is successful (Row 2 and Column 3) as the background difference is now randomly distributed around zero and the lesion change is clearly visible; see Section~\ref{section:data.application} for more details.
The distributions of the corresponding standardized scores are evaluated here theoretically, numerically and via simulations. Theoretically, it is shown that the standardized scores are indeed close to standard normal under a variety of extreme parameter settings. In non-extreme parameter settings, it is shown numerically that the soft assignment methods lead to conservative tail probabilities, making them valid for hypothesis testing purposes. It is also shown that the tail probabilities are not very sensitive to the class probabilities, which is an advantage as these are hard to estimate in practice. 

The second half of the paper addresses a practical challenge in analyzing PET images, namely, the standardization of Gaussian mixture models in spatially heterogeneous data. Brain images typically require using {\it spatial} Gaussian mixtures models, where the class membership probability varies spatially. Prior knowledge of the class membership probability at each voxel is available as templates from brain atlases (for example, SPM~\citep{SPM:05,Ashburner2012}). However, voxels that exhibit large changes in intensity such as lesions, which are precisely the ones we want to detect, are outliers with respect to the background and do not conform to the standard brain tissue templates. A robust estimation algorithm is thus needed to properly estimate the background without being affected by these outliers. 

Our proposed spatial GMM fitting procedure extends the standard spatial GMM in two ways. First, we use the probability maps produced from the baseline scan (Scan 1) by the SPM software, but we update the probability maps to include both scans 1 and 2 as bivariate measurements. Our model thus extends the univariate model of~\cite{SPM:05} implemented in SPM to multivariate cases, allowing multiple scans for the same subject. The usage of bivariate GMM adjusts the background of multiple scans simultaneously, considering correlations between scans from the same subject.

Furthermore, to estimate the mixture model parameters and probability maps, we propose a robust EM algorithm. 
The proposed algorithm uses a robust $M$-estimation step for updating the mean and covariance parameters, which is essential for the application since there are outliers such as tumor lesions present in the observation. The proposed robust EM algorithm is more objective and reliable than the estimation approaches in~\cite{Guo+:14} and~\cite{Qin+:}. In~\cite{Guo+:14}, the background parameters were estimated from the data in manually selected healthy slices, but this was shown in \cite{Qin+:} to be unstable. \cite{Qin+:} used the entire brain volume but iterated the background estimation with the detection of outliers in an alternate fashion. Instead, our robust M-step is directly incorporated into the EM algorithm, and is based on statistical theory of robust estimation which is shown in our simulations to give remarkably accurate results. It is worth mentioning that another option for robust estimation may be to use a GMM with extra components for the outliers. However, a prior template map would not be available for the outliers because lesions do not occur at predictable spatial locations. Moreover, prior templates are obtained from healthy subjects who have no lesions present at all.

Both univariate and bivariate simulations are used to evaluate the tail probabilities after standardization of the Gaussian mixture when all the model parameters are estimated via the proposed robust EM algorithm. We show that the proposed robust EM algorithm is accurate without outliers but robust when outliers are present, and show that the obtained standardized scores from background adjustment methods with soft assignment have tail probabilities that are either very close to or more conservative than standard normal. In addition, we observe that background adjustment methods based on estimated parameters surprisingly have slightly better performances than the one with true parameters. 

For brain images, the proposed approach can be applied directly by practitioners since the calculated scores and
resulting $p$-values provide immediate reference for inferences about the
change in disease status. We emphasize that although this paper is motivated by PET image analysis in
oncology, the concepts of standardization of multivariate Gaussian mixtures and robust EM
algorithm are generic, and can be used in other contexts. The Matlab toolbox {\tt RB-SGMM-BA} is available online in Matlab Central to implement the proposed methods. Supplementary materials contain all proofs and additional simulation studies.

\section{Standardization of multivariate Gaussian mixtures} 
\label{section:standardization} 

Let $y$ be a random $p$-dimensional vector ($p \geq 1$) generated by a Gaussian mixture model (GMM) with $K$ components: 
\begin{equation} \label{GMM} f(y) =
\sum_{k = 1}^{K} \pi_k \phi (y | \bmu_k, \bSigma_k), \end{equation} where each $\pi_k > 0$ and $ \sum_{k= 1}^K \pi_k = 1$, and $\phi(\cdot; \mu_k, \Sigma_k)$ is the probability
density function of a $p$-dimensional Gaussian variable with mean $\mu_k$ and covariance
matrix $\Sigma_k$ i.e. $N(\mu_k, \Sigma_k)$. Let the allocation variable $c \in \{1,
2,\ldots, K\}$ mark the class from which $y$ is generated, and let $s_k = \Ind(c
= k)$, where $\Ind(\cdot)$ is the indicator function. Then we can represent the model~\eqref{GMM} using the latent variables $c$'s as $P(c = k) =
\pi_k$ and $y | c \sim N(\bmu_{c}, \bSigma_{c})$, or
equivalently using the indicators $s_k$ and ${\bf{s}} = \{s_{1}, \ldots, s_{K}\}$: \begin{equation} \label{eq:SGMM}
{\bf{s}} \sim \mathrm{Multinomial}(1;
\pi_{1}, \ldots, \pi_{K}), \quad y | {\bf{s}} \sim N\left(\sum_{k = 1}^K s_k
\bmu_k, \sum_{k = 1}^K s_k \bSigma_k\right). \end{equation}
\subsection{Standardization methods}

Let $\phi_k(\cdot) = \phi(\cdot|\bmu_k, \bSigma_k)$. Then the posterior probability
$w_k = P(c = k | y) = P(s_k = 1 | y)$ that $y$ belongs to the $k$th class
is given by \begin{equation} \label{eq:w.general} w_k = \frac{\pi_k
	\phi_k(y)}{\sum_{k = 1}^K \pi_k \phi_k(y)} \end{equation} according to the Bayes'
theorem. The posterior probability $w_k$ is often referred to in the literature as the
{\it membership weight} or {\it responsibility} of class $k$ for $y$. The following two
methods are commonly used in practice to recover the latent labels $s_k$'s:
\begin{itemize} 
	\item Hard assignment: assign $\ts_k = 1$ if $k = \underset{k'}{\arg
		\max} \; {{w}_{ik'}}$; otherwise, $\ts_k = 0$.
	\item Soft assignment: assign $\ts_k = w_k$. 
\end{itemize}

Based on the membership weights, we wish to adjust the mean and covariance of each
observation $y$ with the hope that its resulting distribution will be close to a
multivariate standard normal. To achieve this, consider the following representation of
model~\eqref{GMM}. Let $Z \sim N(0, I_p)$, where $I_p$ is the $p
\times p$ identity matrix. Conditional on $\bf{s}$, the observation $y$ is multivariate normal with mean $\sum_{k = 1}^K s_k \bmu_k$ and covariance $\sum_{k = 1}^K s_k \bSigma_k$, which is identically distributed as the
random variable
\begin{equation} \label{eq:y.form} 
Y = \left(\sum_{k = 1}^K s_k \Sigma_k^{1/2}\right)Z + \sum_{k = 1}^K s_k \mu_k, 
\end{equation} 
where $\Sigma^{1/2}$ is the principal square root matrix of a positive definite matrix $\Sigma$ such that $\Sigma^{1/2}$ is positive definite and $\Sigma^{1/2} \Sigma^{1/2} = \Sigma$.  Since the $s_k$ are 0-1 indicators, we can rewrite \eqref{eq:y.form} as $Z = \left(\sum_{k = 1}^K s_k \Sigma_k^{-1/2}\right) \left(y - \sum_{k = 1}^K s_k \mu_k\right). $ If the latent labels $s_k$'s were known, this equation would provide an exact linear transformation to a multivariate standard normal distribution. Since the latent labels $s_k$'s are not known, we use the following plug-in transformation via the estimated labels $\ts_k$: 
\begin{equation} \label{eq:Ti} 
T^{(1)} = \left(\sum_{k = 1}^K \ts_k \Sigma_k^{-1/2}\right) \left(y - \sum_{k = 1}^K \ts_k \mu_k\right). 
\end{equation} 
This will be our transformation of choice in Section~\ref{sec:repara.theory} to Section~\ref{sec:oracle}. In the same spirit, one
may also consider to calculate the combination of covariance matrices first and then
invert, leading to the transformation \begin{equation} \label{eq:Ti.type2} T^{(2)} =
\left(\sum_{k = 1}^K \ts_k \Sigma_k\right)^{-1/2} \left(y - \sum_{k = 1}^K \ts_k \mu_k\right).
\end{equation} Other similar expressions are also possible. In particular,~\cite{Guo+:14}
proposed to use the marginal covariance of $Y$, i.e., \begin{equation}
\label{eq:Ti.Guo} T^{(3)} = \tSigma^{-1/2} (y - \tmu), \end{equation} where
$\tmu = \sum_{k = 1}^K \ts_k \bmu_k$ and $\tSigma =
\sum_{k = 1}^K \ts_k [\bSigma_k + (\bmu_k - \tmu) (\bmu_k - \tmu)^T]. $

The three transformations~\eqref{eq:Ti},~\eqref{eq:Ti.type2} and~\eqref{eq:Ti.Guo} yield
exact multivariate standard normal if the latent labels $\ts_k$ are
replaced by the true values $s_k$. They are equivalent if hard assignment is
used in the estimation of latent labels, but not so if soft assignment is used. For
this reason, whenever we need to distinguish between soft and hard assignment in what follows, we use the generic notations $T_S^{(1)}, T_S^{(2)}$ and
$T_S^{(3)}$ respectively for the three transformations with the soft assignment, and use
$T_H$ for the transformation with the hard assignment.



The goal of the rest of this section is to determine parameter scenarios under which the
distributions of the standardized scores $T^{(1)}$, $T^{(2)}$ and $T^{(3)}$ are close to multivariate standard normal.
We focus on the transformation $T^{(1)}$ because it is the easiest to analyze theoretically. This is sufficient because, as we shall see later in the simulations, the other two transformations $T^{(2)}$ and $T^{(3)}$ perform similarly.

Throughout this section, we consider the situation $K = 2$ for simplicity, although the same ideas apply to a higher number of classes.

\subsection{Reparametrization of the standardized scores} 
\label{sec:repara.theory} 

When $K = 2$, the standardized
score $T^{(1)}$ in equation~\eqref{eq:Ti} can be written as \begin{align}
 T^{(1)}  & = (\ts_{1} \bSigma_1^{-1/2} + \ts_{2} \bSigma_2^{-1/2})(Y
- \ts_{1} \bmu_1 - \ts_{2} \bmu_2) \\ & = 	\begin{cases} 		(\ts_{1} I + \ts_{2} \tau)(Z
+ \ts_2 \Delta_1) & s_1 = 1; \\ 		(\ts_{1} \tau^{-1}  + \ts_{2} I)(Z - \ts_1 \tau
\Delta_1) & s_1 = 0, \\ 		\end{cases}  \end{align} where $\Delta_1 =
\Sigma_1^{-1/2}(\mu_1 - \mu_2)$ and $\tau = \Sigma_2^{-1/2} \Sigma_1^{1/2}$. Let the
ratio between likelihoods of the two components be $r(y) = 2 \log(\phi_1(y)/\phi_2(y))$
and let $\pi_0 = 2 \log(\pi_2/\pi_1)$. Then the posterior probabilities
in~\eqref{eq:w.general} become \begin{equation} \label{eq:w.form.2} w_1 = \frac{1}{1 +
	\exp(-\frac{1}{2}(r(Y) - \pi_0))};  \quad w_2 = \frac{1}{1 + \exp(\frac{1}{2}(r(Y) -
	\pi_0))}. \end{equation} Noting that $|\tau| = |\Sigma_2|^{-1/2} |\Sigma_1|^{1/2}$, i.e.
$\log|\tau| = -(\log|\Sigma_2| - \log|\Sigma_1|)/2$, we have that \[r(Y)  = -2 \log|\tau|
+ 	\begin{cases} 	 (\tau Z + \tau \Delta_1)^T (\tau Z + \tau \Delta_1) - Z^T Z & s_1 =
1; \\ Z^T Z - (\tau^{-1}Z - \Delta_1)^T (\tau^{-1}Z - \Delta_1) & s_1 = 0,\\ 	
\end{cases}\] or equivalently, \begin{equation} \label{eq:r.multivariate} \begin{split}
r(Y)  = -2 \log|\tau| + [(\tau Z + \tau \Delta_1)^T (\tau Z + \tau \Delta_1) - Z^T Z]
\cdot \Ind(s_1 = 1) \\ + [Z^T Z - (\tau^{-1}Z - \Delta_1)^T (\tau^{-1}Z - \Delta_1)]
\cdot \Ind(s_1 = 0). \end{split} \end{equation}

For hard assignment, $\ts_1 = \Ind(r(Y) > \pi_0) = \Ind(w_1 > 1/2)$ according to
equation~\eqref{eq:w.form.2}, and $\ts_2 = 1 - \ts_1$. We thus can represent $T_H$ as follows: 
$T_H = 	Z$ if $s_1 = \ts_1$;  $T_H = 
\tau^{-1}Z - \Delta_1$ if $\ts_1 = 1, s_1 = 0$; and $T_H = \tau Z + \tau \Delta_1$ if $\ts_1 = 0,
s_1 = 1, $
i.e. \begin{equation}
\label{eq:t.z.multivariate} T_H = Z \cdot \Ind(\ts_1 = s_1) + (\tau^{-1}Z - \Delta_1)
\cdot \Ind(\ts_1 > s_1) + 	 (\tau Z + \tau \Delta_1) \cdot \Ind(\ts_1 < s_1).
\end{equation} In contrast, the soft assignment estimates the labels according to the
posterior probabilities, i.e. $\ts_1 = w_1$ and $\ts_2 = w_2$.

The full model depends on the parameter space $\Theta = \{(\mu_1, \mu_2, \Sigma_1,
\Sigma_2, \pi_1, \pi_2): \mu_1, \mu_2 \in \mathbb{R}^p, \Sigma_1, \Sigma_2 \in
\mathbb{R}^{p \times p}, \pi_1, \pi_2 \in [0, 1], \pi_1 + \pi_2 = 1\}$ of dimension $2p +
2p(p+1)/2 + 1 = p^2 + 3p +1$. However, the transformation $T^{(1)}$ in~\eqref{eq:Ti} only
depends on the full parameter space through a lower dimensional subset $(\Delta_1, \tau,
\pi_0)$ of dimension $p + p(p+1)/2 + 1 = (p^2 + 3p + 2)/2$. This
appealing property does not generally hold for the other two transformations $T^{(2)}$ and $T^{(3)}$ in~\eqref{eq:Ti.type2} and~\eqref{eq:Ti.Guo}. For example, since $\Sigma_1$ and $\Sigma_2$ are not always simultaneously {diagonalizable}, the term $\left( \ts_1 \Sigma_1 + \ts_2 \Sigma_2 \right)^{-1/2}$ in the transformation $T_S^{(2)}$ cannot decompose into the simpler form as it is for $T^{(1)}$.

\subsection{Approximate normality}

While our goal is to compare the tail probabilities between the standardized score $T^{(1)}$
and a standard multivariate normal $Z$, it can be seen that the former approximates the
latter over the entire domain under several extreme parameter scenarios. 

Let 
$\Theta_0 = \{\mu_1 = \mu_2, \Sigma_1 = \Sigma_2\} \cup \{\pi_1 = 1 \text{ or } \pi_2 =
1\} = \{\Delta_1 = 0, \tau = I_p\} \cup \{|\pi_0| = +\infty\}$. 
The set $\Theta_0$ trivially leads to perfect standardization so that $T^{(1)} = Z$, as the mixture model becomes one single component. This holds for any values of the parameters inside $\Theta_0$, so identifiability is not important in this case.
The following theorem states that approximate normality also holds when the parameters
are either close to or far from the set $\Theta_0$. Let $\|\cdot\|_2$ denote the
Euclidean norm in the case of a vector or the Frobenius norm in the case of a matrix.

\begin{theorem} \label{th:asy.multivariate} For both hard and soft assignments, if
	$(\Delta_1, \tau, \pi_0) \in \Theta_0^c$ are bounded, then we have the following
	results:
	
	1) For fixed $(\Delta_1, \tau)$, assume there is a sequence of parameters $(\pi_1,
	\pi_2)$ such that $\pi_1 \rightarrow 1$ or $\pi_2 \rightarrow 1$, then $T^{(1)}
	\overset{p}{\rightarrow} Z$.
	
	2) For fixed $(\pi_1, \pi_2)$, assume that there is a sequence of parameters $(\Delta_1,
	\tau)$ such as $\|\Delta_1\|_2 \rightarrow +\infty$ and $\|\tau\|_2$ is bounded, then
	$T^{(1)} \overset{a.s.}{\rightarrow} Z$.
	
	3) If there is a sequence of parameters $(\Delta_1, \tau)$ such as $\|\Delta_1\|_2
	\rightarrow 0$ and $\|\tau - I_2\|_2 \rightarrow 0$, then $T^{(1)}
	\overset{a.s.}{\rightarrow} Z$.
	
	4) If there is a sequence of parameters $(\Delta_1, \tau)$ and a vector $a$ such as $a^T
	\Delta_1 \rightarrow 0$ and $\|\tau - A\|_2 \rightarrow 0$ where $A a = a$, then $a^T
	T^{(1)} \overset{a.s.}{\rightarrow} a^T Z$. \end{theorem}

\begin{remark} \label{remark:as.pi} In Theorem~\ref{th:asy.multivariate} (1), we can obtain a.s. convergence for $T^{(1)}$ if $\pi_1$
	goes to 0 or 1 fast, for instance, if $\sum_n \pi_1^{(n)} \wedge \pi_2^{(n)}< \infty$ by the
	Borel–-Cantelli lemma. \end{remark}

Theorem~\ref{th:asy.multivariate} states that the standardized score will be
approximately standard multivariate normal (with convergence in probability or almost
surely) if one of the following occurs in a limiting sense: (1) one of the mixture components is dominant; (2) the mean vectors of the
two components are well separated relative to the covariance matrices; (3) the mean
vectors and covariance matrices of the two components are close to each other; (4) the
mean vectors and covariance matrices of the two components are close to each other along
a particular direction, in which case normality is obtained on the corresponding contrast
along that direction. 

\begin{remark} \label{remark:Theta0} Items (1), (2) and (4) in
	Theorem~\ref{th:asy.multivariate} discusses $\theta \in \Theta_0^c$ and describe
	scenarios where $\theta \rightarrow \theta_0$ for some $\theta_0 \in \Theta_0$.
	Therefore, the results in Theorem~\ref{th:asy.multivariate} ensure that the standardized
	score $T^{(1)}$ is continuous at $\Theta_0$. \end{remark}

{
	\begin{remark} \label{remark:equal.var}
		Note that it is a special case when the two covariance matrices are the same,  i.e. $\tau = I_p$. In this case, we have $\Delta_1 = \Delta_2$ and the results in Theorem~\ref{th:asy.multivariate} still hold.
	\end{remark}
}

\subsection{Explicit distribution in the case of hard assignment}

To further understand the behavior of the standardized scores in scenarios other than those considered in Theorem~\ref{th:asy.multivariate}, it is helpful to have an explicit formula for the distribution of $T^{(1)}$. This is also useful in the numerical evaluations in Section \ref{sec:oracle} below.

Let $F_X(\cdot)$ denote the CDF of a random variable $X$, $\Phi(\cdot)$ be the CDF of standard normal and $\Phi_p(\cdot)$ be the distribution function of a $p$-dimensional standard normal $Z$, i.e., $\Phi_p(A) = P(Z \in A)$ for a Borel set $A$.  Then  Theorem~\ref{th:cdf.multivariate} gives the exact CDF of a contrast on $T_H$.

\begin{theorem}
	\label{th:cdf.multivariate}
	Define the two maps $g$ and $h$ as $g: \mathbb{R}^p \rightarrow \mathbb{R}^p, g(x) = \tau x + \tau \Delta_1$ and $h: \mathbb{R}^p \rightarrow \mathbb{R}^p, h(x) = - x^T x + (\tau x + \tau \Delta_1)^T(\tau x + \tau \Delta_1) - 2 \log|\tau|$.
	For a given vector $a \in \mathbb{R}^p$ such that $\|a\|_2 = 1$ and any $t \in \mathbb{R}$, we have
	\begin{align}
	F_{a^T T_H}(t) - \Phi(t) =  \; & \pi_1 [\Phi_p(g^{-1}(R_2) \cap R_3) - \Phi_p(R_2 \cap R_3)] \\ & + \pi_2 [\Phi_p(g(R_2) \cap g(R_3^c)) - \Phi_p(R_2 \cap g(R_3^c))],
	\end{align}
	where $R_2 = \{x: a^T x < t \}$, $R_3 = \{x: h(x) < \pi_0\}$ and $g^{-1}$ is the inverse map of $g$.
\end{theorem}


The set $R_3$ in Theorem~\ref{th:cdf.multivariate} involves the quadratic form $h(x)$ in $x$ if $\tau \neq I_p$. When $p = 1$, we actually can solve this quadratic equation $h(x) = \pi_0$ explicitly as follows. Without loss of generality, we consider the case when $\tau \geq 1$.
\begin{lemma}
	\label{lemma:decision.region}
	Assume that the dimension $p$ = 1, $\theta \in \Theta_0^c$ and $\tau \geq 1$.
	Let $c_0 = (\tau^2 - 1) (\pi_0 +  2 \log \tau) + \Delta_2^2 $ and $c^+_0 = \max(c_0, 0)$.
	Then the set $R_3 = \{h(x) < \pi_0 \} = (a_{-}(\theta), a_+(\theta))$, where
	\begin{numcases}{\quad \quad (a_-(\theta), a_+(\theta)) = }
	\label{eq:a.pm.case1}
	\left(\frac{- \tau \Delta_2 - \sqrt{c_0^+}}{\tau^2 - 1} , \frac{- \tau \Delta_2 + \sqrt{c_0^+}}{\tau^2 - 1} \right)& \hspace{-15pt} \text{if} $\tau > 1$ \\
	\label{eq:a.upper}
	\left(-\infty, \frac{\pi_0}{2\Delta_2} - \frac{\Delta_2}{2}\right) & \hspace{-75pt} \text{if} $\tau = 1$ and $\Delta_2 > 0$ \\
	\label{eq:a.lower}
	\left(\frac{\pi_0}{2\Delta_2} - \frac{\Delta_2}{2}, +\infty\right) & \hspace{-75pt} if $\tau = 1$ and $\Delta_2 <0.$
	\end{numcases}
\end{lemma}

\begin{remark}
	The situation when $\tau = 1$ means that the two variances are the same.
	The definitions of $a_{\pm}(\theta)$ in~\eqref{eq:a.upper} and~\eqref{eq:a.lower} can actually be obtained by applying L'Hospital's Rule to the expressions in~\eqref{eq:a.pm.case1} when $\tau \rightarrow 1+$ (limit from above). This suggests a continuous behavior of the set $R_3$ when the variances of different components change from the heterogeneous to homogeneous.
\end{remark}
For simplicity of notation, we now drop the argument $\theta$ when referring to $a_-(\theta), a_+(\theta)$. The following Theorem~\ref{th:cdf.general} gives the explicit formula for the CDF of $T_H$, which is obtained by combining Theorem~\ref{th:cdf.multivariate} and Lemma~\ref{lemma:decision.region}.

\begin{theorem}
	\label{th:cdf.general}
	Assume that the dimension $p$ = 1, $\theta \in \Theta_0^c$ and $\tau \geq 1$.  We have
	\begin{align}
	\nonumber
	F_{T_H}&(t) = \Phi(t) + \pi_1 [\Phi((b_- \vee (t \wedge b_+))/\tau - \Delta_1) - \Phi(a_- \vee (t \wedge a_+))] \\
	\label{eq:cdf.tau.big}
	& + \pi_2 [\Phi( \tau (t \wedge a_-) + \Delta_2 ) + \Phi(\tau (t \vee a_+) + \Delta_2) - \Phi(t \wedge b_-) - \Phi(t \vee b_+)],
	\end{align}
	for any $t \in \mathbb{R}$, where $b_-= \tau a_- + \Delta_2$, $b_+ = \tau a_+ + \Delta_2$, and the operator $\wedge$ and $\vee$ are the min and max operator, i.e. for any $a, a' \in \bR$, $a \wedge a' = \min(a, a'), $ and $a \vee a' = \max(a, a'). $
\end{theorem}

\subsection{Numerical evaluation of tail probabilities} \label{sec:oracle}

From Theorem \ref{th:asy.multivariate} we learn that the scenarios in which the
standardized score may be far from standard multivariate normal are those where the
mixture components are not close to each other nor far from each other. In this section,
we study some of these scenarios numerically. We focus on tail probabilities, which
are most important for statistical testing.

Let the vector $a$ be the contrast of interest and let $\alpha$ be the size of the test.
If the decision threshold is set as $t_{\alpha} = \Phi^{-1}(\alpha/2)$, according to the
standard normal distribution, then we are interested in whether the true size of the test
is below or above $\alpha$. The true size of a two-sided hypothesis test is $P(|a^T
T^{(1)}| \geq t_{\alpha}) =  F_{a^TT}(t_{\alpha} ) + 1 - F_{a^TT}(-t_{\alpha})$. Let
$R(\alpha) = P(|a^T T^{(1)}| \geq t_{\alpha}) /\alpha$ be the ratio between the true size
of the test and the size based on the standard normal distribution:  $R(\alpha) = 1$
means that the size is exact; $R(\alpha) < 1$ means that the test is conservative;
$R(\alpha) > 1$ means that the test is invalid.

In this section we study the relative size $R(\alpha)$ for various combinations of model
parameters $(\Delta_1, \tau, \pi_0)$. We may also write $R(\alpha; \Delta_1, \tau,
\pi_0)$ to emphasize the dependence of this ratio on the parameters $(\Delta_1, \tau,
\pi_0)$. {As a reduction on the set of parameters to be evaluated, the following Lemma~\ref{lemma:size.symmetric} states that the relative size
	$R(\alpha; \Delta_1, \tau, \pi_0)$ is symmetric with respect to the sign of $\Delta_1$; thus changes in $\Delta_1$ need only be evaluated in one direction.}
\begin{lemma} \label{lemma:size.symmetric} 
	For any $t \in \mathbb{R}$, we have $$P(|a^T T^{(1)}| \geq t; \Delta_1, \tau, \pi_0) = P(|a^T T^{(1)}| \geq t; -\Delta_1, \tau,
	\pi_0).$$
\end{lemma}

A theoretical expression for $R(\alpha)$ for soft assignment is difficult to
obtain. Instead, we use numerical evaluation to investigate the tail probabilities via Monte Carlo simulation. For
simplicity, we focus on the bivariate case $p = 2$. Similar to the data analysis, we use
$a^T = (1, -1)/\sqrt{2}$ as the contrast of interest, measuring the normalized difference
between the two coordinates of $T^{(1)}$. Since the distribution of $a^T T^{(1)}$ depends
only on the parameters $(\Delta_1, \tau, \pi_0)$, we assume without loss of generality
$\mu_2 = 0$ and $\Sigma_2 = I$ 
{when generating the data $Y$. We use the following parameter settings to investigate the tail probabilities:
	\renewcommand{\arraystretch}{0.5} 
	\begin{align} & \text{Case 1:} \quad\; \;\mu_1 = \kappa_1 \begin{pmatrix}1\\1\end{pmatrix}, \quad\quad\;\;\Sigma_1 =
	\kappa_2 \left(\begin{array}{cc} 1 & \rho  \\ \rho & 1  \end{array}\right); \quad \mu_2 =
	0, \quad \Sigma_2 = I. \\
	& \text{Case 2:} \quad \mu_1 = \kappa_1
	\begin{pmatrix}2/\sqrt{5} \\ 1/\sqrt{5} \end{pmatrix}, \quad \Sigma_1 = \kappa_2
	\left(\begin{array}{cc} 1 & \rho  \\ \rho & 1  \end{array}\right); \quad \mu_2 = 0, \quad
	\Sigma_2 = I. 
	\end{align}
}
In the above setting, the parameter $\rho$
controls the correlation of the bivariate normal in the first class, while $\kappa_1$ and $\kappa_2$ control how the two classes differ from each other. 

\begin{figure}[ht!] 
	\centering \includegraphics[trim=150 310
	150 320, clip, width=0.98\linewidth]{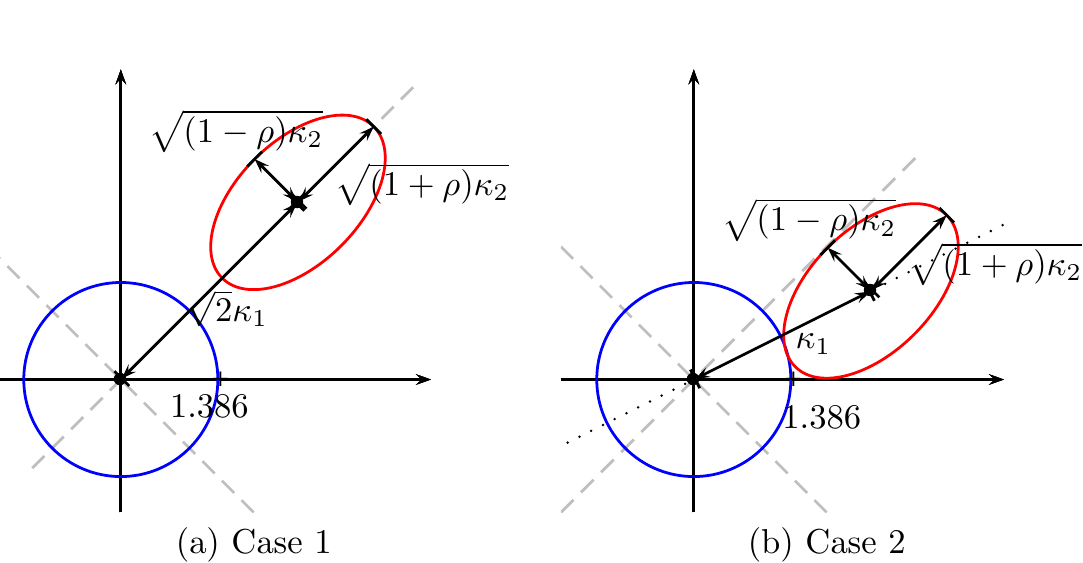} \caption{50\% density contours of the
		two bivariate mixture components used for numerical evaluation of tail probabilities of
		the standardized score when $\rho$ = 0.5. The red ellipse represents the density contour
		of the first class, while the blue circle centered at the origin represents the density
		contour of the second class. The two plots use $\kappa_2 = 1.923$ (both cases), $\kappa_1 = 2.773$ (Case 1) and $\kappa_1 = 3.100$ (Case 2) for demonstration.} \label{figure:setting} 
\end{figure}

Figure~\ref{figure:setting} illustrate schematically the density contours of the two cases when $\rho = 0.5$. In Case 1, the vector $\Delta_1$ is orthogonal to the contrast vector $a$, while in Case 2 it is not.

Figures~\ref{fig:case1} and~\ref{fig:case2} below show the relative size $R(\alpha)$ for $\alpha = 0.001$ in the form of heatmaps when $\rho \in \{0, 0.5\}$, $\kappa_2 = (0.1, 1, 10)$ and $\kappa_1$ and $\pi_1$ vary continuously. Note that because of Lemma \ref{lemma:size.symmetric}, it is sufficient to consider $\kappa_1 \ge 0$. In these figures, purple is ideal, indicating a relative size of about 1. Blue indicates a conservative relative size smaller than 1, while red indicates an invalid relative size greater than 1. Black indicates a relative size greater than 2, which may be considered unacceptable.

{
	To see how these parameter combinations fall into the parameter settings in Theorem~\ref{th:asy.multivariate}, we compute
	\begin{align}
	\tau = \Sigma_2^{-1/2} \Sigma_1^{1/2} = \frac{\sqrt{\kappa_2}}{2} \left(\begin{array}{cc} \sqrt{1 + \rho} + \sqrt{1 - \rho} & \sqrt{1 + \rho} - \sqrt{1 - \rho}  \\ \sqrt{1 + \rho} - \sqrt{1 - \rho} & \sqrt{1 + \rho} + \sqrt{1 - \rho}\end{array} \right)
	\end{align} 
	and $\Delta_1 = \Sigma_1^{-1/2}(\mu_1 - \mu_2)$ as: 
	\begin{align} 
	\text{Case 1: } & 
	\Delta_1 = \Sigma_1^{-1/2}(\mu_1 - \mu_2) = \frac{\kappa_1}{\sqrt{\kappa_2}(1 - \rho)} \begin{pmatrix}1\\1\end{pmatrix} \\
	\text{Case 2: } &
	\Delta_1 = \frac{\kappa_1}{2\sqrt{5\kappa_2}}\begin{pmatrix}  -1/\sqrt{1 - \rho} + 3/\sqrt{1 + \rho} \\ 1/\sqrt{1 - \rho} + 3/\sqrt{1 + \rho} \end{pmatrix}. 
	\end{align} 
	
	As predicted by Theorem~\ref{th:asy.multivariate}, the relative sizes are all close to 1 for extreme values of $\pi_1$ close to 0 or 1 in all subplots (satisfying conditions in Theorem~\ref{th:asy.multivariate} (1)), large $\kappa_1$ with respect to $\kappa_2$ (satisfying conditions in (2) as $\|\Delta_1\|_2$ is large), and $\kappa_2 = 1$ and $\rho = 0$ (satisfying conditions in (4) as $\tau = I_2$ and $a^T \Delta_1 = 0$).  
}

{When comparing hard and soft assignment,} Figure~\ref{fig:case1} shows that, in Case 1, hard assignment leads to unacceptable relative sizes regardless of the mixture proportion $\pi_1$, especially when $\kappa_1$ is small. Soft assignment, however, corrects this and leads to conservative relative sizes in all the cases shown. A similar pattern is observed in Figure~\ref{fig:case2} for Case 2, except that soft assignment leads to relative sizes slightly greater than 1 for some parameter combinations. More combination of parameters studied, including evaluation of one-sided tail probabilities, lead to similar observations and thus are not included due to space limitations.

\begin{figure} 
	\vspace*{-0.15in} \centering 
	\begin{tabular}{*{6}{c}} & & $\kappa_2 = 0.1$
		& $\kappa_2 = 1$ & $\kappa_2 = 10$ &  \\ \begin{sideways} \rule[0pt]{0.35in}{0pt} Hard
			($\rho = 0$) \end{sideways} &  \begin{sideways} \rule[0pt]{0.6in}{0pt} $\pi_1$
		\end{sideways} 
		& \includegraphics[trim=25 40 100
		30,clip,width=0.25\textwidth]{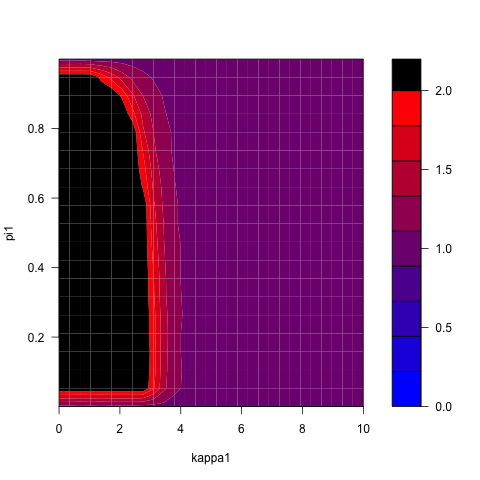} &
		\includegraphics[trim=25 40 100
		30,clip,width=0.25\textwidth]{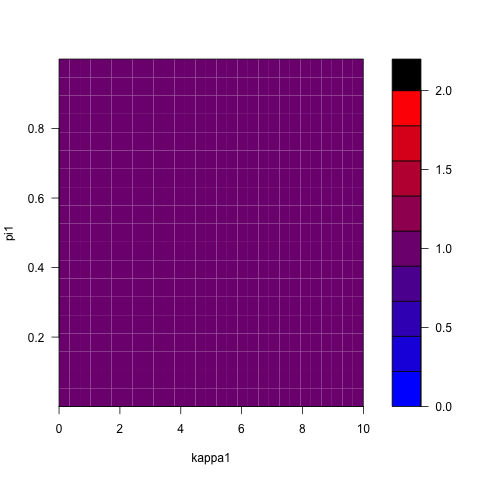} &
		\includegraphics[trim=25 40 100
		30,clip,width=0.25\textwidth]{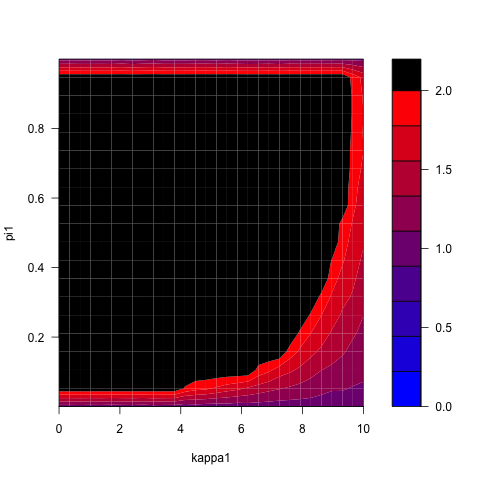} &
		\includegraphics[trim=395 50 25 30,clip, height =
		0.21\textheight]{{plot.dir/TailProb-update/case1.rho.0/two-tail.hard.10.0}.png} \\
		
		
		\begin{sideways} \rule[0pt]{0.35in}{0pt} Soft ($\rho = 0$) \end{sideways} &
		\begin{sideways} \rule[0pt]{0.6in}{0pt} $\pi_1$ \end{sideways} & \includegraphics[trim=25
		40 100
		30,clip,width=0.25\textwidth]{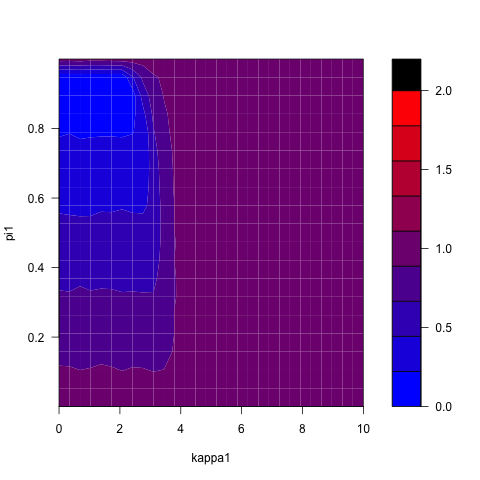} &
		\includegraphics[trim=25 40 100
		30,clip,width=0.25\textwidth]{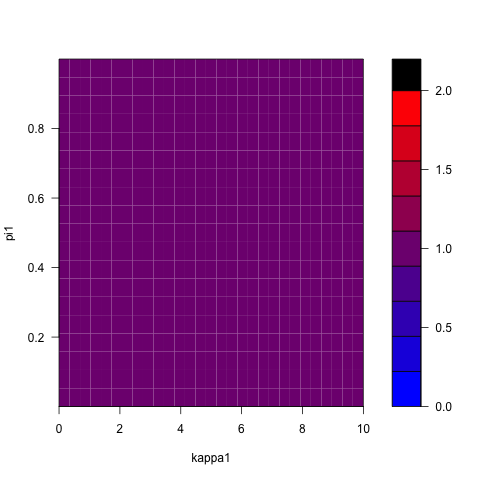} &
		\includegraphics[trim=25 40 100
		30,clip,width=0.25\textwidth]{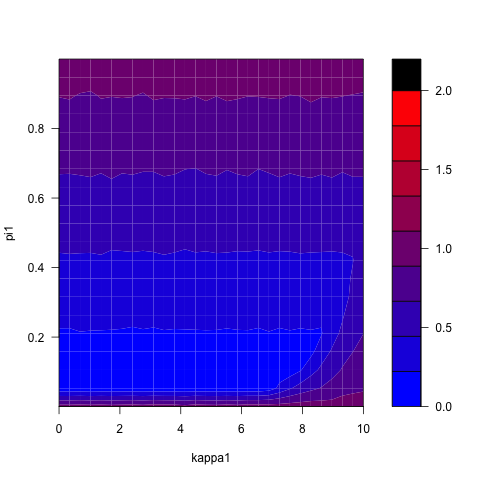} &
		\includegraphics[trim=395 50 25 30,clip,height =
		0.21\textheight]{{plot.dir/TailProb-update/case1.rho.0/two-tail.soft.10.0}.png} \\

		\begin{sideways} \rule[0pt]{0.35in}{0pt} Hard ($\rho = 0.5$) \end{sideways} &
		\begin{sideways} \rule[0pt]{0.6in}{0pt} $\pi_1$ \end{sideways} & \includegraphics[trim=25
		40 100
		30,clip,width=0.25\textwidth]{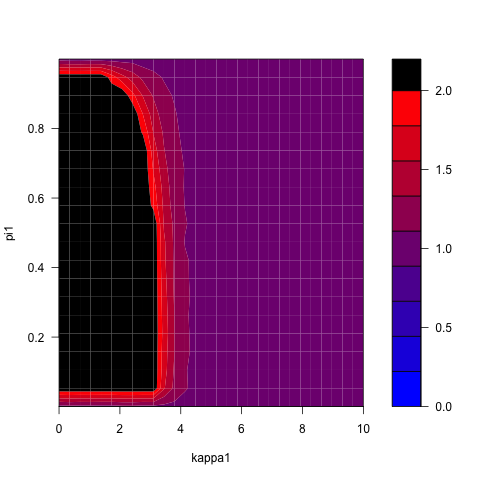} &
		\includegraphics[trim=25 40 100
		30,clip,width=0.25\textwidth]{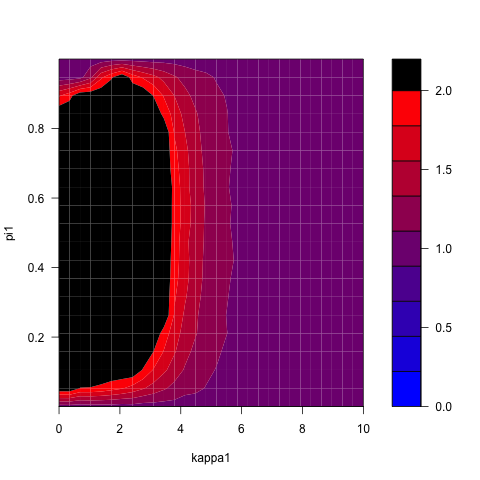} &
		\includegraphics[trim=25 40 100
		30,clip,width=0.25\textwidth]{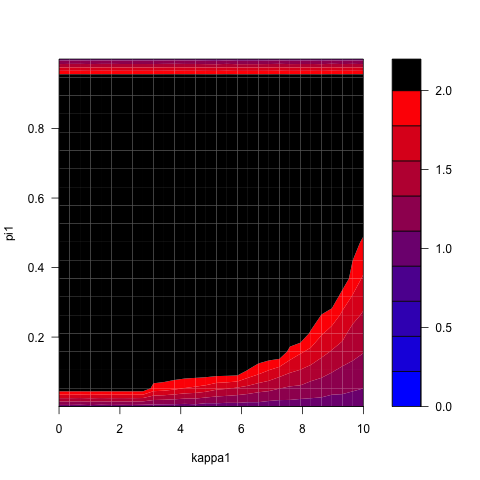} &
		\includegraphics[trim=395 50 25 30,clip,height =
		0.21\textheight]{{plot.dir/TailProb-update/case1.rho.5/two-tail.hard.10.0}.png} \\ 
		\begin{sideways} \rule[0pt]{0.35in}{0pt} Soft ($\rho = 0.5$) \end{sideways} &
		\begin{sideways} \rule[0pt]{0.6in}{0pt} $\pi_1$ \end{sideways} & \includegraphics[trim=25
		40 100
		30,clip,width=0.25\textwidth]{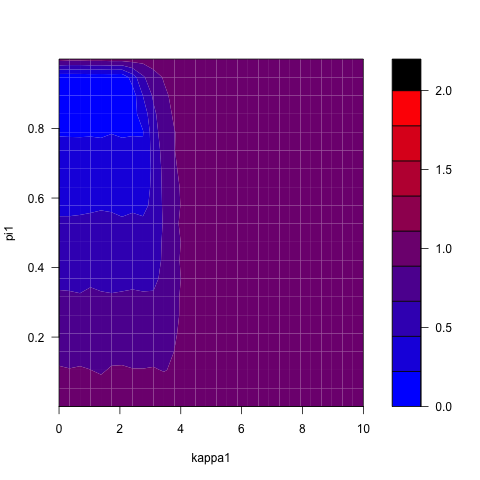} &
		\includegraphics[trim=25 40 100
		30,clip,width=0.25\textwidth]{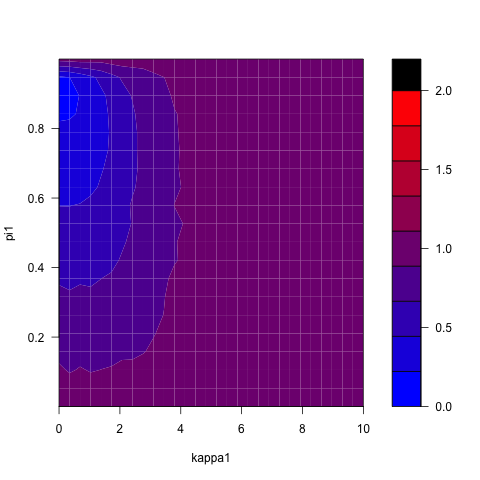} &
		\includegraphics[trim=25 40 100
		30,clip,width=0.25\textwidth]{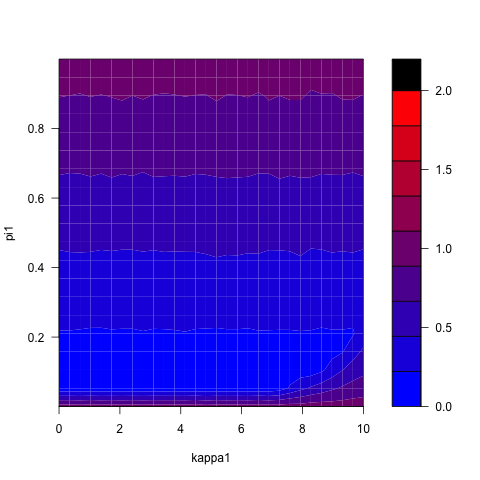} &
		\includegraphics[trim=395 50 25 30,clip,height =
		0.21\textheight]{{plot.dir/TailProb-update/case1.rho.5/two-tail.soft.10.0}.png} \\ & &
		$\kappa_1$ &  $\kappa_1$ &  $\kappa_1$ & \\ 
	\end{tabular} 
	\caption{Heatmaps of
		$R(\alpha)$ (two-side tails) where $\alpha = 0.001$ for Case 1. } \label{fig:case1}
\end{figure}

\begin{figure} \vspace*{-0.15in} \centering \begin{tabular}{*{6}{c}} & & $\kappa_2 = 0.1$
		& $\kappa_2 = 1$ & $\kappa_2 = 10$ &  \\ \begin{sideways} \rule[0pt]{0.35in}{0pt} Hard
			($\rho = 0$) \end{sideways} &  \begin{sideways} \rule[0pt]{0.6in}{0pt} $\pi_1$
		\end{sideways} & \includegraphics[trim=25 40 100
		30,clip,width=0.25\textwidth]{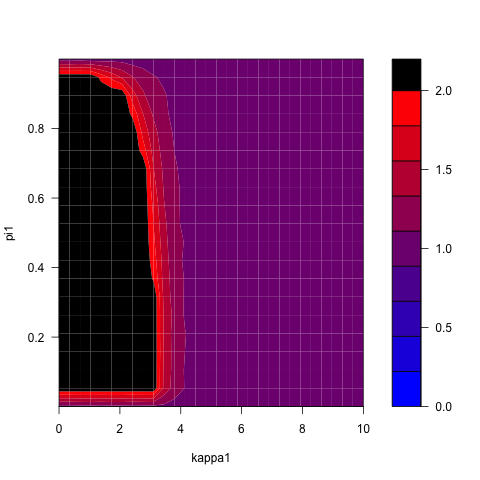} &
		\includegraphics[trim=25 40 100
		30,clip,width=0.25\textwidth]{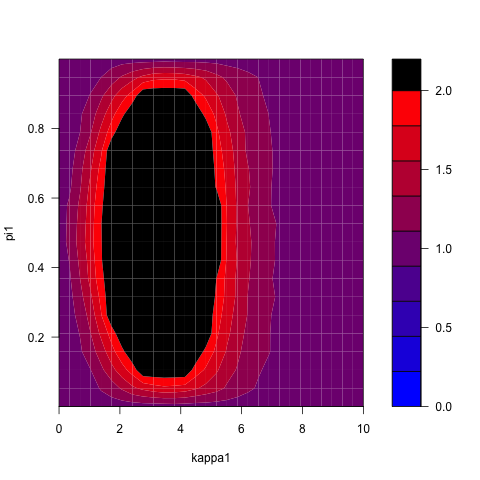} &
		\includegraphics[trim=25 40 100
		30,clip,width=0.25\textwidth]{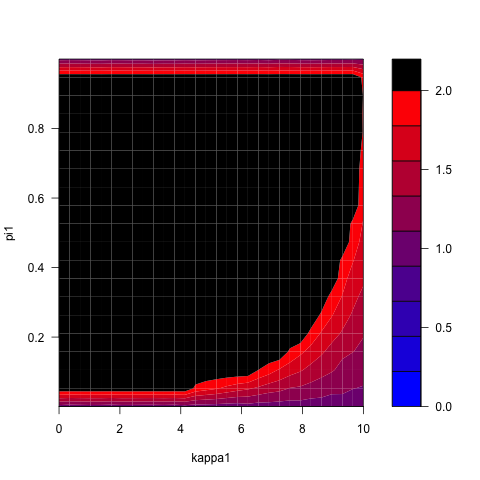} &
		\includegraphics[trim=395 50 25 30,clip,height =
		0.21\textheight]{{plot.dir/TailProb-update/case4.rho.0/two-tail.hard.10.0}.png} \\
		
		
		\begin{sideways} \rule[0pt]{0.35in}{0pt} Soft ($\rho = 0$)  \end{sideways} &
		\begin{sideways} \rule[0pt]{0.6in}{0pt} $\pi_1$ \end{sideways} & \includegraphics[trim=25
		40 100
		30,clip,width=0.25\textwidth]{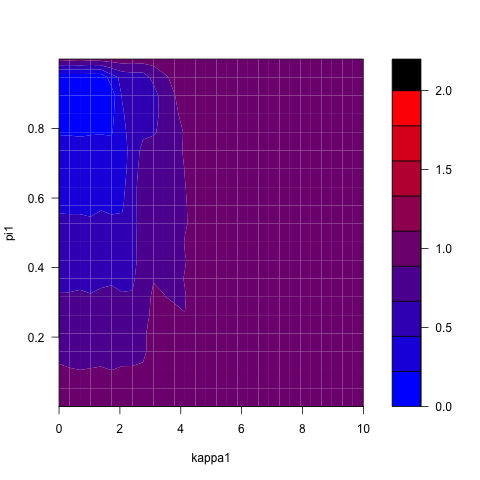} &
		\includegraphics[trim=25 40 100
		30,clip,width=0.25\textwidth]{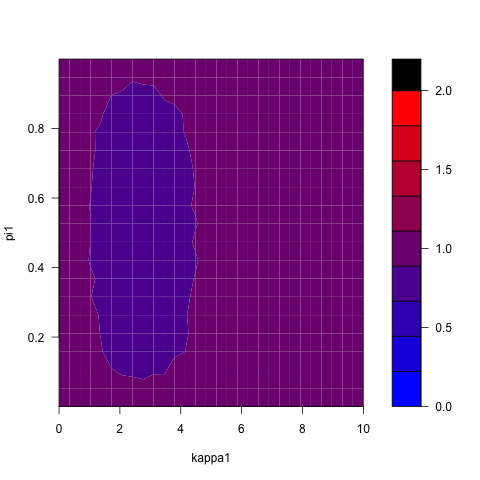} &
		\includegraphics[trim=25 40 100
		30,clip,width=0.25\textwidth]{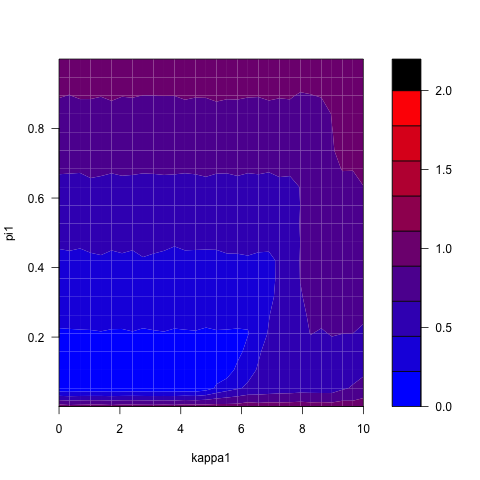} &
		\includegraphics[trim=395 50 25 30,clip,height =
		0.21\textheight]{{plot.dir/TailProb-update/case4.rho.0/two-tail.soft.10.0}.png} \\ & &
		$\kappa_1$ &  $\kappa_1$ &  $\kappa_1$ & \\
		
		\begin{sideways} \rule[0pt]{0.35in}{0pt} Hard ($\rho = 0.5$) \end{sideways} &
		\begin{sideways} \rule[0pt]{0.6in}{0pt} $\pi_1$ \end{sideways} & \includegraphics[trim=25
		40 100
		30,clip,width=0.25\textwidth]{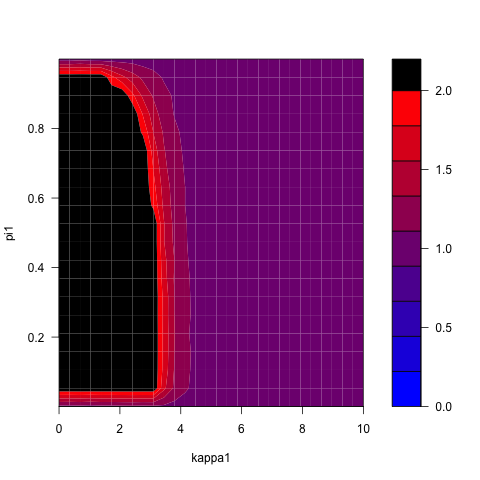} &
		\includegraphics[trim=25 40 100
		30,clip,width=0.25\textwidth]{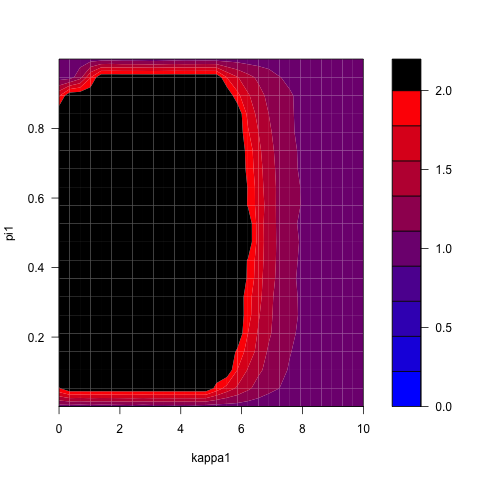} &
		\includegraphics[trim=25 40 100
		30,clip,width=0.25\textwidth]{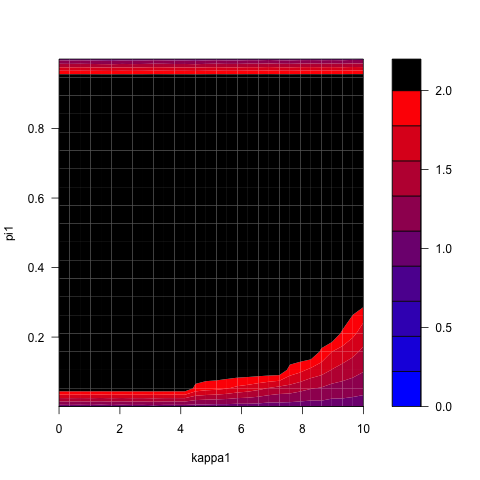} &
		\includegraphics[trim=395 50 25 30,clip,height =
		0.21\textheight]{{plot.dir/TailProb-update/case4.rho.5/two-tail.hard.10.0}.png} \\ 
		\begin{sideways} \rule[0pt]{0.35in}{0pt} Soft ($\rho = 0.5$)  \end{sideways} &
		\begin{sideways} \rule[0pt]{0.6in}{0pt} $\pi_1$ \end{sideways} & \includegraphics[trim=25
		40 100
		30,clip,width=0.25\textwidth]{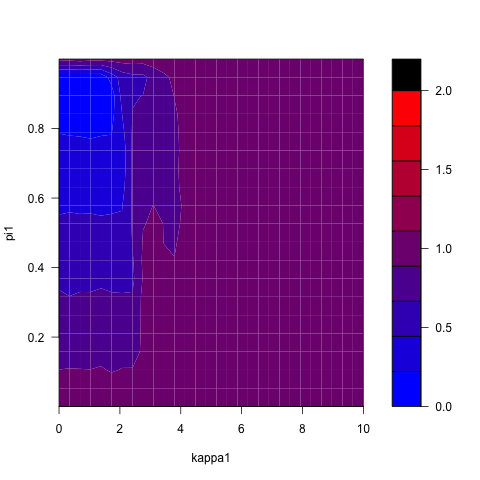} &
		\includegraphics[trim=25 40 100
		30,clip,width=0.25\textwidth]{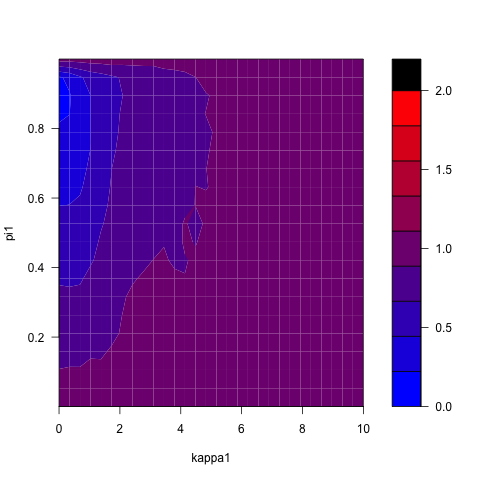} &
		\includegraphics[trim=25 40 100
		30,clip,width=0.25\textwidth]{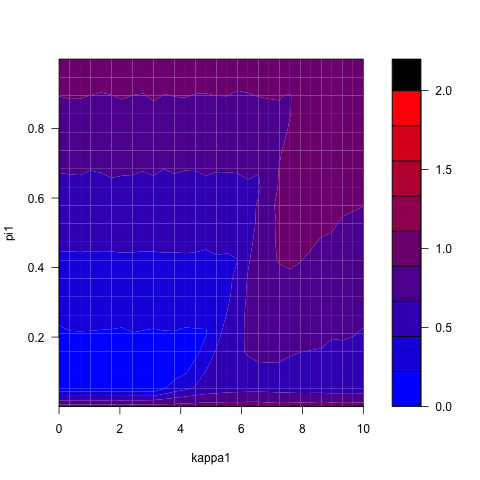} &
		\includegraphics[trim=395 50 25 30,clip,height =
		0.21\textheight]{{plot.dir/TailProb-update/case4.rho.5/two-tail.soft.10.0}.png} \\ & &
		$\kappa_1$ &  $\kappa_1$ &  $\kappa_1$ & \\ 
	\end{tabular} 
	\caption{Heatmaps of $R(\alpha)$ (two-side tails) where $\alpha = 0.001$ for Case 2.} \label{fig:case2}
\end{figure}

\section{Spatial and robust fitting of Gaussian mixture models} 
\label{sec:SGMM.EM}

To fit the PET data as a Gaussian mixture model requires two particular features. One is to incorporate a spatial component into the model, so that the prior probabilities vary according to a spatial pattern defined by the tissue types. These spatial patterns can be incorporated into the EM algorithm via pre-defined spatial templates. The second feature is to make the M step in the EM algorithm robust to outliers so that estimation of the background is not affected by tumor voxels. The tumor voxels cannot be modeled as a separate component in the mixture such as in because they are relatively few in volume and because they do not correspond to any particular spatial pattern known a-priori.

\subsection{Spatial Gaussian mixture models} 
\label{section:model} 

Most existing approaches about spatial GMM are based on the Markov random field on $\pi_{ik}$ such as the method of iterated conditional modes in~\citep{Besag1986a}, discussions in Chapter 13 of~\cite{mclachlan2004finite}, and many other developments along this line~\citep{Hebert1998, Nguyen2012a, Vehtari2012a}. However, in our applications template maps of the class membership probability are available from brain atlases~\citep{Ashburner2012}, which makes it more sensible to adopt an approach that utilizes the existing spatial templates. 

Let the observations be $\{y_{i}\}_{i =
	1}^n$, where $y_i$ is a $p$-dimensional vector ($p \geq 1$) at the $i$th location and $n$
is the total number of voxels. For 2-dimensional or higher dimensional images where the
location index has more than one direction, we can vectorize the location to have this
single index $i$. We assume that the $y_i$'s are independently generated by a GMM in~\eqref{GMM} but with a spatial 
mixture probabilities: \begin{equation} \label{SGMM} f(y_i) =
\sum_{k = 1}^{K} \pi_{ik} \phi (y_i | \bmu_k, \bSigma_k), \end{equation} where $ \sum_{k
	= 1}^K \pi_{ik} = 1$ for any $i$ and $\pi_{ik} > 0$. We refer to model~\eqref{SGMM} as a Spatial
Gaussian Mixture Model (SGMM). 

The traditional Gaussian Mixture Model (GMM) assumes that $\pi_{ik} = \pi_k$ for all $i$.
In contrast, SGMM considers the spatial information by using different mixture
probabilities at each location. In order to incorporate spatial information, we assume that the probability
maps are generated by some given prior probability maps $b_{ik}$: 
\begin{equation}
\label{eq:b2pi} 
\pi_{ik} = {\gamma_k b_{ik}} \left\{{\sum_{j = 1}^K \gamma_j b_{ij}} \right\}^{-1},
\end{equation} 
with the identifiability constraint $\sum_{k = 1}^K \gamma_k = 1$, which is also used by~\cite{SPM:05}. The
probability maps $b_{ik}$ are often referred to as templates. Spatial templates provide a
natural way to incorporate prior knowledge from previous studies about the probability of
each location $i$ belonging to each class. In the case of the brain, the templates
represent reference probability maps for brain tissue types like GM, WM and CSF. 
Such templates, constructed from segmentations of images of healthy subjects, are available within the software SPM. Other ways to introduce the probability map $\pi_{ik}$ are possible but may have more parameters to estimate, such as the one based on Markov random field~\citep{sanjay1998bayesian, chen2001markov,Soffientini2016} 

The usage of templates also reduces the dimension of the parameters in the model, which
otherwise would be nearly proportional to the number of voxels $n$. Under the
constraint~\eqref{eq:b2pi}, the parameters of model~\eqref{eq:SGMM} become $\theta =
(\mu_k, \sigma_k, \gamma_k)_{k = 1}^K$, with dimension $K[p + p(p+1)/2] + (K-1)$. For
$p=2$ and $K=3$, the dimension is 17. It can be seen again that the traditional GMM is a
special case of SGMM by letting $b_{ik} = b_k$ for all $i$ in~\eqref{eq:b2pi}, i.e. the
traditional GMM uses constant template maps.

Given all parameters, the standardized scores at each location $i$ can be obtained using the methods introduced in Section~\ref{section:standardization}. 
In Section~\ref{section:EM}, we propose a robust expectation-maximization (EM) algorithm to obtain maximum likelihood estimators (MLE) in model~\eqref{SGMM}.  The tail probabilities and performances of the robust EM algorithm are investigated through simulation in Section~\ref{sec:simulation}. 

\subsection{Robust EM algorithm} 
\label{section:EM} 
The EM algorithm introduced by~\cite{Dempster+Laird+Rubin:77}
and its invariants are popular to obtain the maximum likelihood estimators (MLE) in a
mixture model; see~\cite{McLachlan2007} for a comprehensive treatment. Since the model of SGMM~\eqref{GMM} uses spatial mixture probabilities via
the parameterization~\eqref{eq:b2pi}, the conventional EM does not apply directly and modifications are needed to make the EM algorithm work, which results in a
generalized EM algorithm. Furthermore, voxels that exhibit large changes in intensity such as lesions,
which are precisely the ones we want to detect, are outliers with respect to the
background. A robust estimation procedure is thus necessary to properly estimate the background without being affected by these outliers. As robust estimators, $M$-estimators in mixture models have been well developed in the literature~\citep{Campbell:84, McL+Bas:88,Maronna+Martin+Yohai:06}. The basic idea is to reduce the weights of abnormal observations while keeping close to full weights for the others. We shall use $M$-estimates in the M-step of an EM algorithm to achieve robust fitting of SGMM. 

Let $\bs{y} = (y_1, \ldots, y_n)$, then the log-likelihood function is given by
\begin{equation} \label{logL} \ell(\theta; \bs{y}) =\sum_{i=1}^n \log \left\{\sum_{k =
	1}^{K} \pi_{ik} \phi_k(y_i | \bmu_k, \bSigma_k)\right\}. \end{equation} If the latent
labels $\bs{s} = (s_{11},\ldots,s_{1K}; \ldots; s_{n1},\ldots,s_{nK})$ are observed,
the joint log-likelihood is 
\begin{equation}
\label{eq:logL.joint} 
\ell(\theta; \bs{y, s}) = \sum_{i=1}^n
\sum_{k=1}^K s_{ik} [\log(\pi_{ik}) + \log(\phi(y_i | \bmu_k, \bSigma_k))].
\end{equation} 
Given an intermediate estimate $\theta^t$, we first calculate the
conditional expectation of $\ell(\theta; \bs{y, s})$ in equation~\eqref{eq:logL.joint} denoted as
$Q(\theta |\theta^{t} )$ (the E-step), and then maximize this conditional expectation with robustness (the robust M-step). For the robustness step, we use the Mahalanobis distance between observations and the estimated mean vectors and covariance matrices to determine whether an observation is abnormal or not.
Given the distance, observations are weighted according to the weight function $u(s) =
\psi(s)/s$ where $\psi(s) = \min(s, k_1(p))$ is Huber's $\psi$-function~\citep{Huber:64, Maronna:76} with a tuning constant $k_1(p)$ depending on the dimension $p$. 
We next present the detailed generalized EM algorithm. 

\begin{itemize} \item E-step. Conditional Expectation on $\bs{y},
	\theta^{t}$: \begin{align} Q(\theta |\theta^{t} ) = &\sum_{i=1}^n \sum_{k=1}^K \E (s_{ik}
	| \bs{y}, \theta^{t}) [\log(\pi_{ik}) + \log(\phi(y_i | \bmu_k, \bSigma_k))] \\ = &
	\sum_{i=1}^n \sum_{k=1}^K w_{ik}^t [\log(\pi_{ik}) + \log(\phi(y_i | \bmu_k,
	\bSigma_k))], \end{align} where 
	\begin{equation} \label{Estep}
	w_{ik}^t = \E (s_{ik} | \bs{y}, \theta^{t})
	= \frac{\pi_{ik}^t \phi(y_i | \bmu_k^{t}, \bSigma_k^{t}) }{\sum_{k=1}^{K} \pi_{ik} ^t
		\phi(y_i | \bmu_k^{t}, \bSigma_k^{t})}, \quad\text{and}\quad\pi_{ik}^t =
	\frac{\gamma_k^t b_{ik}}{\sum_{k = 1}^K \gamma_k^t b_{ik}}. \end{equation}
	
	\item Update $\gamma_k$:  
	\begin{equation}
	\label{eq:update.gamma}
	\gamma_k^{t+1} = \sum_{i = 1}^n \left\{w_{ik}^t  \left({\sum_{i = 1}^n \frac{b_{ik}}{\sum_{j = 1}^k \gamma_j^t b_{ij}}}\right)^{-1}\right\}.
	\end{equation}
	\item Robust M-step. 	
	
	Update the mean vector: 	
	\begin{equation} 	
	\label{Mstep.Robust.mu} 	
	r_{ik}^{t, 1} = \sqrt{(y_i - \mu_k^{t})^T \Sigma_k^{-1} (y_i - \mu_k^{t })}; \quad  \bmu_k^{t + 1} =
	\frac{\sum_{i = 1}^n w_{ik}^{t} u(r_{ik}^{t, 1}) y_i }{ \sum_{i = 1}^n w_{ik}^{t}
		u(r_{ik}^{t, 1})}. 	
	\end{equation} 	
	
	Update the covariance matrix: 	
	\begin{align} 	
	r_{ik}^{t, 2} & = \sqrt{(y_i - \mu_k^{t +1})^T \Sigma_k^{-1} (y_i - \mu_k^{t + 1})}; \\ \bSigma_k^{t + 1} & = \frac{\sum_{i =
			1}^n w_{ik}^{t} u^2(r_{ik}^{t,2})(y_i - \bmu_k^{t + 1}) (y_i - \bmu_k^{t + 1})^T}{\sum_{i
			= 1}^n w_{ik}^{t} u^2(r_{ik}^{t,2})}. 	
	\end{align} 
\end{itemize}

The algorithm is terminated when the relative change of the log-likelihood
function in~\eqref{logL} {with respect to the previous iteration becomes smaller than a
	given tolerance or a maximum number of iterations is reached}. 	
The tuning parameter $k_1(p)$ in the robust M-step depends on the proportion of contaminated data in the observation. We use $k_1(p) = \sqrt{\chi^2_{p, \alpha}}$ where $\chi^2_{p, q}$ is the $q$th quantile of the $\chi^2_p$ distribution as used in~\cite{Dev+Gna+Ket:81}. In the provided toolbox
{\tt RB-SGMM-BA}, the default values for the tolerance, maximum number of iterations and $q$ are $(10^{-5}, 1000, 0.99)$, respectively. 

\begin{remark}[Comparison with the the conventional EM]
	The EM algorithm in the conventional GMM without spatial templates uses $\pi_{ik}^t = \pi_k^t$ and $\pi_{k}^{t} =
	{\sum_{i = 1}^n w_{ik}^{t}}{/n}$ in the E-step and does not update  $\gamma_k$. The M-step without robustness may update the mean vector and covariance matrix by 
	\begin{equation} \label{Mstep}
	\bmu_k^{t + 1} = \frac{\sum_{i = 1}^n w_{ik}^t y_i }{ \sum_{i = 1}^n w_{ik}^t},
	\qquad
	\bSigma_k^{t + 1} = \frac{1}{\sum_{i = 1}^n w_{ik}^t} \sum_{i = 1}^n w_{ik}^t (y_i -
	\bmu_k^{t + 1}) (y_i - \bmu_k^{t + 1})^T, 
	\end{equation}
	which can be viewed as a special case of the proposed robust EM algorithm where the weight function $u(s) = 1$ for all $s$. 
\end{remark}

\begin{remark}[Derivation of the robust EM algorithm]
	\label{remark:EM.derivation} 
	We just need to derive the update of $\gamma_k$ as the E-step follows the conventional EM but has voxel-varying $\pi_{ik}$ and $\pi_{ik}^t$, and the M-step is an application of $M$-estimation. The update of $\gamma_k$ is obtained by solving $\partial Q(\theta | \theta^t) / \partial \gamma_k= 0.$ Specifically, 
	\begin{align}
	Q(\theta |\theta^{t} )& = \sum_{i=1}^n \sum_{k=1}^K w_{ik}^t {\log(\pi_{ik})} + \text{constant} \\
	& = \sum_{i=1}^n \sum_{k=1}^K w_{ik}^t \log(\gamma_k b_{ik}) - \sum_{i=1}^n \sum_{k=1}^K w_{ik} \log\left(\sum_{j = 1}^K \gamma_j b_{ij}\right) + \text{constant} \\
	& = \sum_{i=1}^n  w_{ik}^t \log(\gamma_k) - \sum_{i=1}^n \log\left(\sum_{j = 1}^K \gamma_j b_{ij}\right) + \text{ constant},
	\end{align}  
	where the term \emph{constant} is with respect to $\gamma_k$. Therefore, we obtain that 
	\begin{equation}
	\frac{\partial Q(\theta |\theta^{t} )}{\partial \gamma_k} = \frac{\sum_{i=1}^n  w_{ik}^t }{\gamma_k} - \sum_{i=1}^n \frac{b_{ik}}{\sum_{j = 1}^K \gamma_j b_{ij}},
	\end{equation}
	which is a nonlinear function of $\gamma_k$'s. Formula~\eqref{eq:update.gamma} is thus obtained if we calculate the term $\sum_{j = 1}^K \gamma_j b_{ij}$ using $\gamma_k^{t}$ in the previous iteration. This simplification leads to a generalized EM and has been observed to ensure convergence by~\cite{SPM:05}.  
\end{remark}

It is worth mentioning that other robust estimation of mixture models may be also applicable, such as mixtures of heavy-tailed distributions such as $t$ distributions in~\citep{mclachlan2004finite} or maximizing a transformed likelihood tailored for robustness~\citep{Qin2013}. Supplementary Materials contain a simulation study to compare the proposed robust EM algorithm with a mixture of $t$ distributions in a non-spatial setting. The incorporation of the spatial structures implies non-trivial generalizations of these methods which we view as future research topics.

\section{Simulation} 
\label{sec:simulation}

Section~\ref{section:standardization} investigated the tail probabilities of $T^{(1)}$ when all
parameters are given, a situation that {we refer to as the ``oracle"}. In
this section, we investigate the tail probabilities of $T^{(1)}$ when all the parameters
are estimated via the proposed robust EM algorithm in Section~\ref{section:EM}.

\subsection{Univariate Data} 	 

Although the real motivating data is bivariate, investigation of the univariate case is helpful because it is easier to interpret and provides insight for multivariate cases. Moreover, even for multivariate data, the final quantity of interest is univariate when considering a contrast among the multivariate observations.

\begin{figure}[h!]
	\caption{Simulation settings for the univariate case. The
		three plots are the template maps, transformed template maps and observations. The $x$-axis in each plot is the location $t$.}	 	 
	\label{fig:univariate.setting}
	\centering 
	\begin{tabular}{ccc} 	 		
		$b$ functions & $\pi$ functions & Observations \\ 	 		
		\includegraphics[trim = 30 20 30 20, clip, width =
		0.27\textwidth]{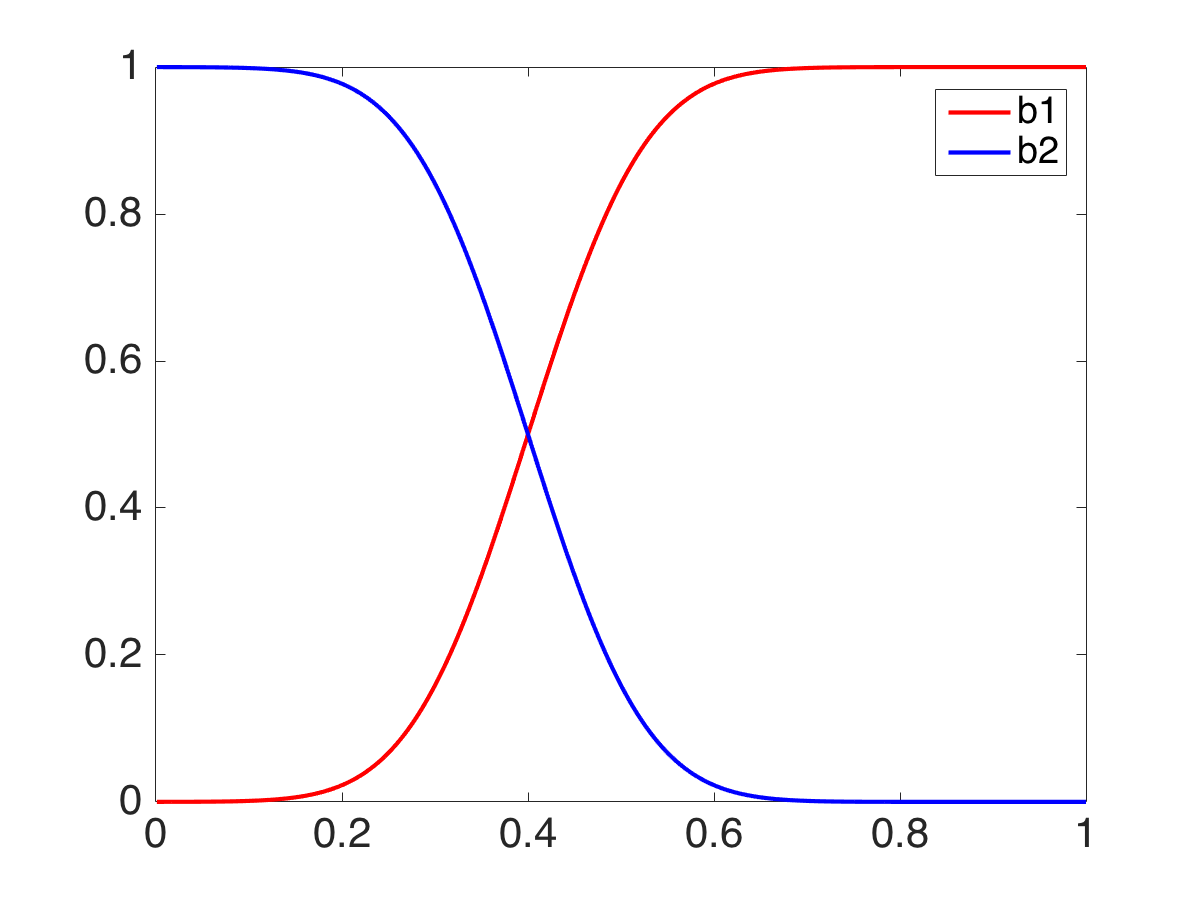} & 	 		
		\includegraphics[trim = 30 20 30 20, clip, width = 0.27\textwidth]{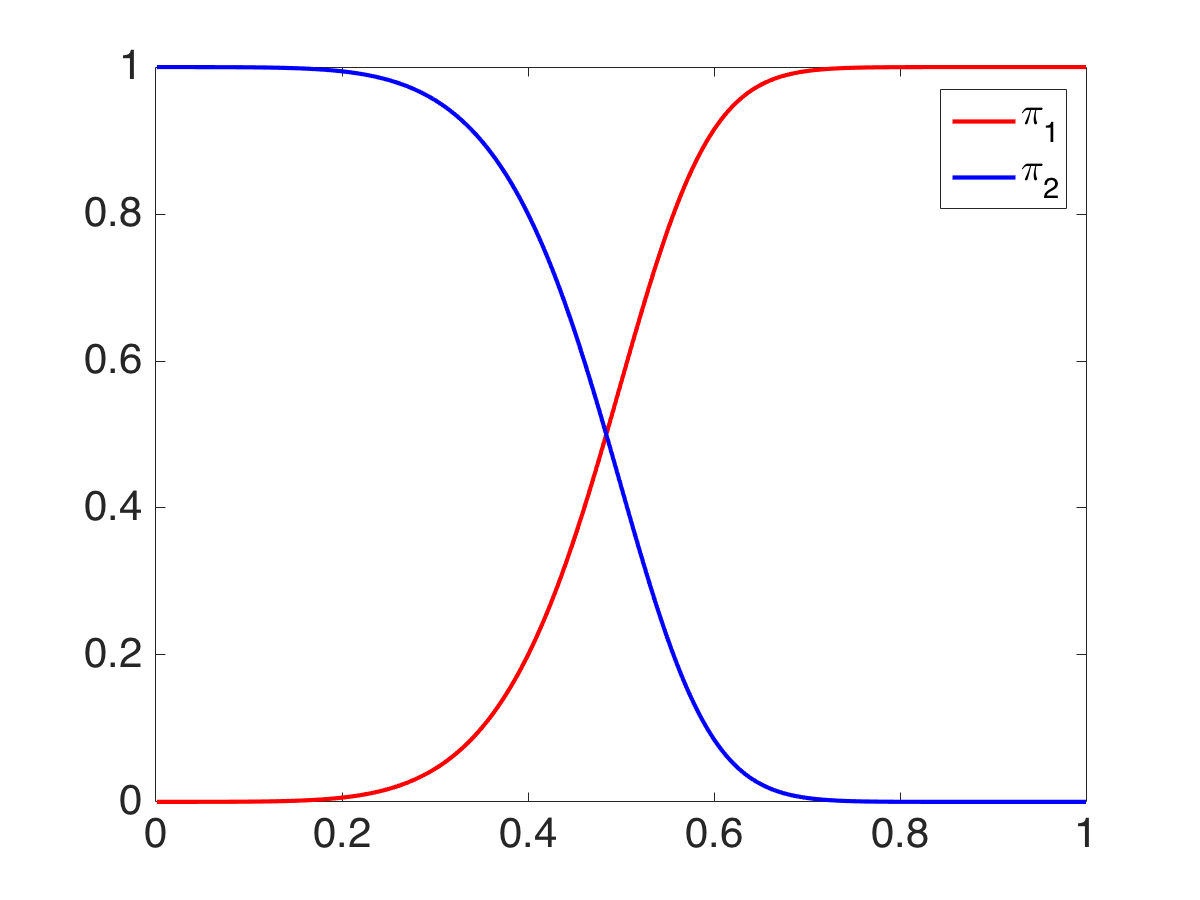} & 	 		
		\includegraphics[trim = 30 20 30 20, clip, width =
		0.27\textwidth]{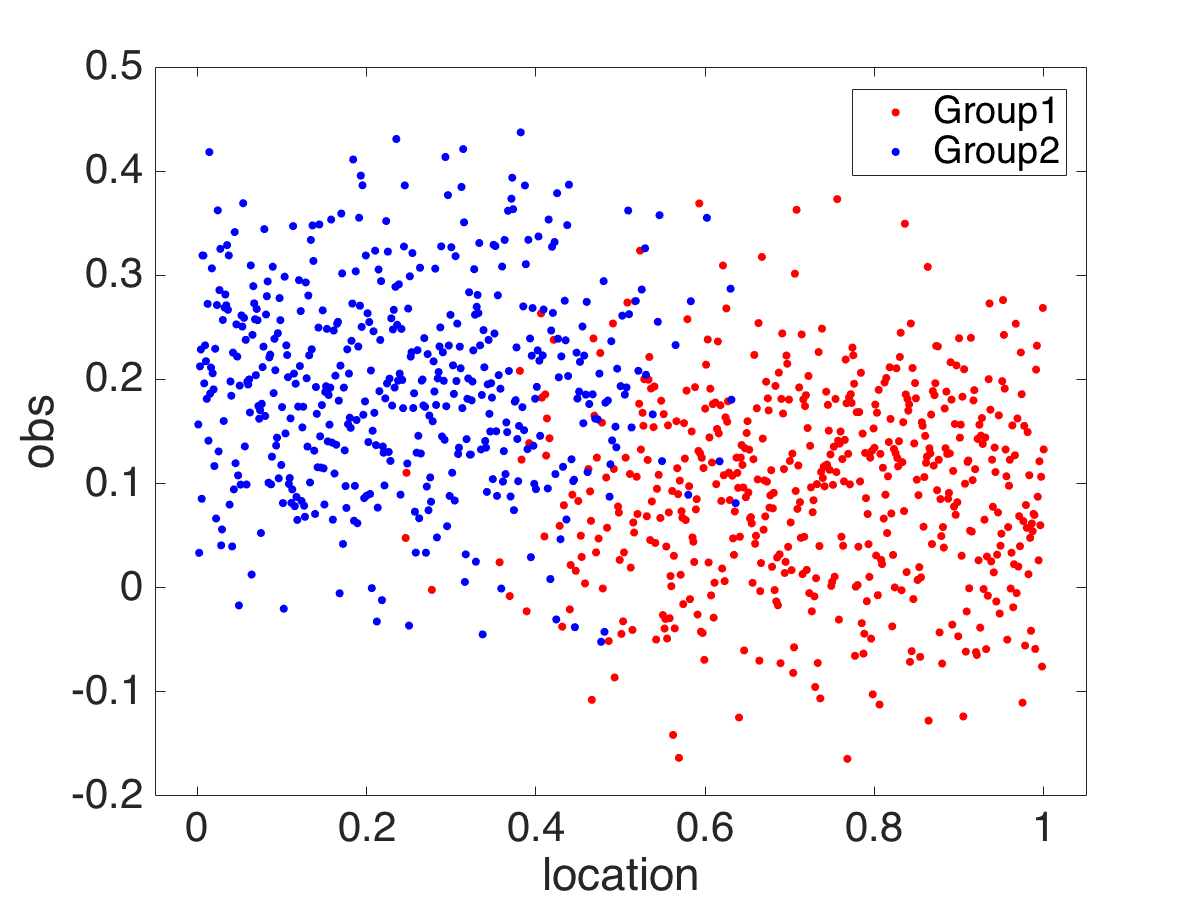} 	 	
	\end{tabular} 
	\vspace{10pt} 	 
\end{figure}

We simulate data according to model~\eqref{GMM} and~\eqref{eq:b2pi} with dimension
$p = 1$. We use $K = 2$ classes and $n = 1000$ locations, where the parameters in the
model are given by $b_1(t) = \Phi(10t - 4),
b_2(t) = 1 - b_1(t), \gamma = (0.2, 0.8), \mu = (0.1, 0.2),$ and $\sigma =
(0.1, 0.1).$ Figure~\ref{fig:univariate.setting} shows the $b$ and
$\pi$ functions, along with one simulated instance of the observations.
{Except for the right and left extremes where $\pi_1$ and $\pi_2$ are close to 0 or 1, the other parameter combinations do not correspond to the favorable parameter combinations of Theorem \ref{th:asy.multivariate}.} 

To estimate the model parameters, we apply the traditional GMM and the proposed
Spatial GMM (without robust adjustments for now). The
results are compared to those using the true parameters, referred to as ``oracle". To
standardize the observations, we use both soft and hard assignments $T_S^{(1)},
T_S^{(2)}, T_S^{(3)}$ and $T_H$. We compare the methods by calculating the relative size
$R(\alpha)$ for $\alpha = 0.01$, based on $10^5$ simulations.

\begin{figure}[h!] 	 	
	\centering
	\caption{Simulated relative size $R(0.01)$ at the right tail
		for the univariate case. Each column corresponds to a different standardization method
		(three soft assignments and one hard assignment), while the three rows correspond to
		three parameter estimation methods (GMM, SGMM and oracle). The $x$-axis in each plot is the location $t$.}
	\label{fig:univariate.right} 	 	\begin{tabular}{ccccc} 	 		& $T_S^{(1)}$ &
		$T_S^{(2)}$ & $T_S^{(3)}$ & $T_H$\\ 	 		\begin{sideways} \rule[0pt]{0.25in}{0pt}
			GMM \end{sideways} & 	 		\includegraphics[trim = 40 20 35 27, clip, width =
		0.2\textwidth]{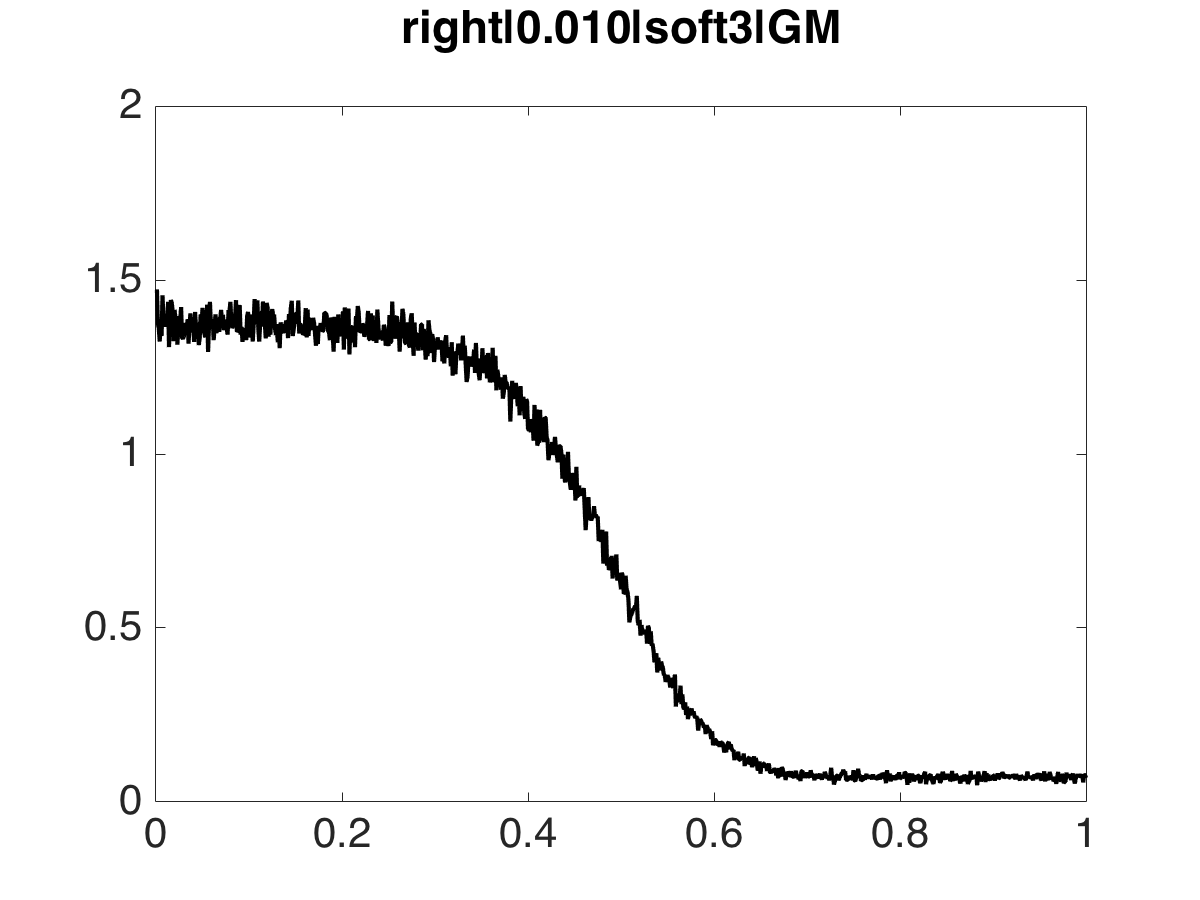} & 	 		
		\includegraphics[trim = 40 20 35 27, clip, width =
		0.2\textwidth]{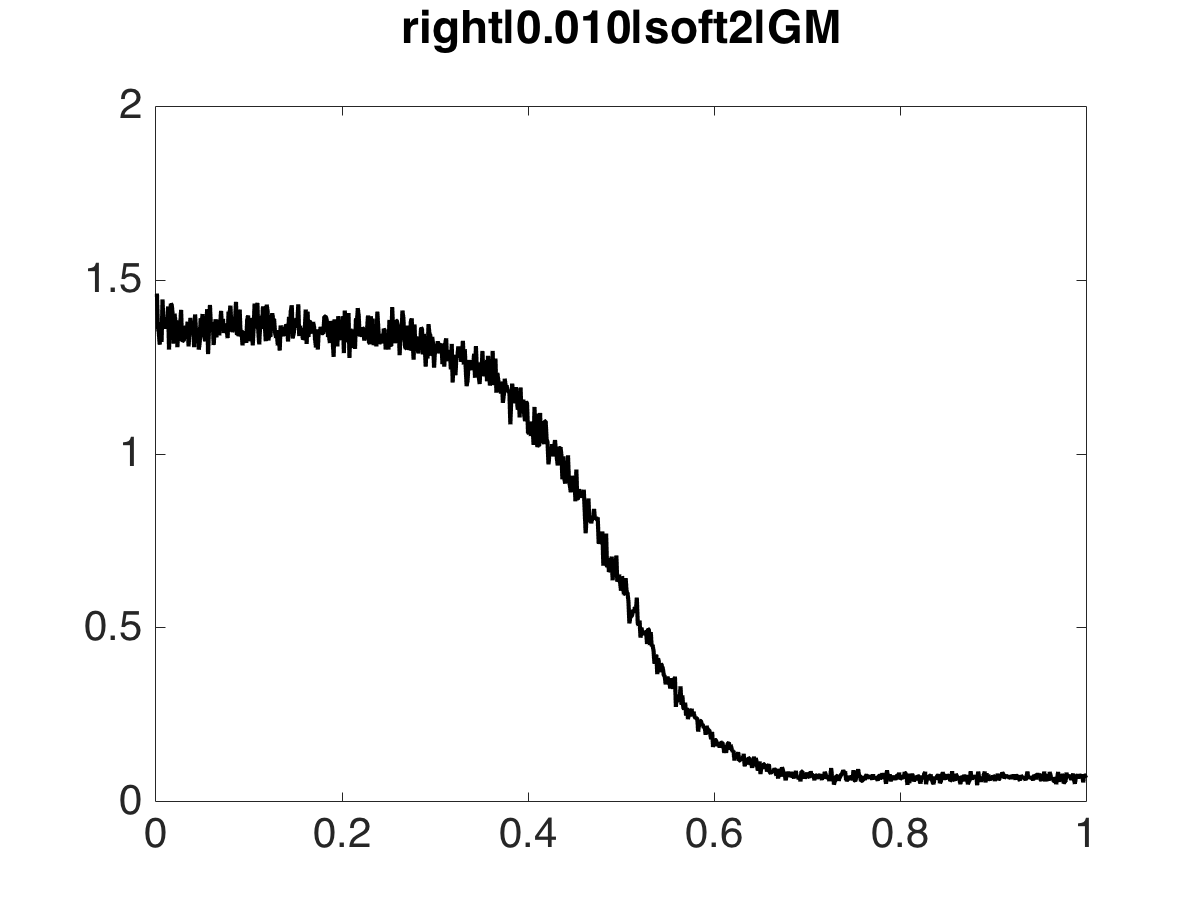} & 	 		
		\includegraphics[trim = 40 20 35 27, clip, width =
		0.2\textwidth]{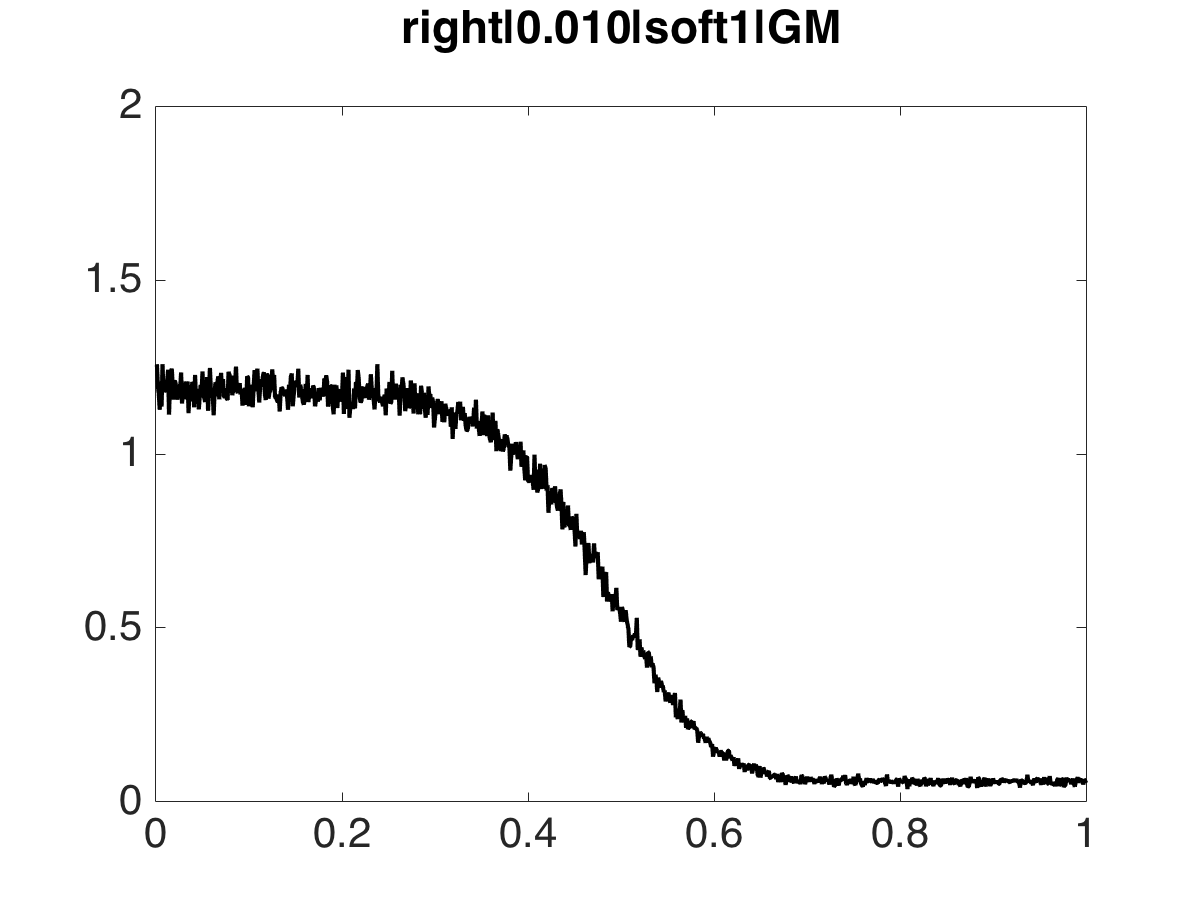} & 	 		
		\includegraphics[trim = 40 20 35 27, clip, width =
		0.2\textwidth]{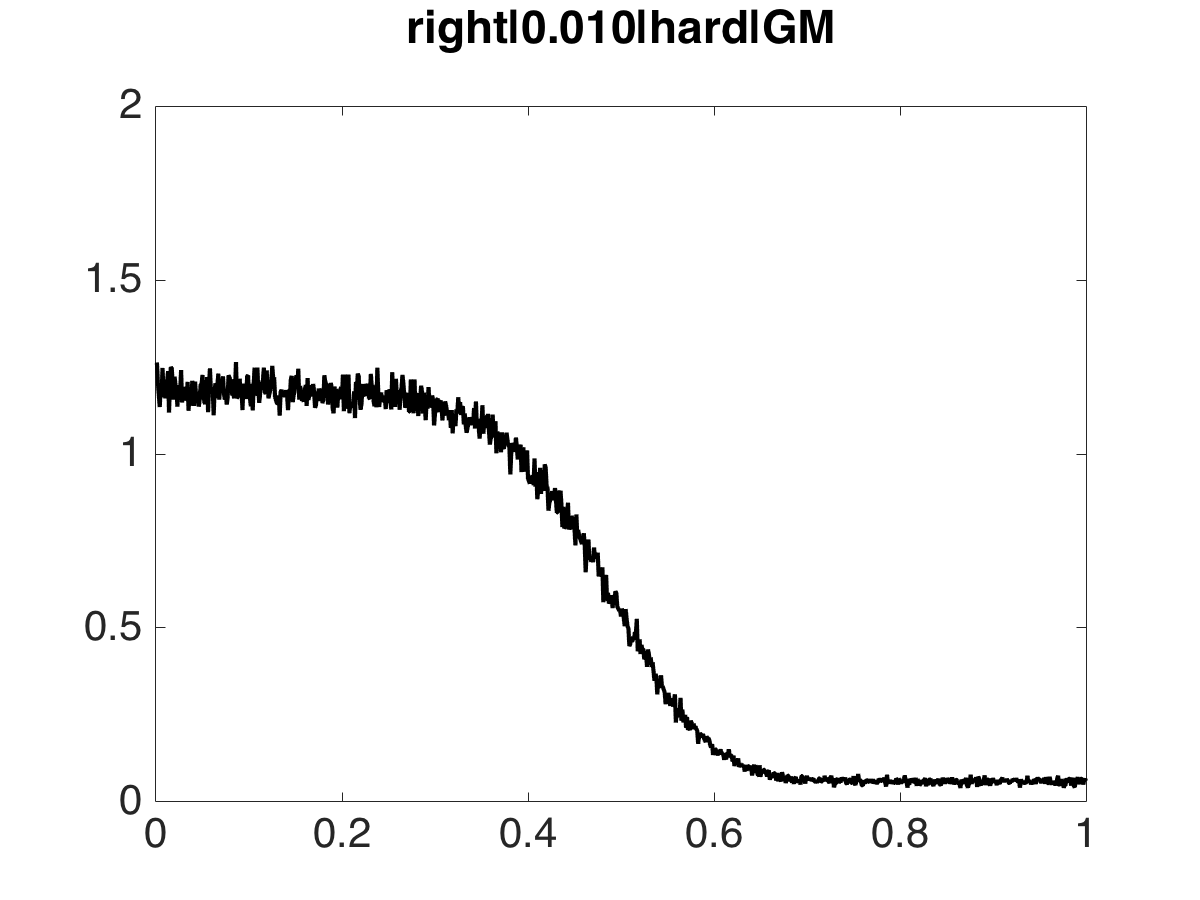} \\ 	 		 \begin{sideways}
			\rule[0pt]{0.2in}{0pt} SGMM \end{sideways} & 	 		\includegraphics[trim = 40 20 35
		27, clip, width = 0.2\textwidth]{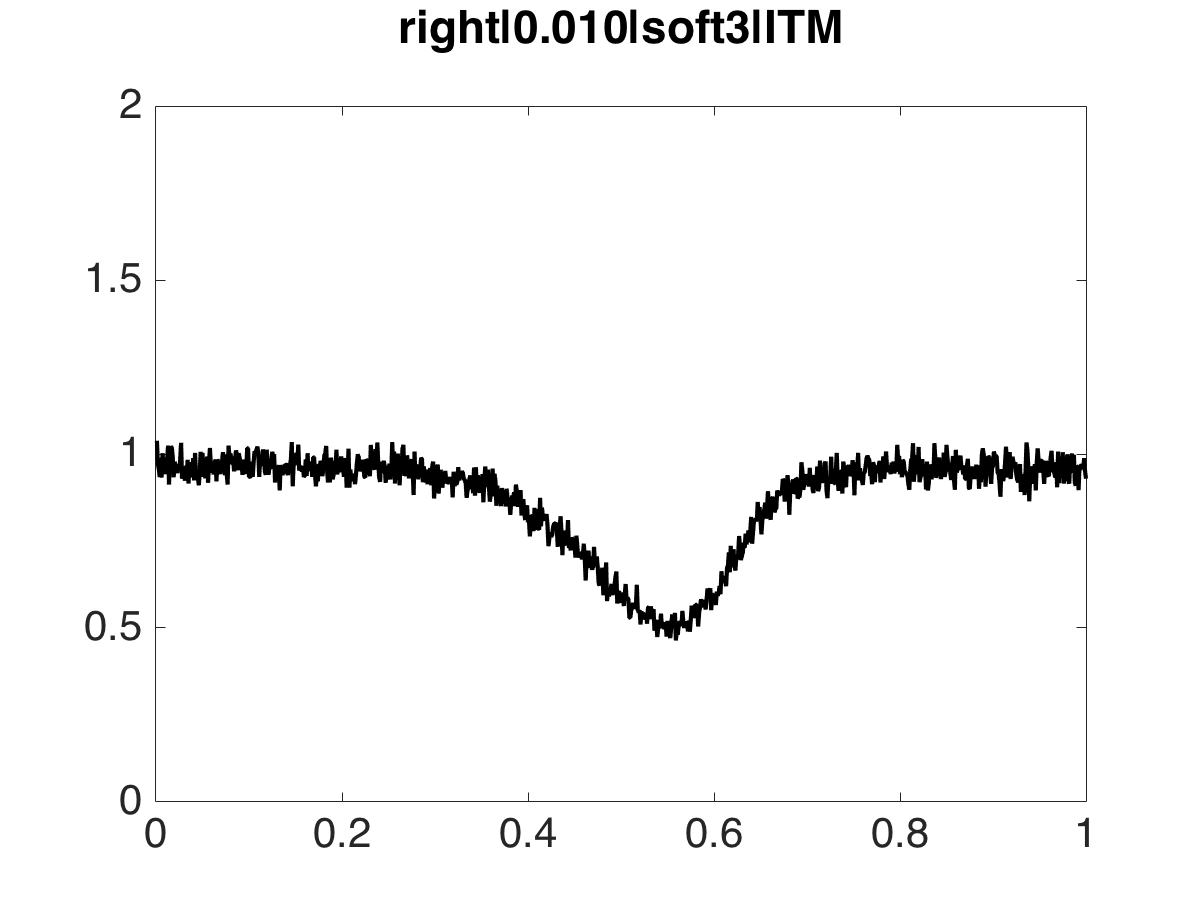} & 	 		 
		\includegraphics[trim = 40 20 35 27, clip, width =
		0.2\textwidth]{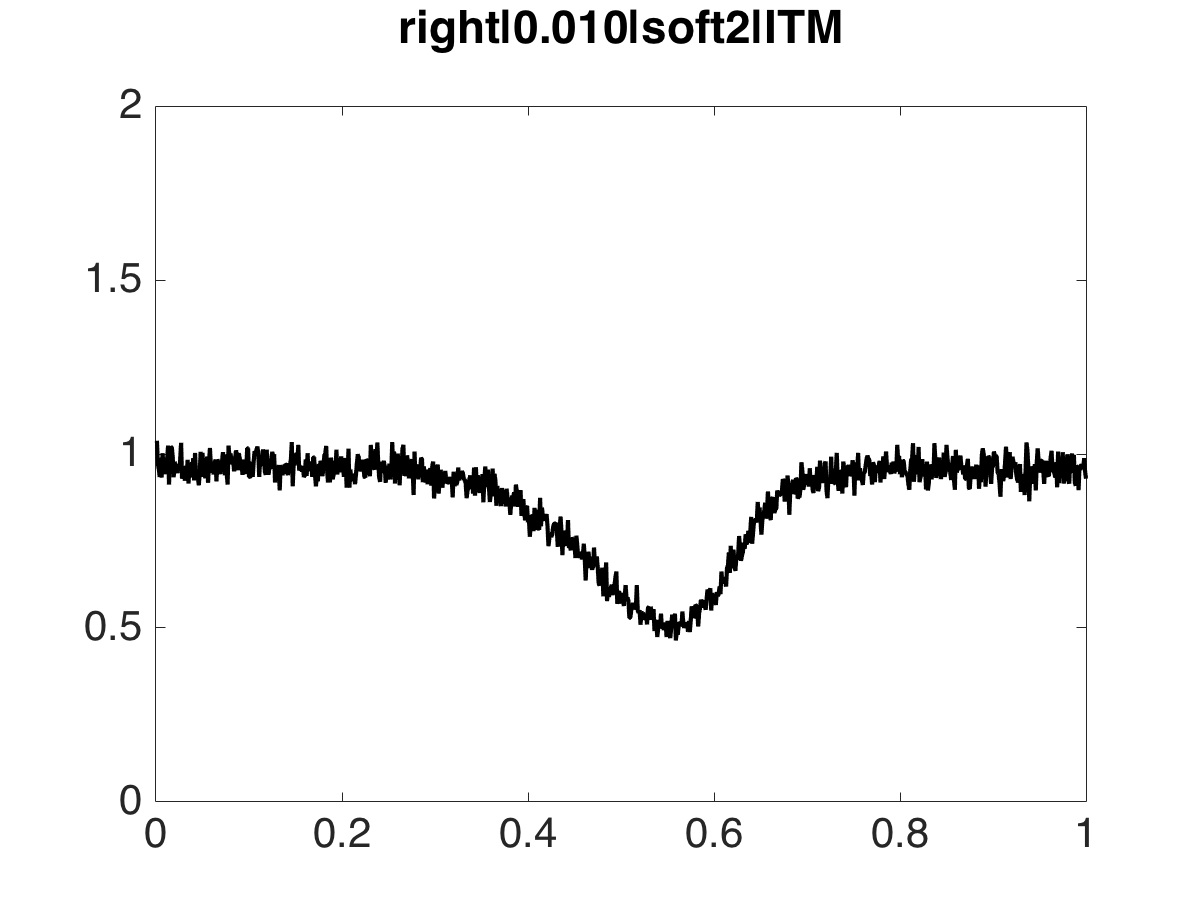} & 	 		
		\includegraphics[trim = 40 20 35 27, clip, width =
		0.2\textwidth]{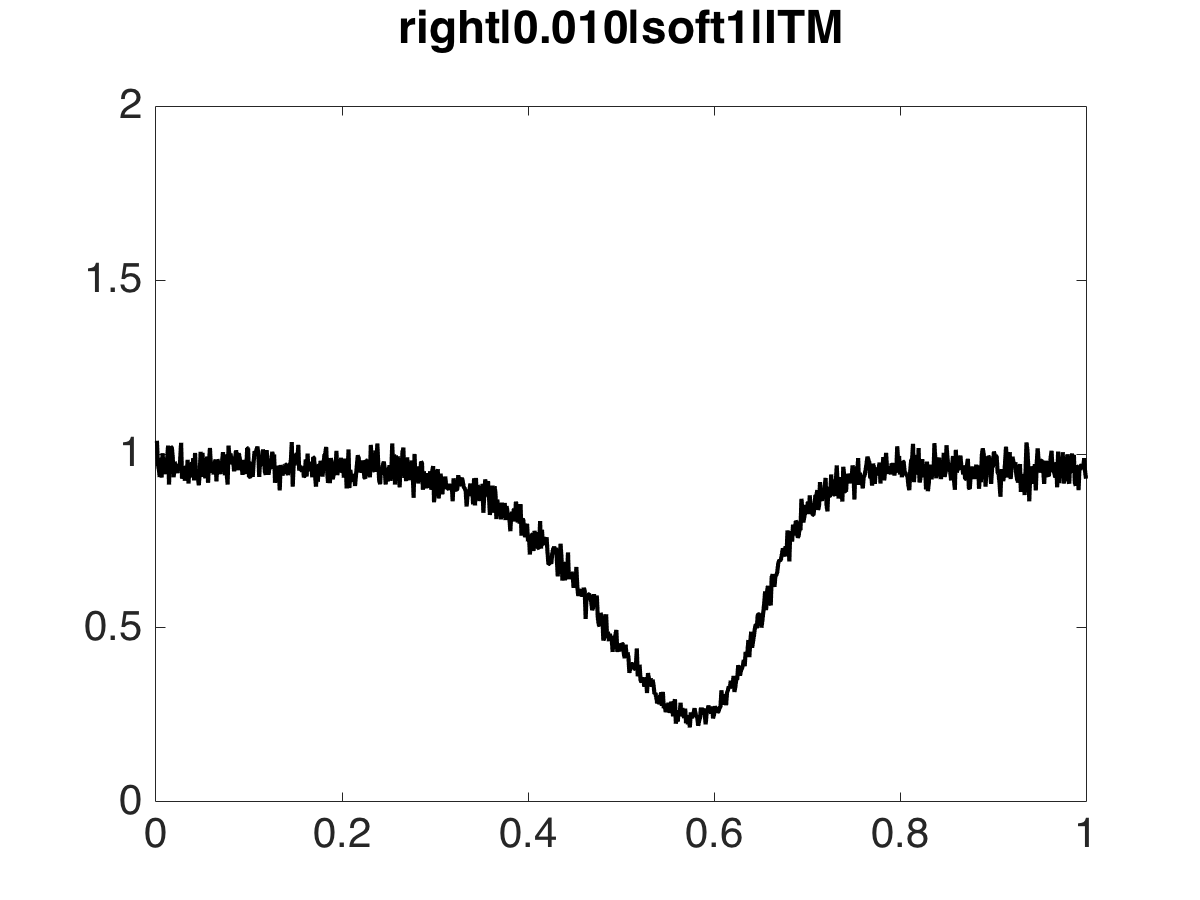} & 	 		
		\includegraphics[trim = 40 20 35 27, clip, width =
		0.2\textwidth]{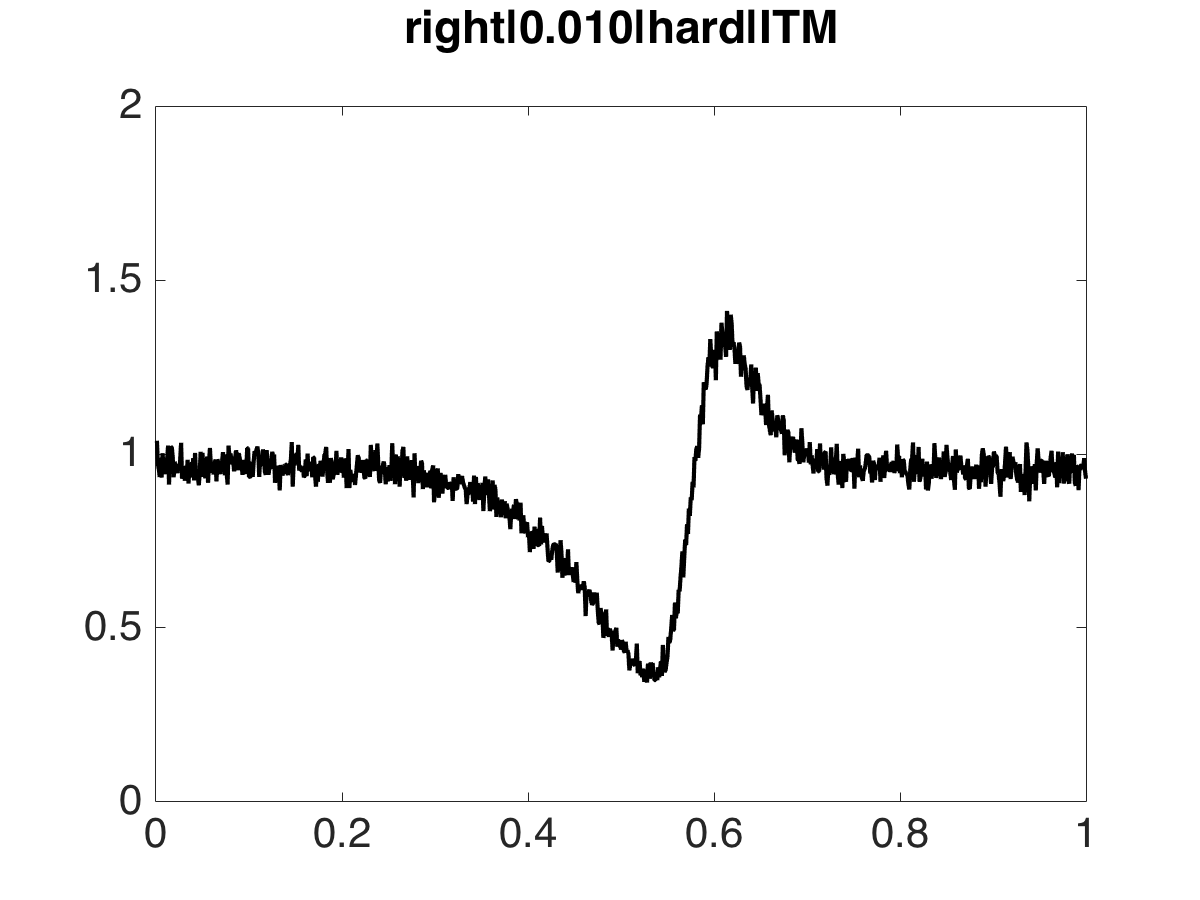} \\ 	 		 \begin{sideways}
			\rule[0pt]{0.2in}{0pt} Oracle \end{sideways}  & 	 		\includegraphics[trim = 40 20
		35 27, clip, width = 0.2\textwidth]{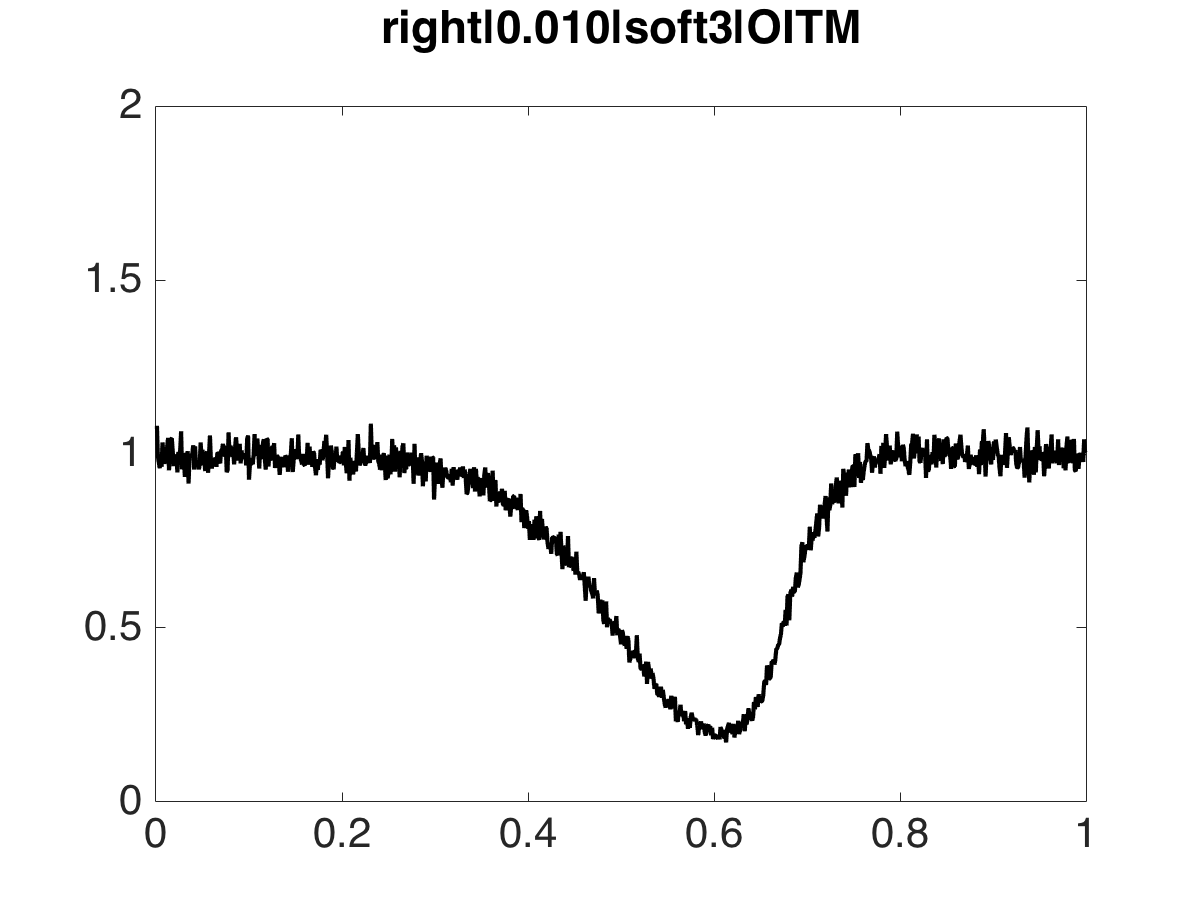} & 	 		 
		\includegraphics[trim = 40 20 35 27, clip, width =
		0.2\textwidth]{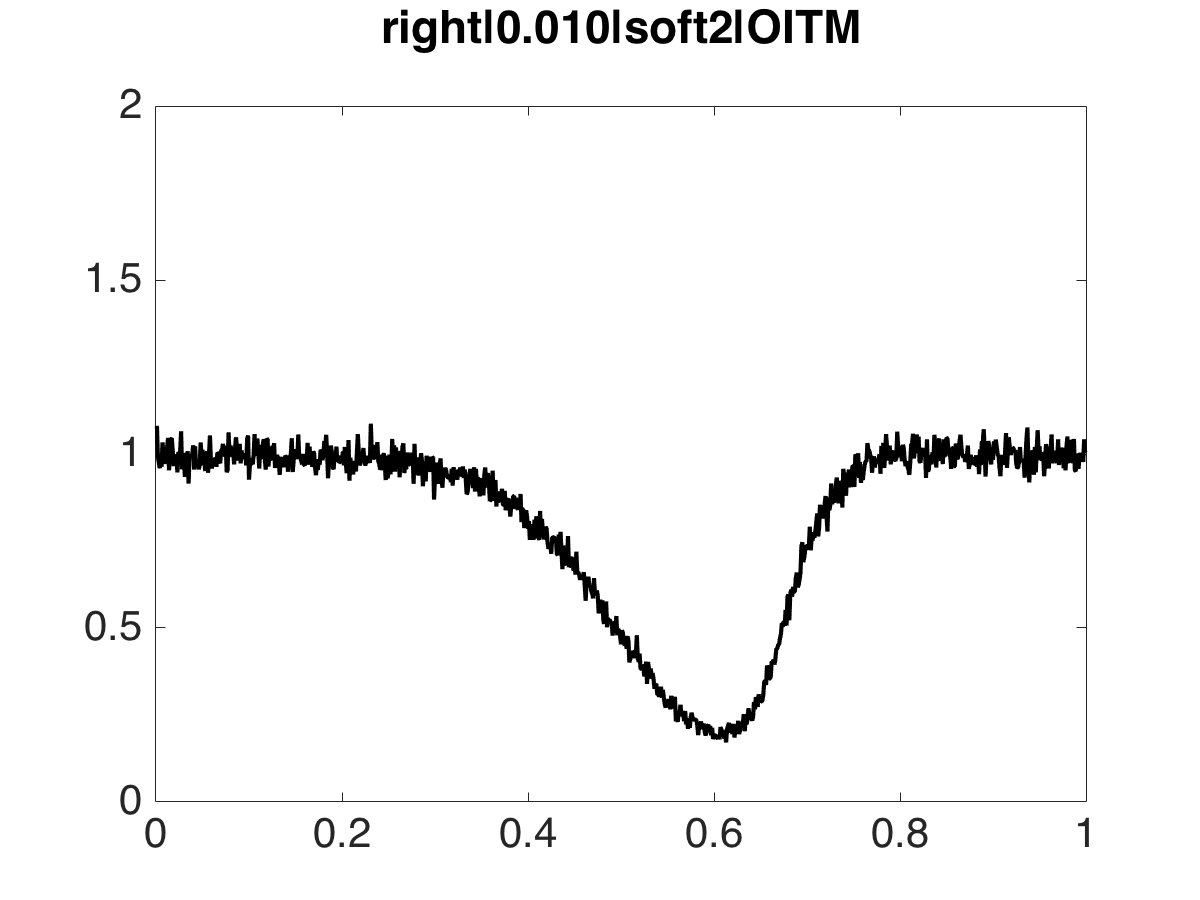} & 	 		
		\includegraphics[trim = 40 20 35 27, clip, width =
		0.2\textwidth]{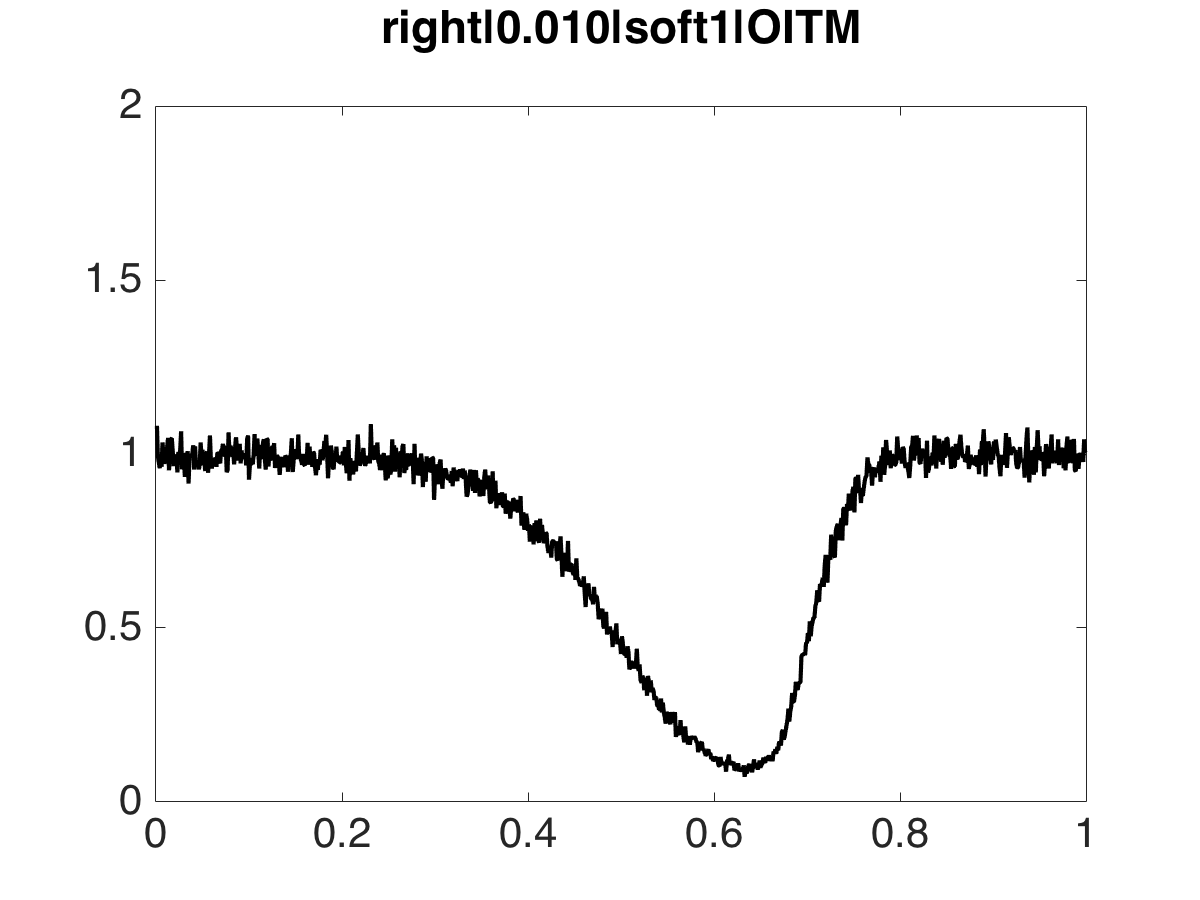} & 	 		
		\includegraphics[trim = 40 20 35 27, clip, width =
		0.2\textwidth]{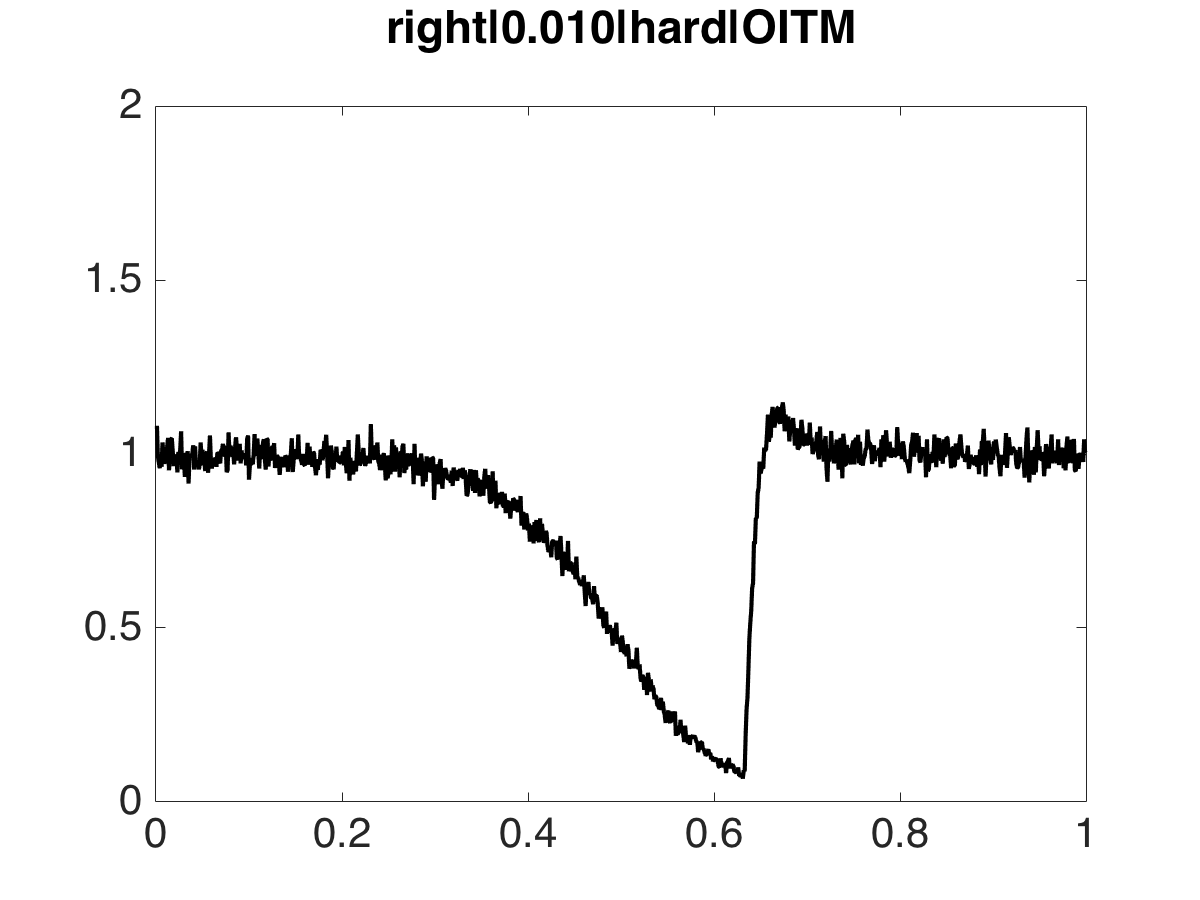} 	 	
	\end{tabular} 	
\end{figure}

Figure~\ref{fig:univariate.right} plots the relative size $R(0.01)$ for the right
tail. We can see that the traditional GMM method leads to relative sizes both greater and
smaller than 1 in different locations. The traditional GMM is heavily affected by the
spatial structure of the $\pi$ functions because it does not consider any spatial
information in the probability templates. In contrast, the proposed SGMM with soft
assignments lead to tail probabilities that are very close to standard normal, except in
the middle locations where there is a strong mixing between the components and the
relative size is conservative.

Furthermore, when the SGMM is used, soft assignment has better performance than hard
assignment in the sense that the hard assignment leads to invalid relative sizes, while
soft assignment does not. We can also see that the oracle method using the true parameter
values offers limited benefit with respect to SGMM with soft assignment, with the latter
performing, surprisingly, even slightly better. This may be due to the fact that the EM
algorithm estimates the component parameters accurately and also adapts to the data set.
The three soft assignment methods perform similarly.	

\begin{figure}[h!]	\caption{Histograms of {standardized} scores across locations {from a
			single simulated realization}. Each column corresponds to a different standardization
		method (three soft assignments and one hard assignment), while the three rows correspond
		to three parameter estimation methods (GMM, SGMM and oracle). {The p-value in
			each plot is computed using a Kolmogorov-Smirnov normality test.}} 	
	\label{fig:univariate.hist} 	\begin{tabular}{*{1}{c@{\hskip 3pt}}*{4}{c@{\hskip 0pt}}}	& $T_S^{(1)}$ & $T_S^{(2)}$ &
		$T_S^{(3)}$ & $T_H$\\ 		\begin{sideways} \rule[0pt]{0.25in}{0pt} GMM \end{sideways} & 		
		\includegraphics[trim = 40 20 35 27, clip, width =
		0.23\textwidth]{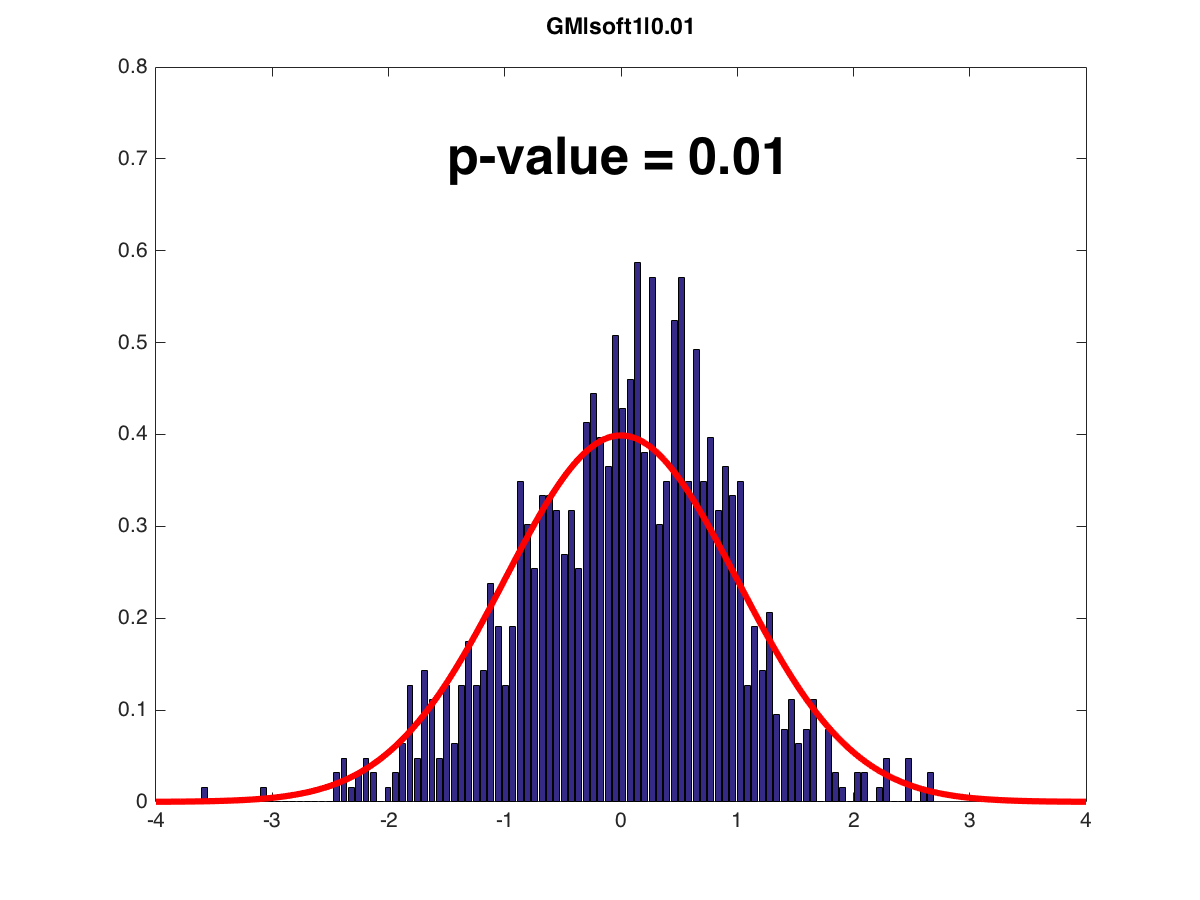} & 		\includegraphics[trim = 40 20 35 27,
		clip, width = 0.23\textwidth]{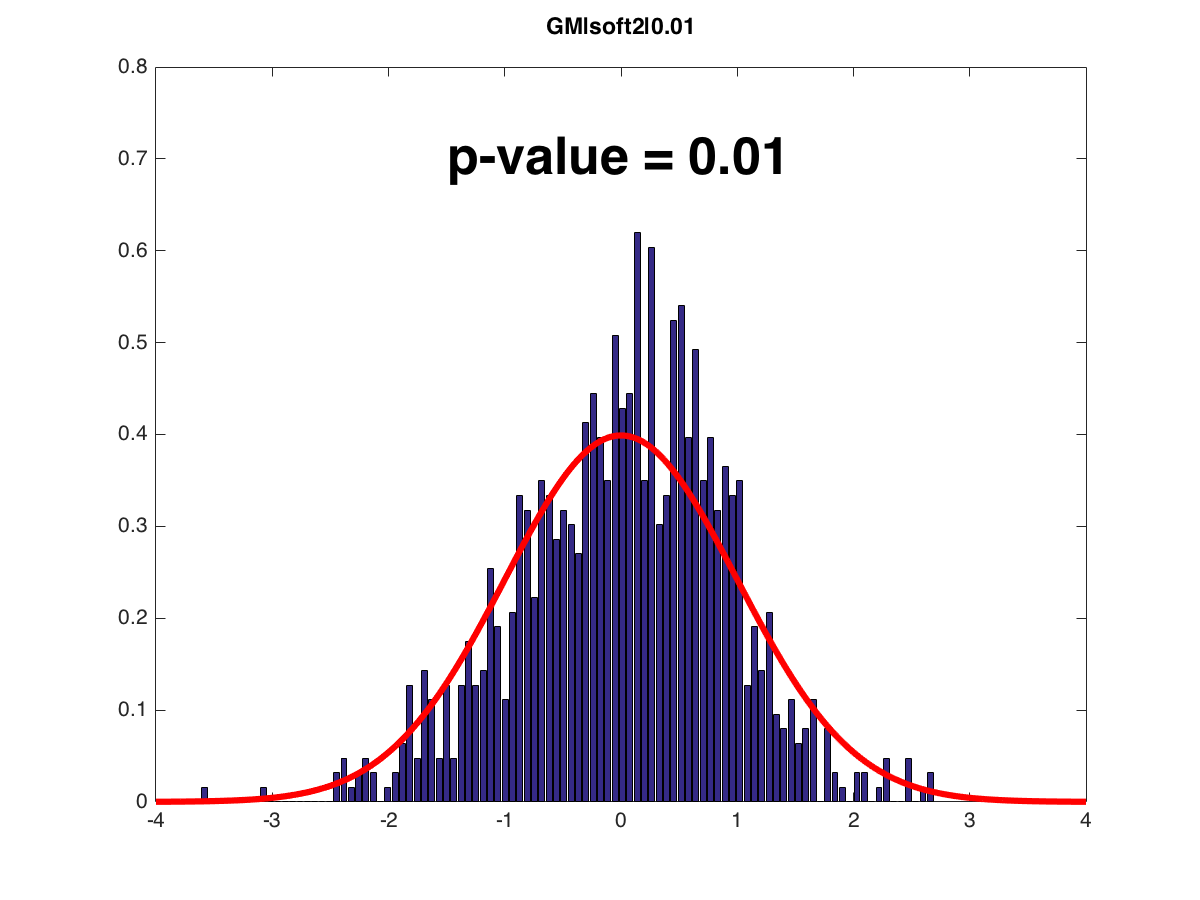} & 		\includegraphics[trim =
		40 20 35 27, clip, width = 0.23\textwidth]{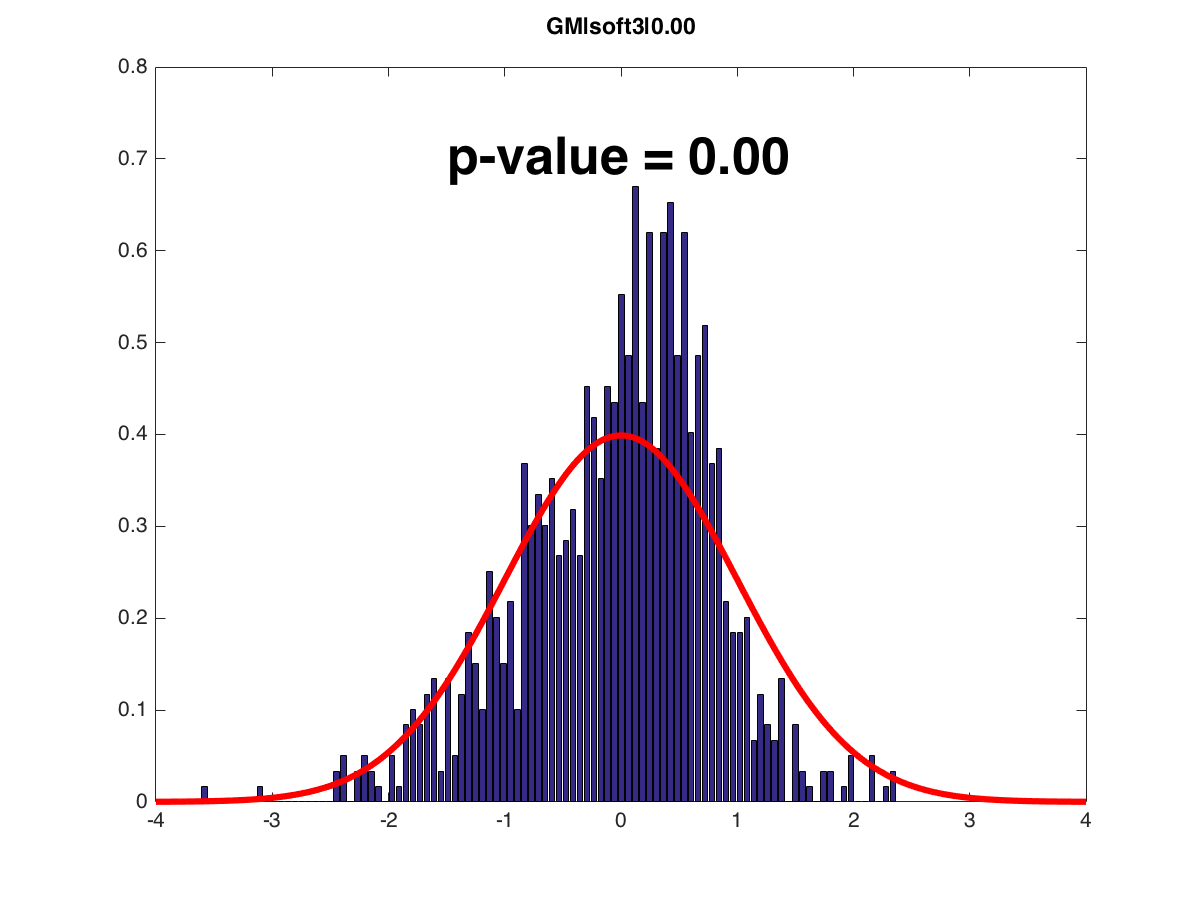} & 		
		\includegraphics[trim = 40 20 35 27, clip, width =
		0.23\textwidth]{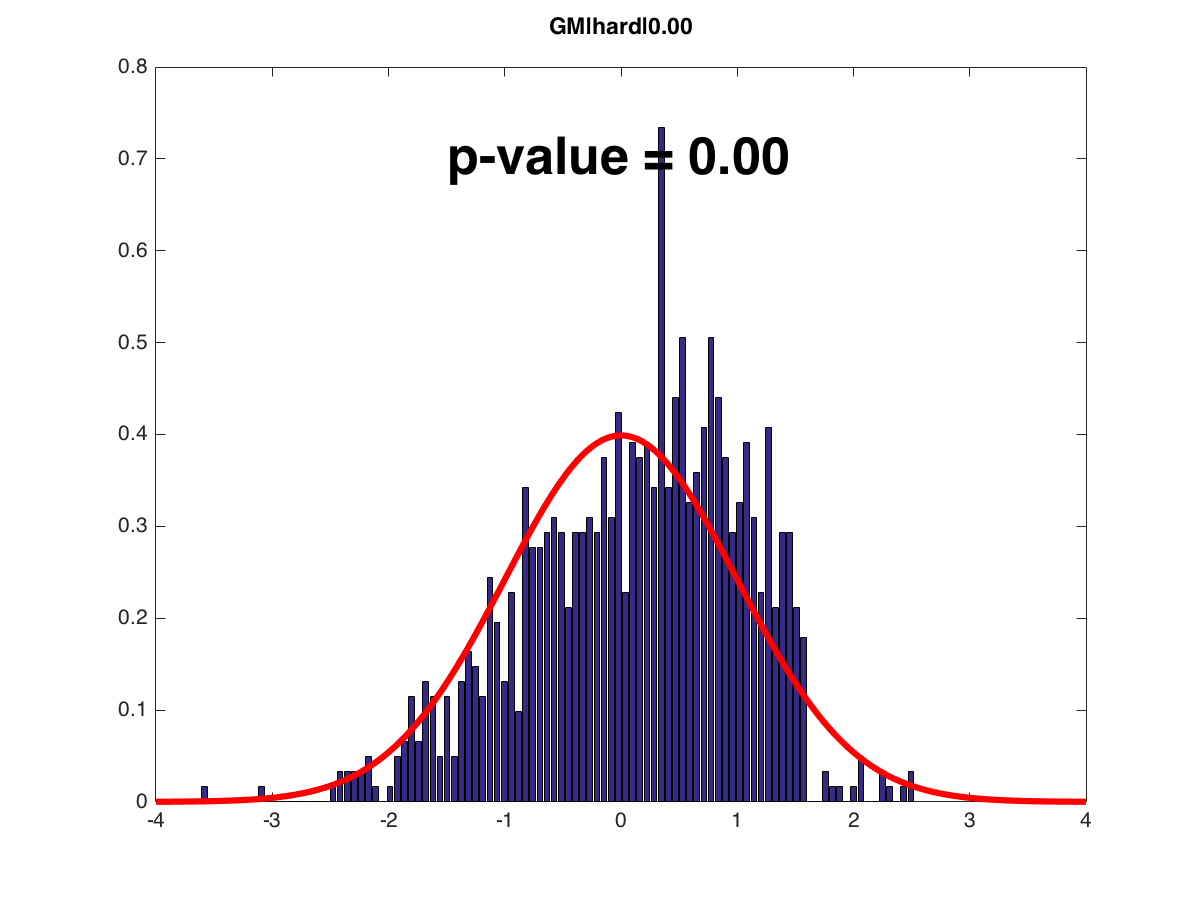} \\ 		\begin{sideways}
			\rule[0pt]{0.2in}{0pt} SGMM \end{sideways} & 		\includegraphics[trim = 40 20 35 27,
		clip, width = 0.23\textwidth]{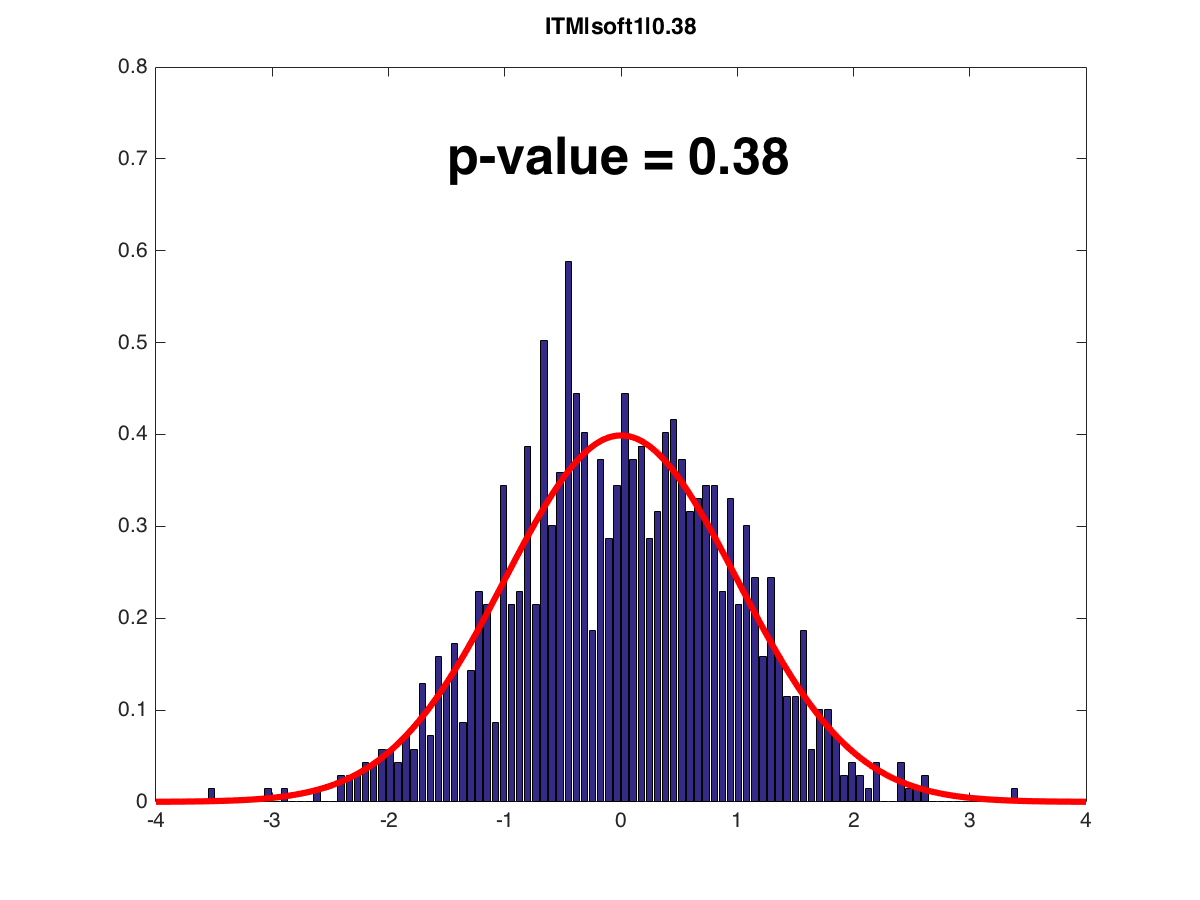} & 		\includegraphics[trim =
		40 20 35 27, clip, width = 0.23\textwidth]{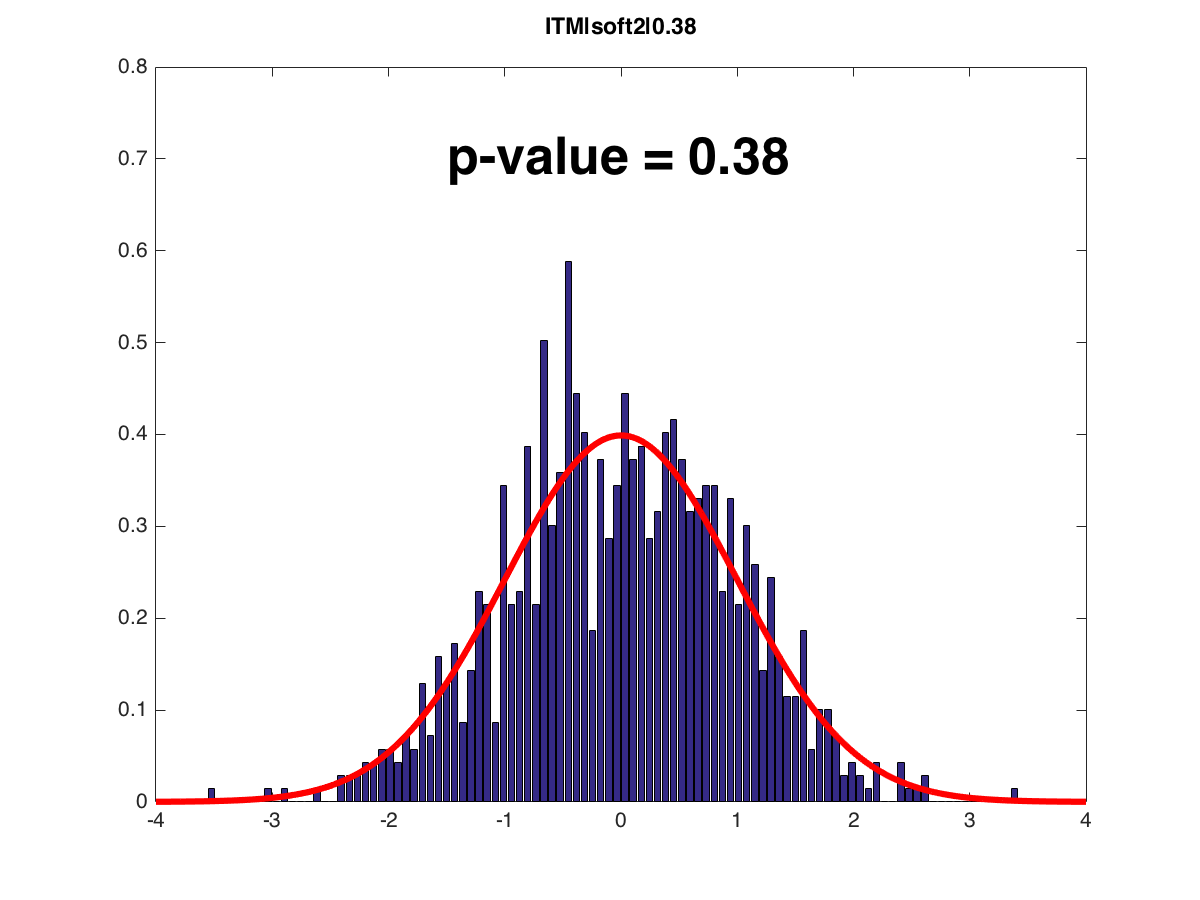} & 		
		\includegraphics[trim = 40 20 35 27, clip, width =
		0.23\textwidth]{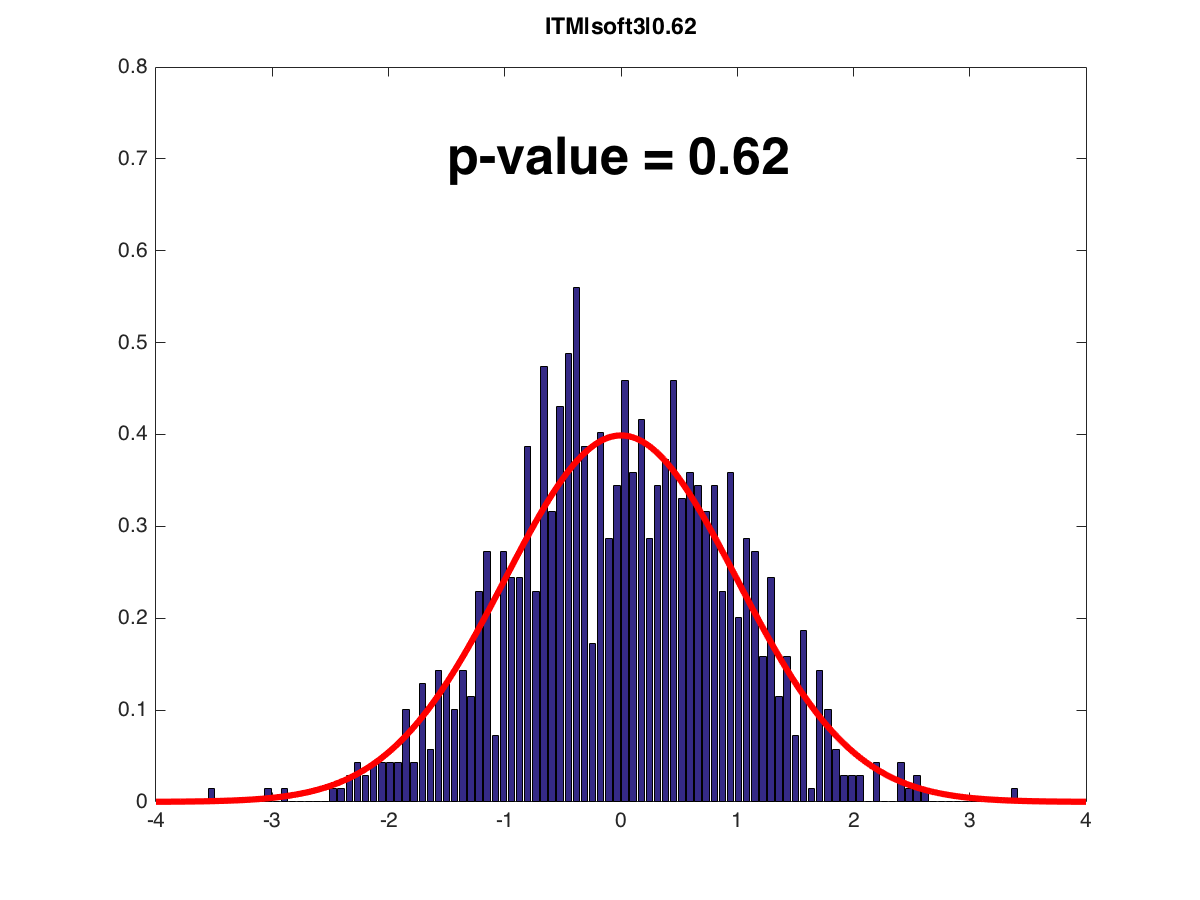} & 		\includegraphics[trim = 40 20 35 27,
		clip, width = 0.23\textwidth]{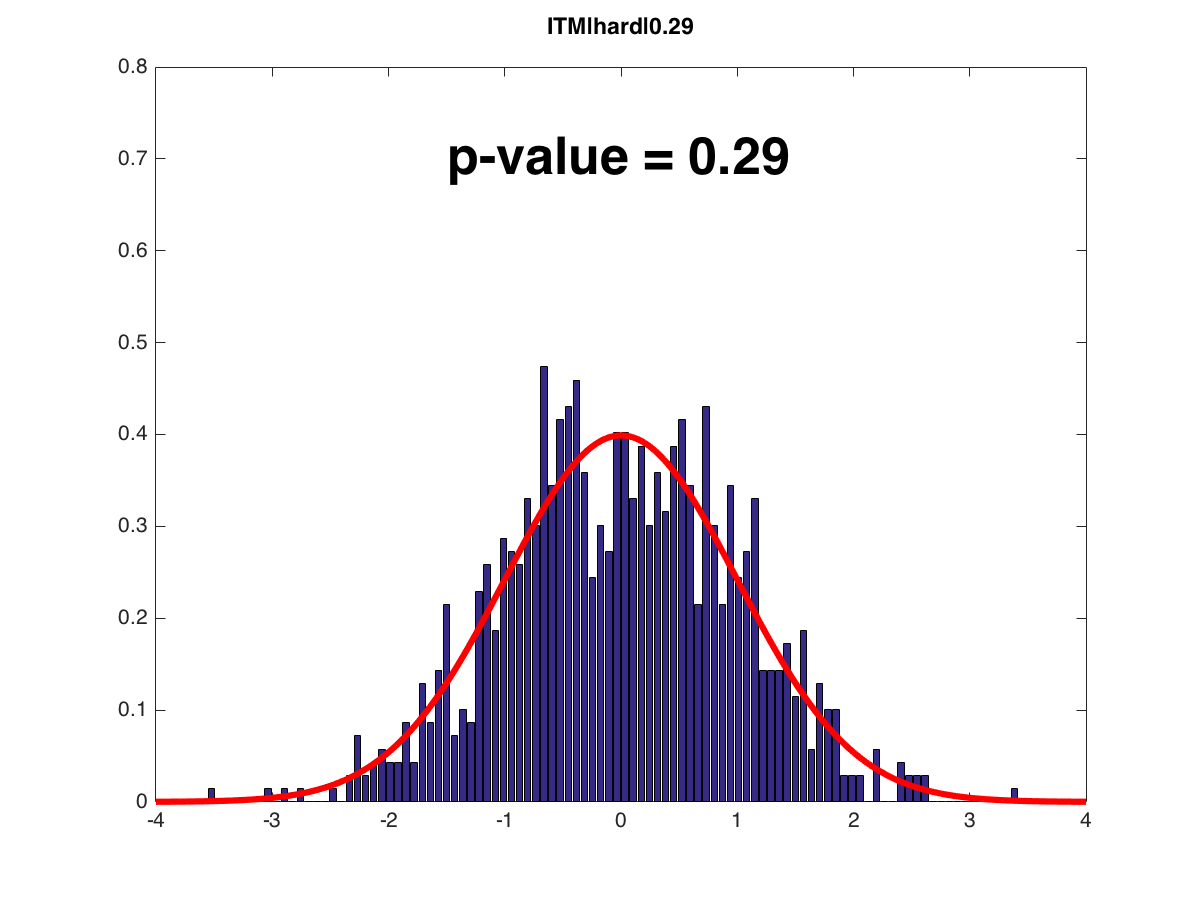} \\ 		\begin{sideways}
			\rule[0pt]{0.2in}{0pt} Oracle \end{sideways}  & 		\includegraphics[trim = 40 20 35
		27, clip, width = 0.23\textwidth]{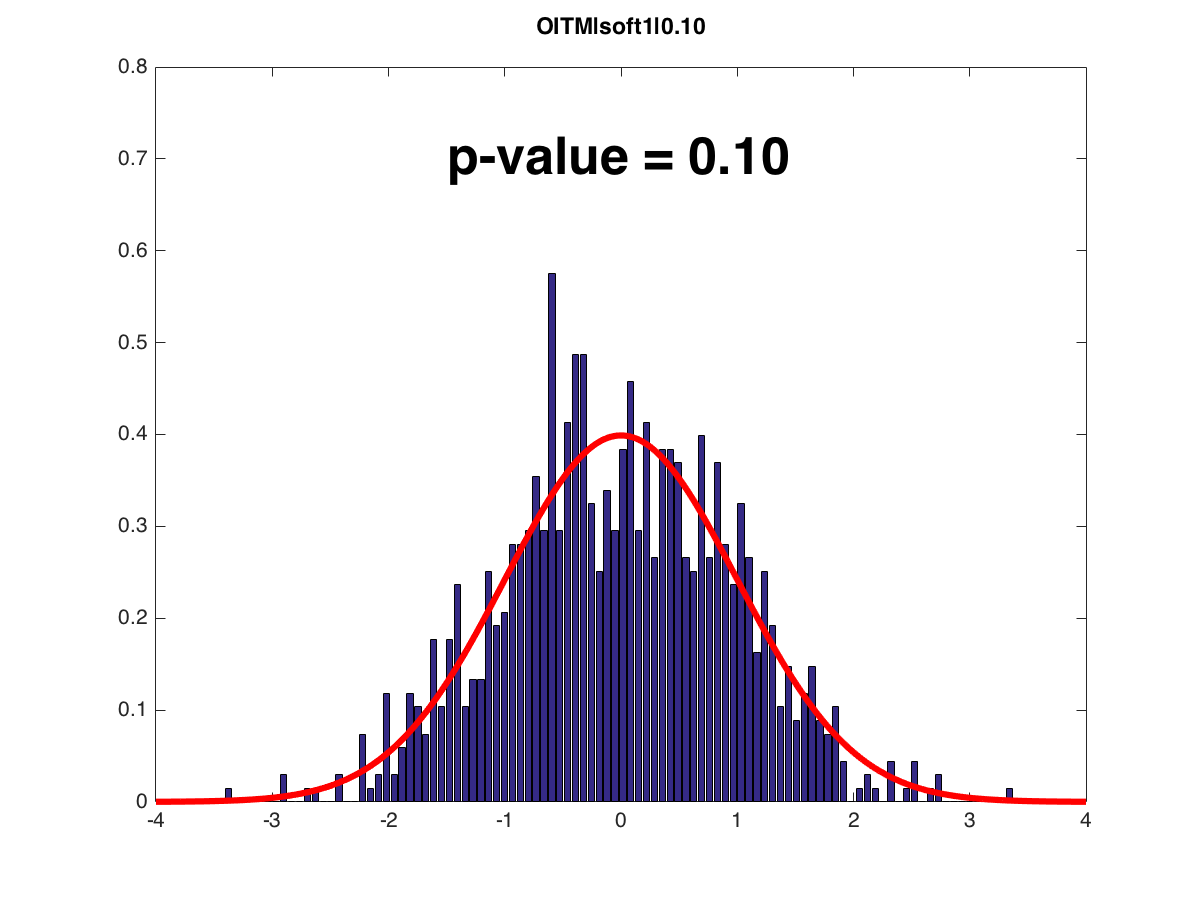} & 		 \includegraphics[trim
		= 40 20 35 27, clip, width = 0.23\textwidth]{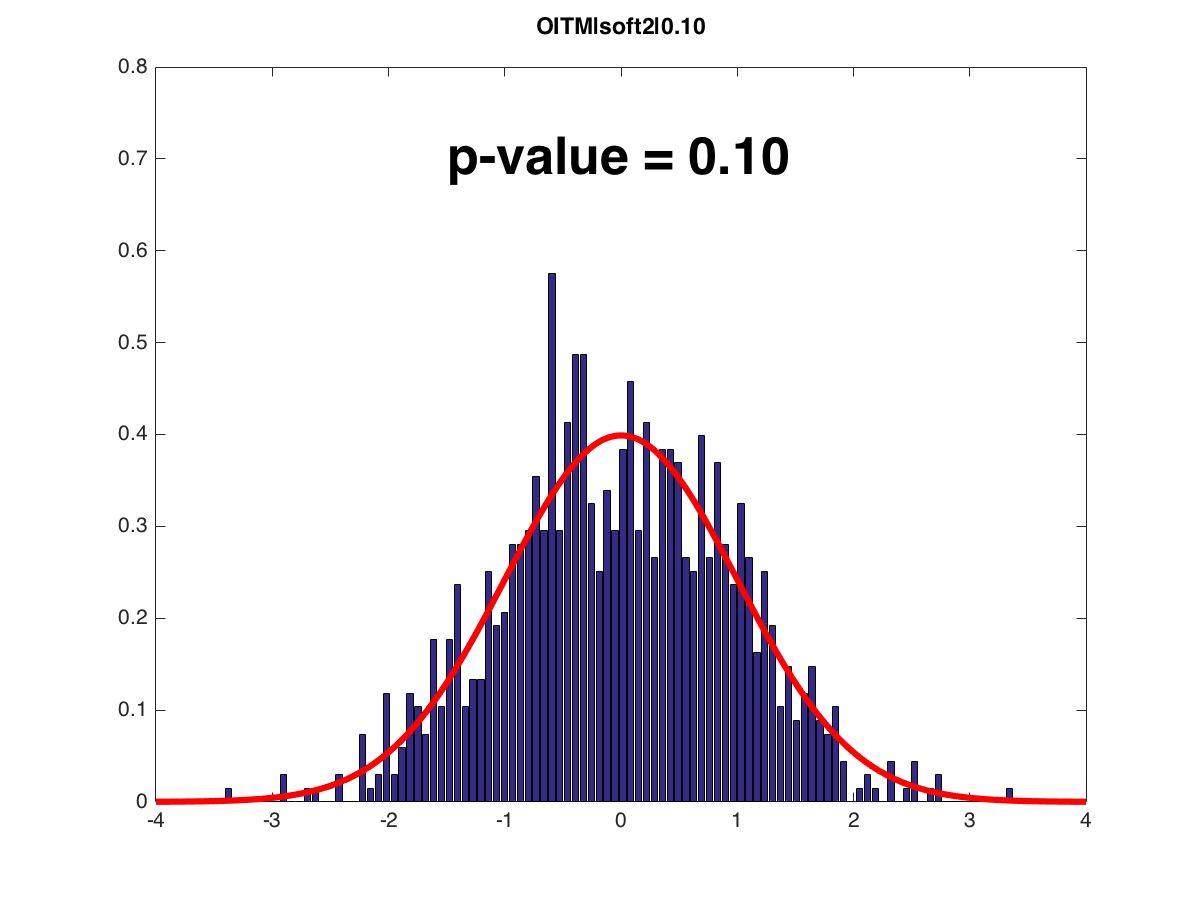} & 		
		\includegraphics[trim = 40 20 35 27, clip, width =
		0.23\textwidth]{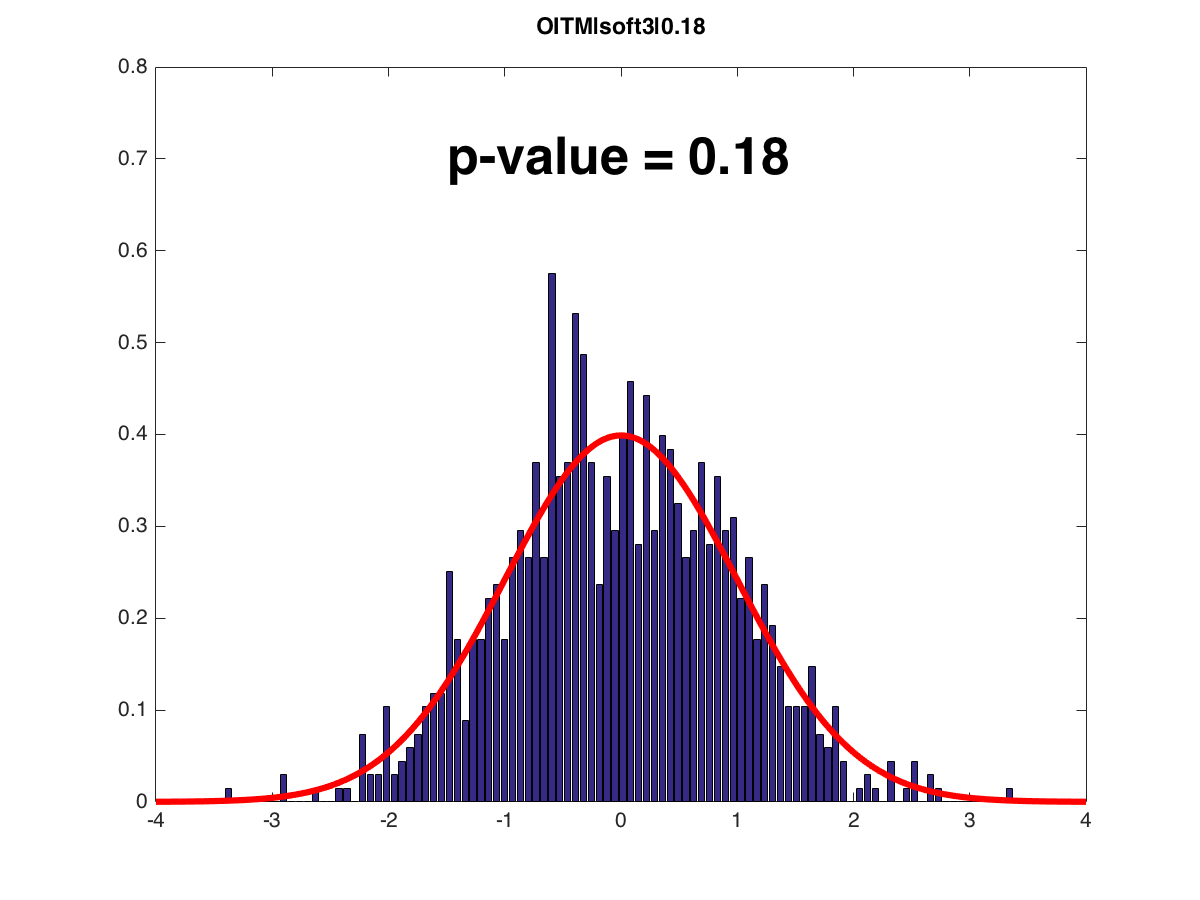} & 		\includegraphics[trim = 40 20 35 27,
		clip, width = 0.23\textwidth]{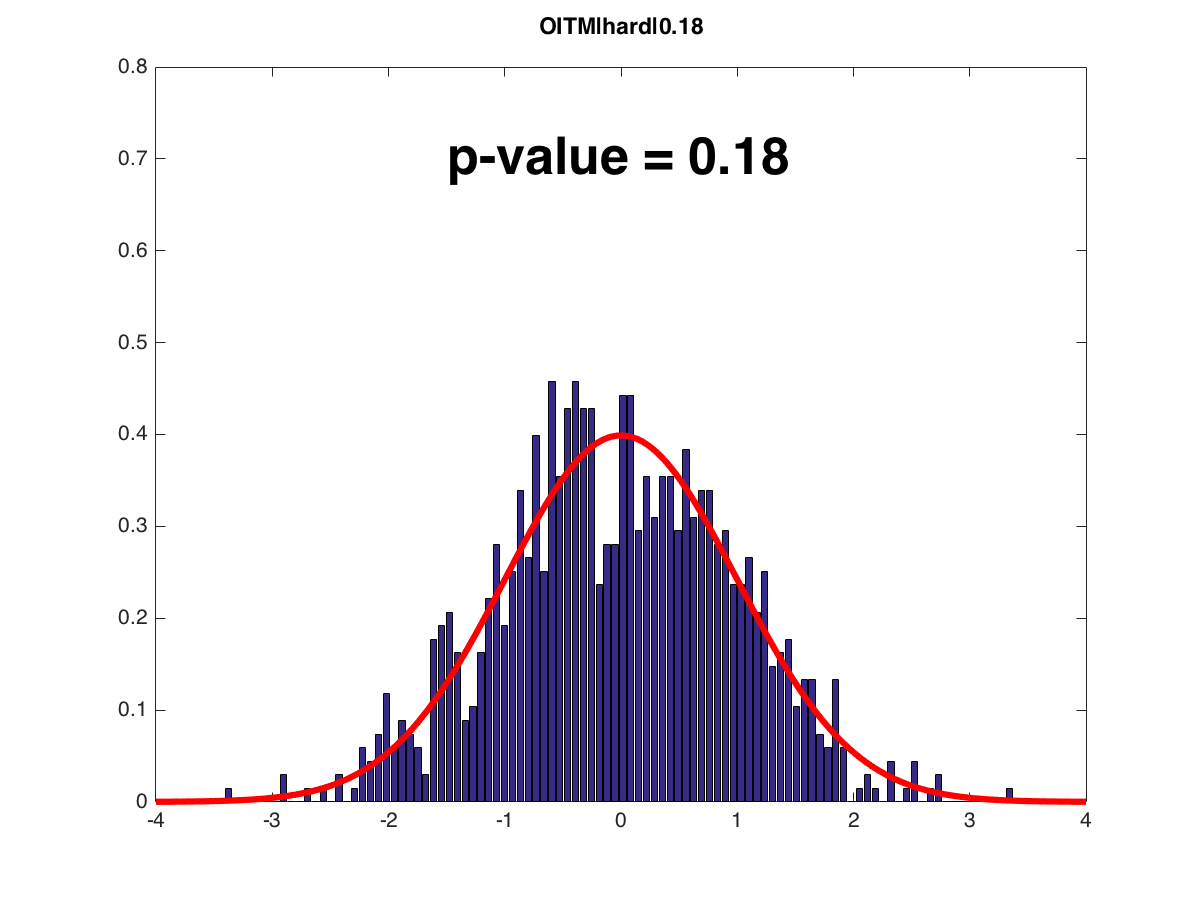} 	
	\end{tabular} 
\end{figure}

To further emphasize the distinction between the distribution of the standardized scores
at {\em each} location presented above from that {\em across} locations,
Figure~\ref{fig:univariate.hist} shows histograms of the latter for a single simulated
realization. Except for the standard GMM method, most histograms match the standard
normal distribution closely and are indistinguishable from it according to the
Kolmogorov-Smirnov normality test. 

\subsection{Bivariate Data}
\label{subsection:bivariate}

\begin{figure} [h!]
	\caption{Simulated observations and background adjustment effects for Scenario A (no lesion). The 1st row shows the simulated original scans and the corresponding contrast, showing a global non-homogeneous background change. The 2nd row
		shows the respective standardized images via background adjustment and their difference.
		The adjusted observations and background difference are now randomly distributed around
		zero.} 	\label{fig:sim.A} 	\begin{tabular}{@{}{c}*{3}{c@{\hskip -1pt}}@{}} 		&
		{Scan 1} \rule{0.75cm}{0pt} & {Scan 2} \rule{0.75cm}{0pt} & Difference
		\rule{0.75cm}{0pt}\\ 		\begin{sideways} \rule[0pt]{0.7in}{0pt} Original
		\end{sideways} 		
		&\includegraphics[trim = 150 40 100 28, clip, width=0.32\linewidth]{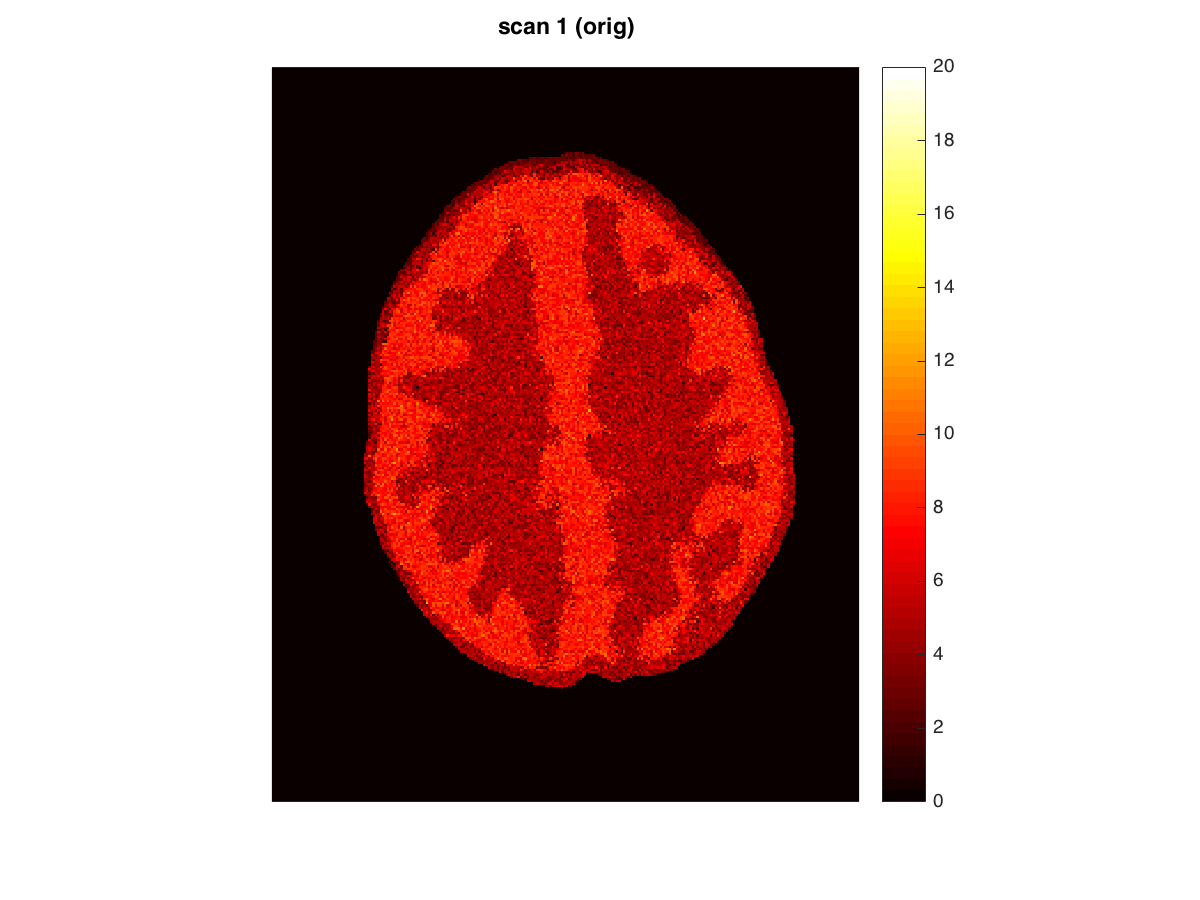} 		& \includegraphics[trim = 150 40 100 28,
		clip,width=0.32\linewidth]{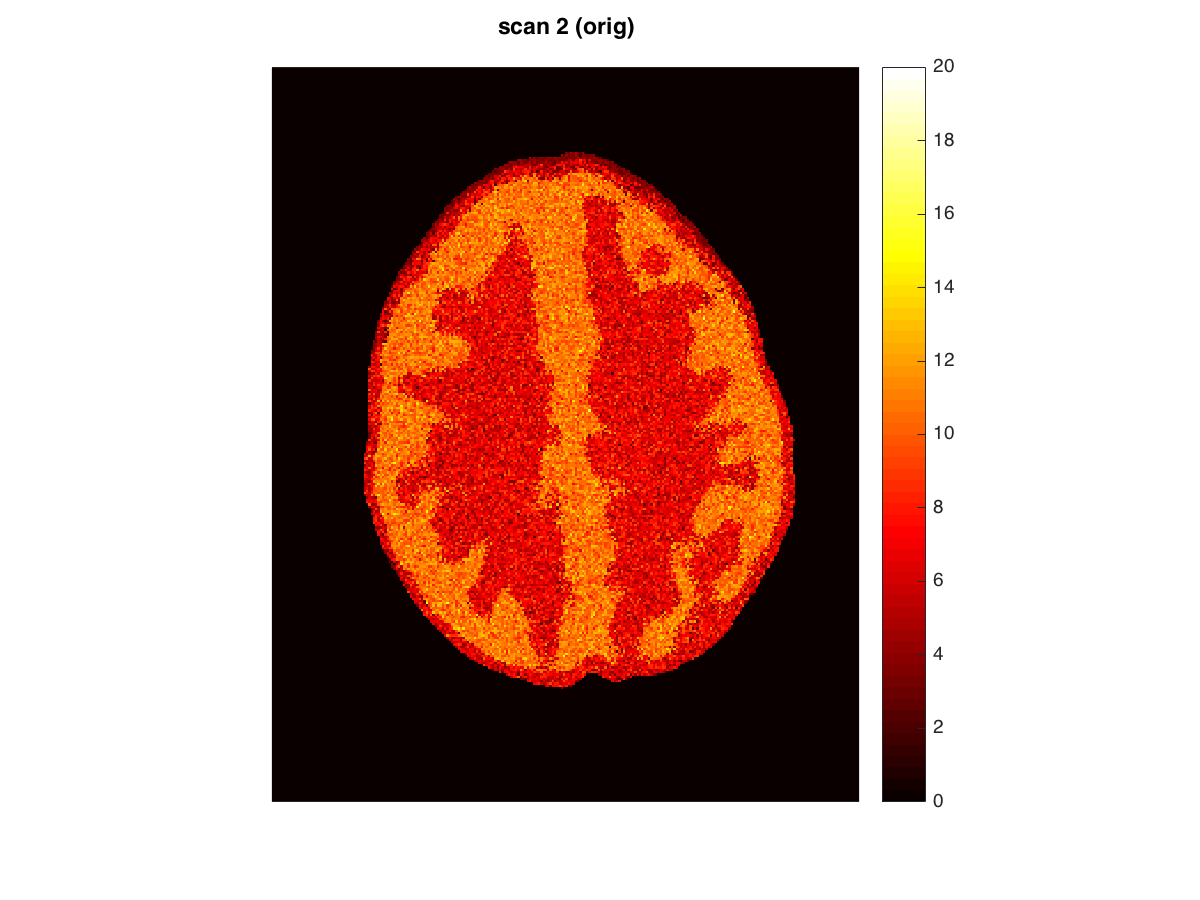} 		& \includegraphics[trim = 150 40 100
		28, clip,width=0.32\linewidth]{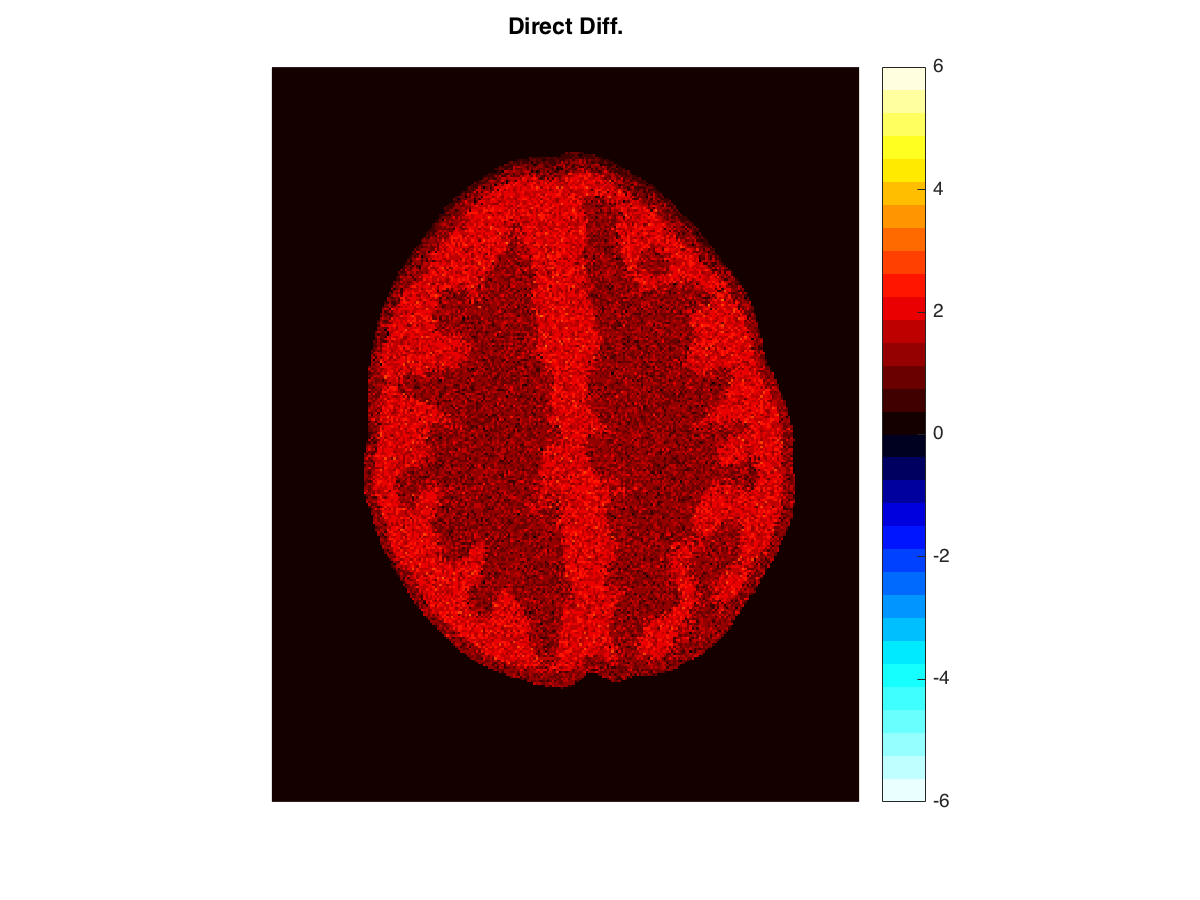} \\ 		\begin{sideways}
			\rule[0pt]{0.25in}{0pt} Background Adjustment \end{sideways} 		
		&\includegraphics[trim = 150 30 100 28, clip,width=0.32\linewidth]{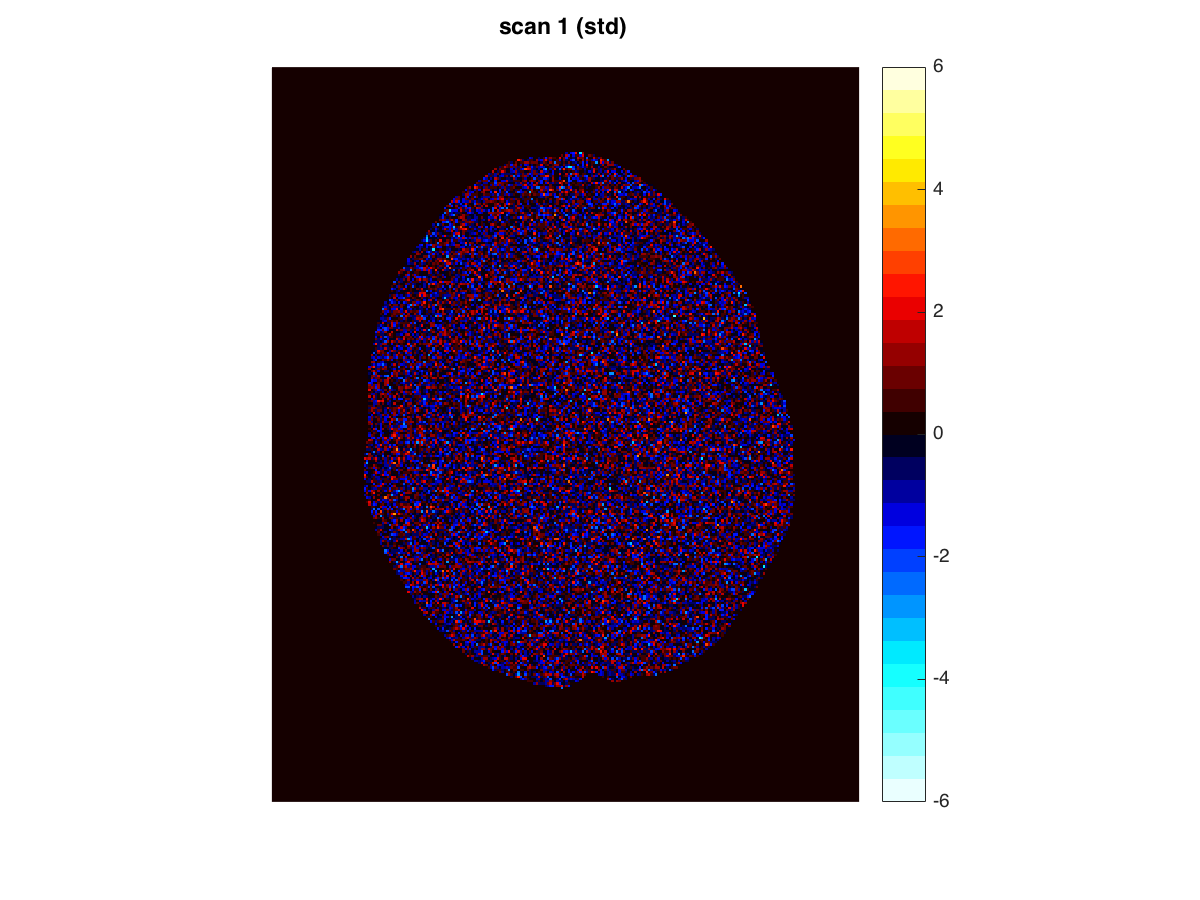} 		 
		& \includegraphics[trim = 150 30 100 28, clip,width=0.32\linewidth]{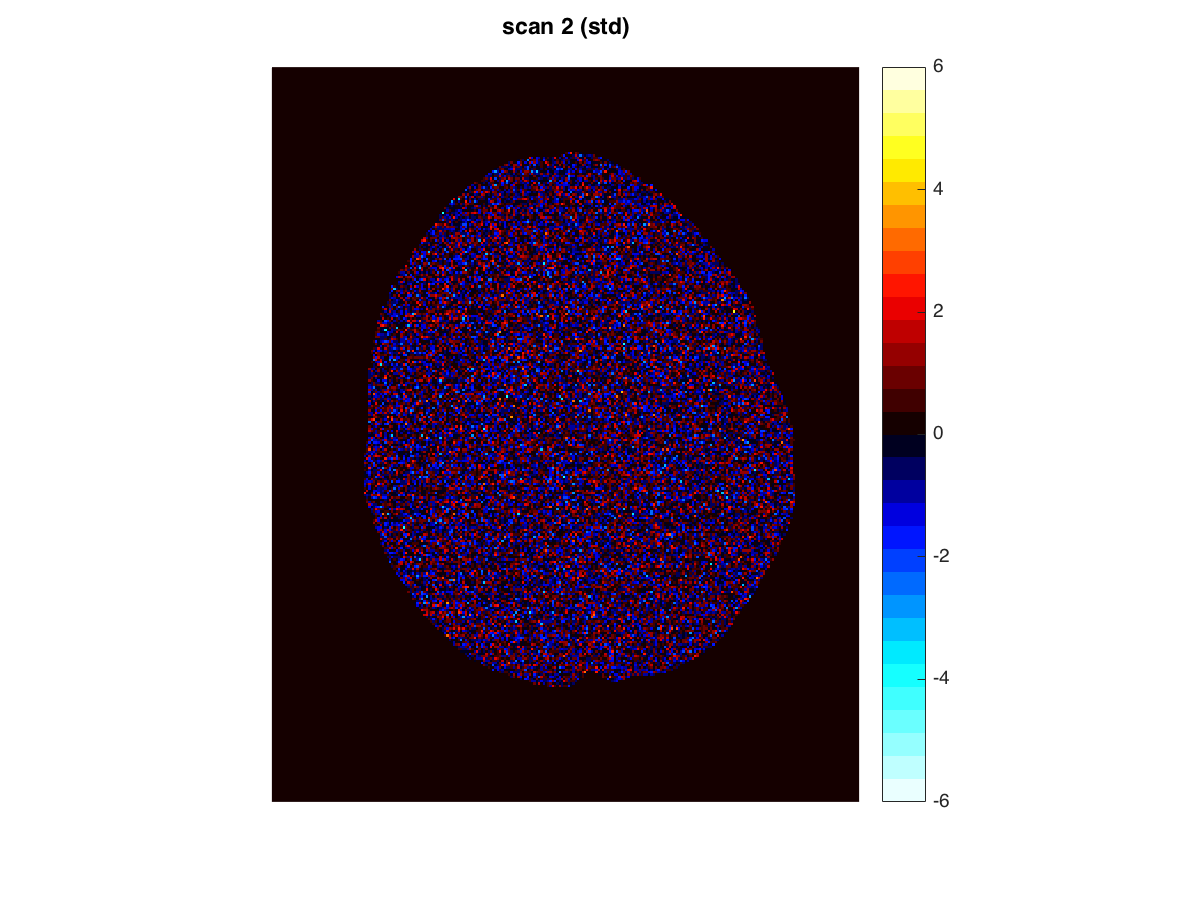} 		 
		& \includegraphics[trim = 150 30 100 28, clip,width=0.32\linewidth]{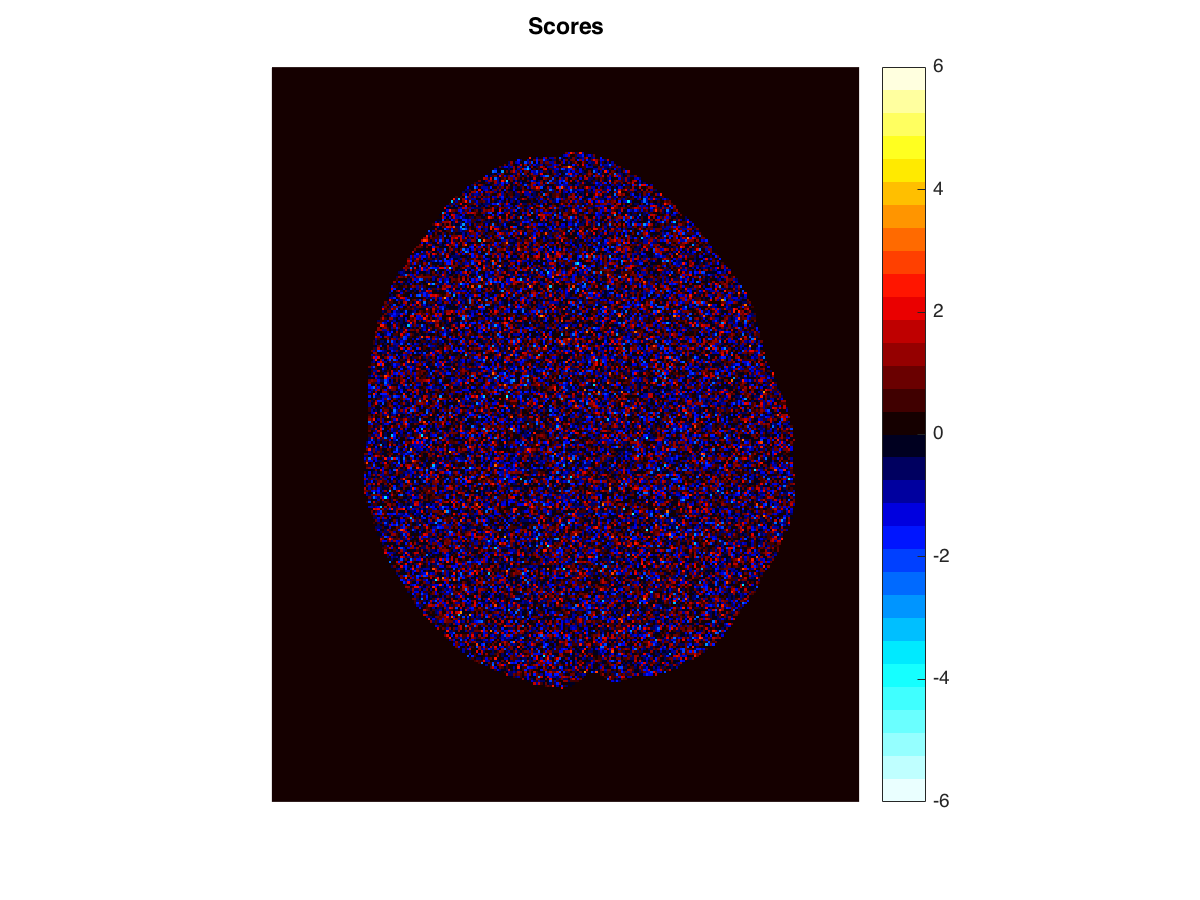} 	 
	\end{tabular} 
\end{figure}

\begin{figure}[h!]
	\caption{Simulated observations and background adjustment effects for Scenario B (lesion). The 1st row shows the simulated original scans and the corresponding contrast, showing a global non-homogeneous background change and little
		change in the lesion. The 2nd row shows the respective standardized images via background adjustment and their difference. The adjusted observations and background difference are
		now randomly distributed around zero and the lesion change is clearly visible.} 	
	\label{fig:sim.B} 	\begin{tabular}{@{}{c}*{3}{c@{\hskip -1pt}}@{}} 		& {Scan 1} \rule{0.75cm}{0pt} & {Scan 2} \rule{0.75cm}{0pt} & Difference
		\rule{0.75cm}{0pt}\\ 		\begin{sideways} \rule[0pt]{0.7in}{0pt} Original
		\end{sideways} 			
		&\includegraphics[trim = 150 30 100 28, clip, width=0.32\linewidth]{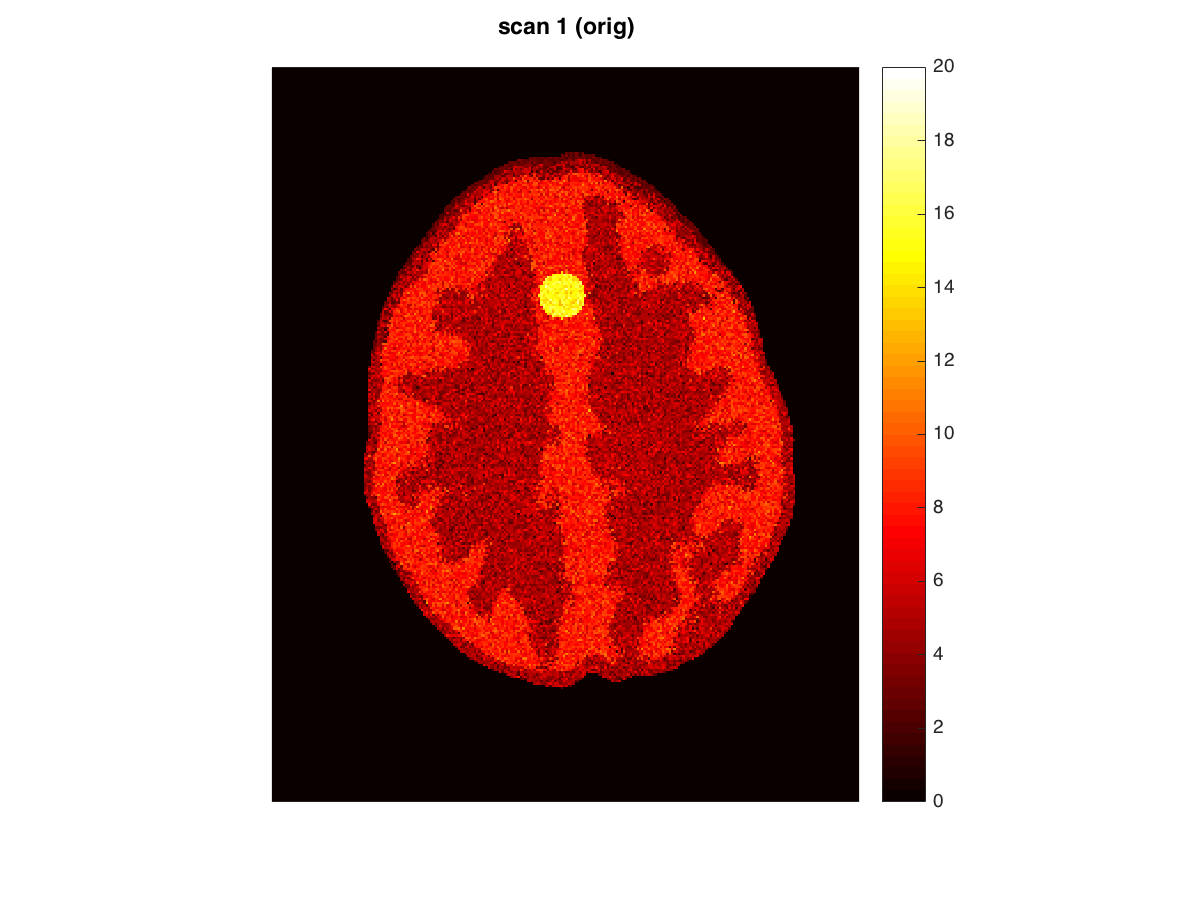} 		& \includegraphics[trim = 150 30 100 28,
		clip,width=0.32\linewidth]{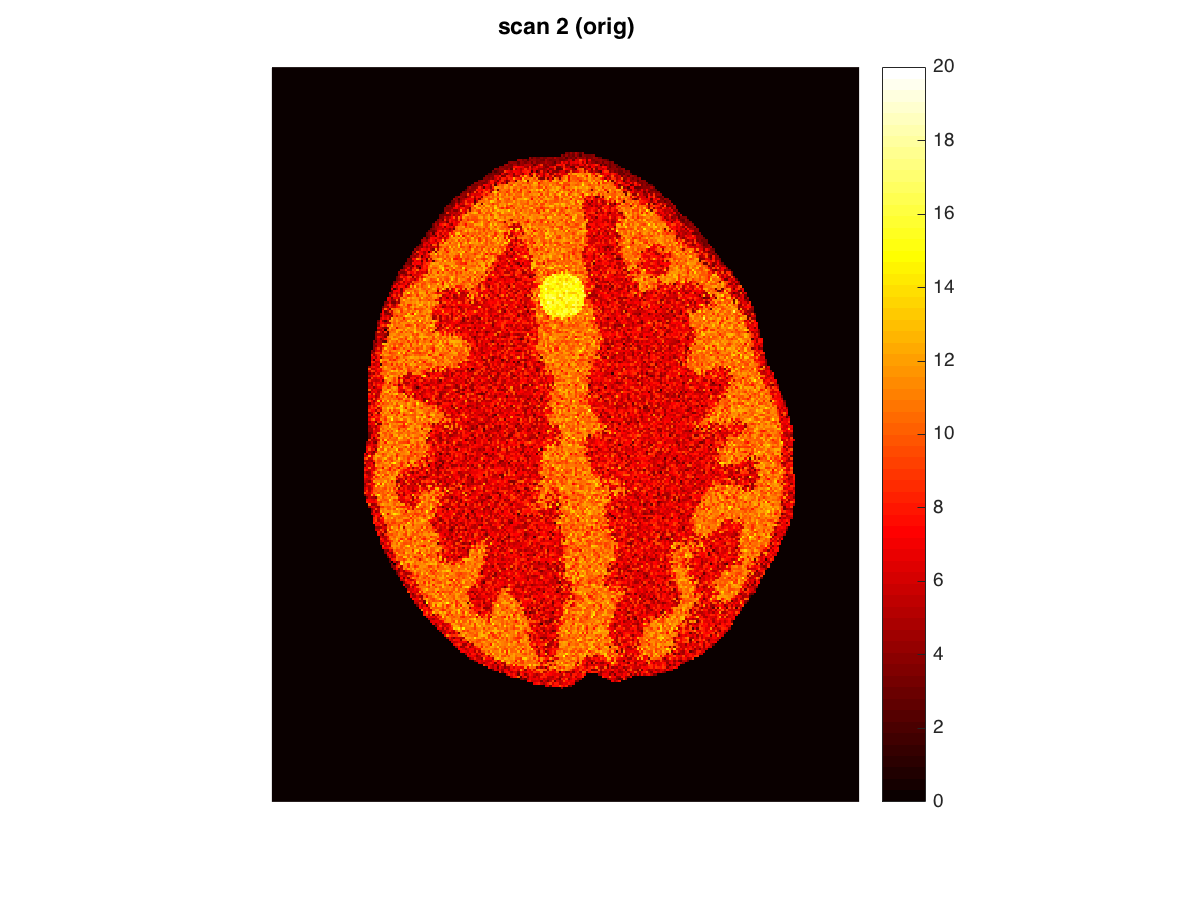} 		& \includegraphics[trim = 150 30 100
		28, clip,width=0.32\linewidth]{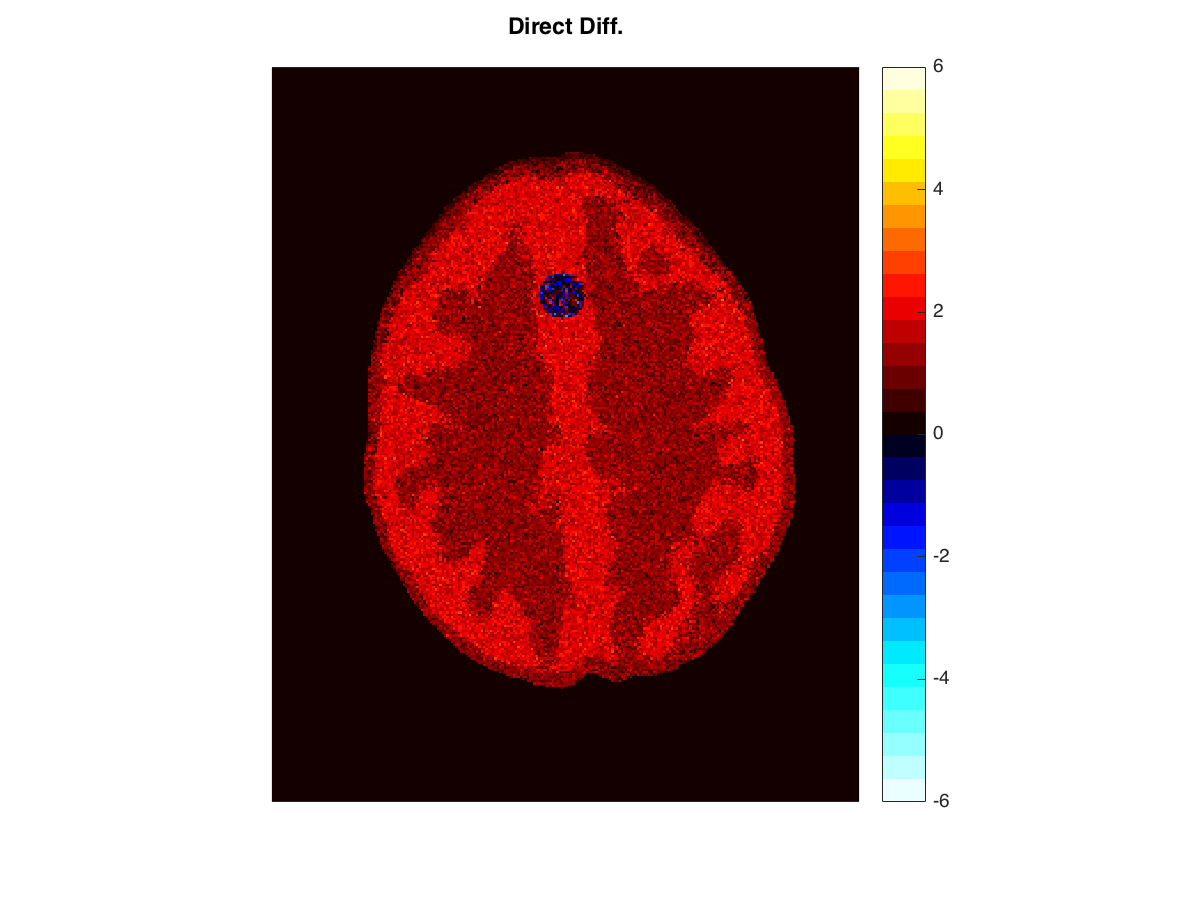} \\ 		\begin{sideways}
			\rule[0pt]{0.25in}{0pt} Background Adjustment \end{sideways} 		
		&\includegraphics[trim = 150 30 100 28, clip,width=0.32\linewidth]{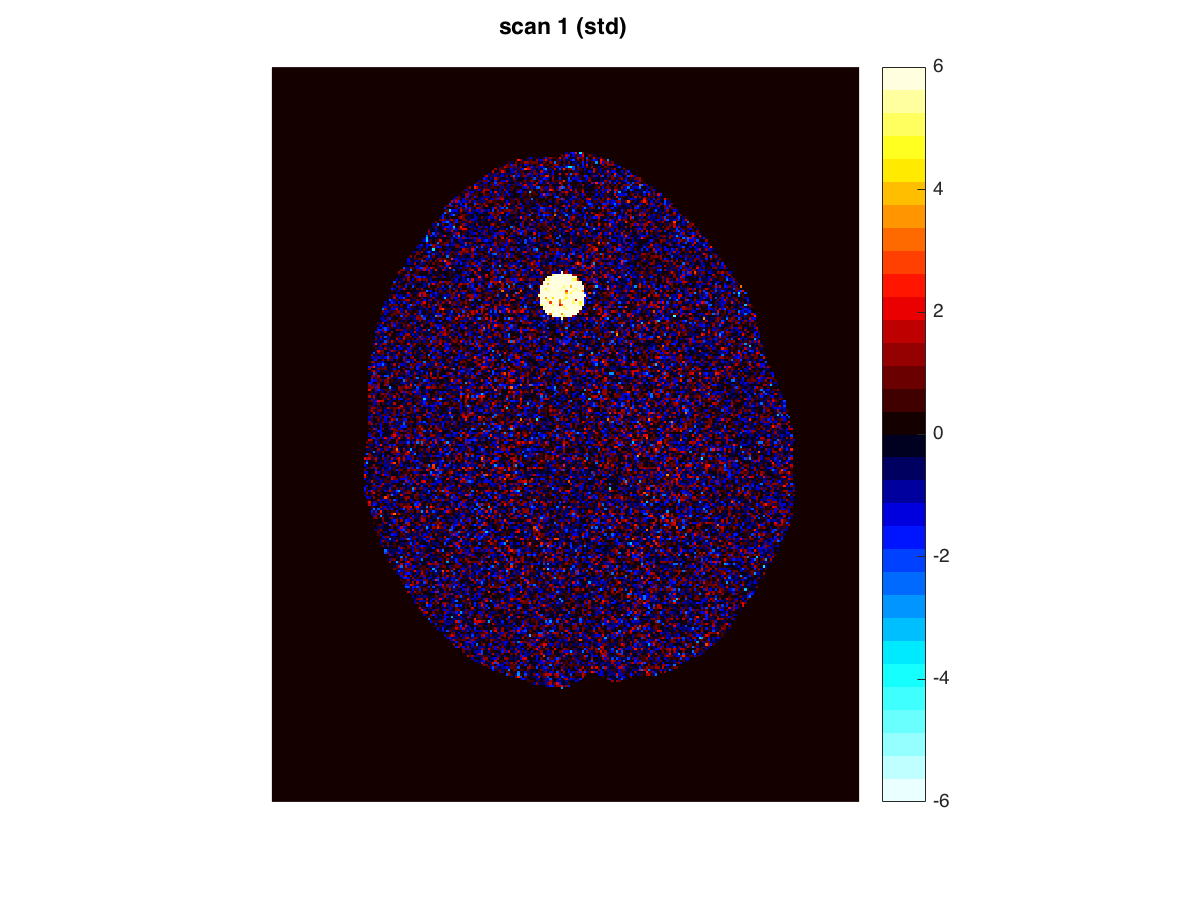} 		 
		& \includegraphics[trim = 150 30 100 28, clip,width=0.32\linewidth]{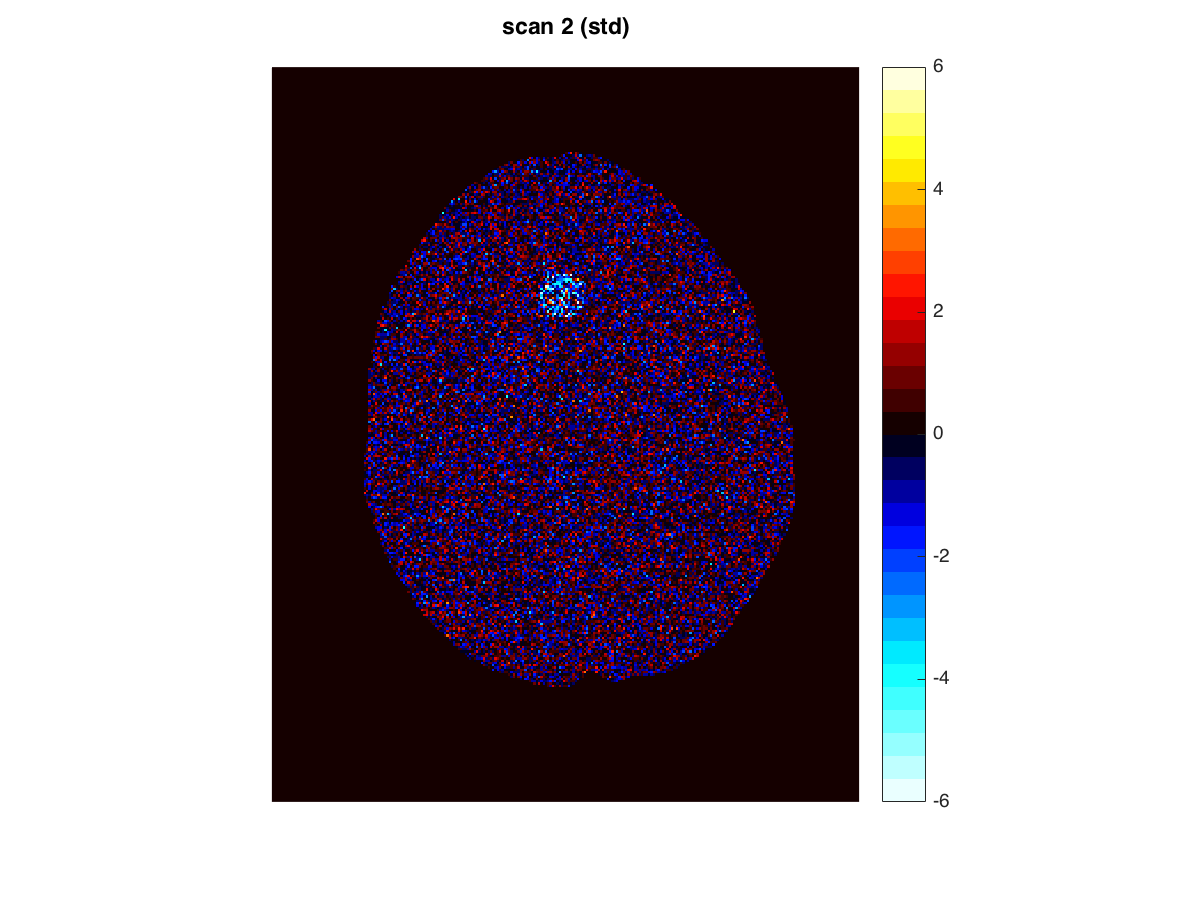} 		
		& \includegraphics[trim = 150 30 100 28, clip,width=0.32\linewidth]{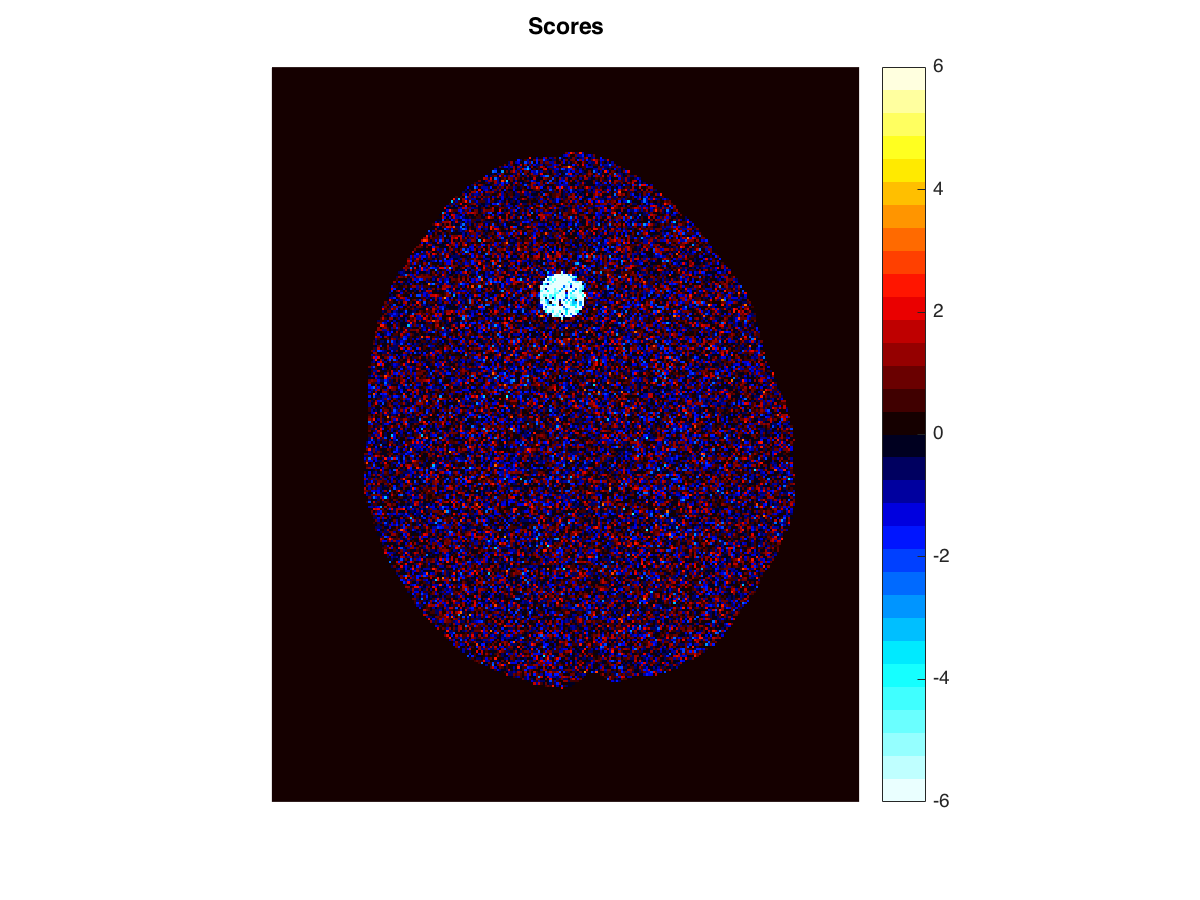} 	
	\end{tabular} 
\end{figure}

In this section, we conduct simulations for bivariate data
($p = 2$), where the two scans correspond to the baseline (BL) and week-1 (W1) scans. We
generate data from model~\eqref{GMM} and~\eqref{eq:b2pi} with $K = 3$, where the three
classes correspond to GM, WM and CSF respectively. We use the probability maps provided by the software SPM as the population membership
probability templates, allowing us to generate PET-like images. The template weights are given as 
$\gamma = (0.94,0.01,0.05)$ and the parameters for each
class are specified as 
\begin{align} \label{eq:data.mean} 
&\mu_1  = (4.91,6.68)^T,  \quad\quad \quad \mu_2  =
(8.04,10.77)^T,  \quad\quad\quad\mu_3 = (2.76, 3.71)^T;  \\
&\Sigma_1  = \left(\begin{array}{ll} 1.23 & 1.63  \\ 1.63 & 2.21
\end{array}\right), \quad\Sigma_2 = \left(\begin{array}{ll} 1.28 & 1.34  \\ 1.34 & 1.61
\end{array}\right), \quad \Sigma_3 = \left(\begin{array}{ll} 0.24  &  0.31\\ 0.31  &
0.44 \end{array}\right). \end{align} These parameters are close to the values that are obtained
by applying the proposed robust EM algorithm to the real data (see
Section~\ref{section:data.application} below). Focusing on a single slice for simplicity,
each simulated PET scan is a 320 by 256 matrix where each pixel is a unit square. We consider two scenarios: (A) a PET image without {lesions} and (B) a PET image with a
lesion. For Scenario B, a lesion in the shape of a circle with radius equal to 10 pixels is added, where
the intensity at each pixel follows a $N(15, 1^2)$ {distribution} independently.
Figures~\ref{fig:sim.A} and~\ref{fig:sim.B} plot an example for each scenario. Background adjustment is performed using soft assignment with the transformation
$T_S^{(1)}$. (Additional simulations considering other lesion sizes and shapes are included in the Supplementary Materials.)

We first investigate the performance of all the three methods (i.e., traditional GMM, SGMM and RB-SGMM) in terms of parameter estimation for both scenarios. We use the
following metrics to evaluate the differences between the estimated parameters
and the true parameters: the Euclidean norm is used for the means and the 2-norm (i.e. maximum singular value of the difference between two matrices) is used for the covariance matrices and probability template map matrices $\pi = \{\pi_{ik}\}$.
The results aggregating 1000 simulated instances are reported in Tables~\ref{table:par.est.AB}. The proposed SGMM and RB-SGMM are almost uniformly better than the traditional GMM for all parameters (except for $\Sigma_2$ in Scenario B where SGMM is slightly worse than GMM). This is expected since SGMM and RB-SGMM use the spatial population probability maps as priors while GMM uses constant probability maps. We also observe that RB-SGMM leads to similar accuracy compared to SGMM when there are no {lesions} in the observation
(Scenario A in Table~\ref{table:par.est.AB}) and is much better than SGMM when there are lesions, i.e., outliers (Scenario B in Table~\ref{table:par.est.AB}). This indicates that the proposed RB-SGMM is
preferred in applications since it is accurate without outliers but robust when outliers
are present.

\begin{table}[!ht]
	\centering
	\caption{
		Performance of the three methods GMM (Gaussian Mixture Models), SGMM (Spatial
		GMM) and RB-SGMM (robust SGMM) in terms of parameter estimation in Scenarios A and B. The norms
		of the differences between the estimated parameters and the true parameters are reported,
		below which is the standard error multiplied by 100 (in parenthesis).} \label{table:par.est.AB}
	\setlength{\tabcolsep}{2pt}
	\resizebox{\textwidth}{!}{
		\begin{tabular}{c|*{7}{c}|*{7}{c}}
			& \multicolumn{7}{c|}{Scenario A} &  \multicolumn{7}{c}{Scenario B}  \\ \hline  
			Method & $\mu_1$ & $\mu_2$ & $\mu_3$ & $\Sigma_1$ & $\Sigma_2$ & $\Sigma_3$ & $\pi$ & $\mu_1$ & $\mu_2$ & $\mu_3$ & $\Sigma_1$ & $\Sigma_2$ & $\Sigma_3$ & $\pi$ \\ \hline \hline
			GMM & 2.42 & 3.39 & 5.46 & 0.80 & 0.54 & 2.23 & 122.56 & 6.26 & 4.57 & 9.63 & 1.29 & 1.13 & 2.24 & 139.44 \\  
			SE $(\times 10^{-2})$ & $(6.53)$ & $(8.58)$ & $(8.80)$ & $(1.37)$ & $(1.48)$ & $(2.21)$ & $(22.81)$ & $(16.32)$ & $(10.48)$ & $(17.40)$ & $(2.51)$ & $(2.39)$ & $(3.36)$ & $(49.69)$ \\ \hline
			SGMM & 0.01 & 0.01 & 0.03 & 0.04 & 0.03 & 0.03 & 1.58 & 0.03 & 0.19 & 0.03 & 0.08 & 1.39 & 0.03 & 1.44 \\  
			SE $(\times 10^{-2})$ & $(0.03)$ & $(0.03)$ & $(0.06)$ & $(0.08)$ & $(0.06)$ & $(0.07)$ & $(2.01)$ & $(0.05)$ & $(0.04)$ & $(0.06)$ & $(0.12)$ & $(0.11)$ & $(0.07)$ & $(1.88)$ \\ \hline
			RB-SGMM & 0.01 & 0.01 & 0.02 & 0.06 & 0.03 & 0.03 & 1.63 & 0.01 & 0.07 & 0.02 & 0.04 & 0.21 & 0.03 & 0.83 \\  
			SE $(\times 10^{-2})$ & $(0.03)$ & $(0.03)$ & $(0.06)$ & $(0.11)$ & $(0.07)$ & $(0.07)$ & $(2.01)$ & $(0.03)$ & $(0.04)$ & $(0.05)$ & $(0.09)$ & $(0.10)$ & $(0.07)$ & $(1.38)$ \\ 
	\end{tabular} }
\end{table}

\begin{figure} 	\centering 	
	\caption{Bivariate density contour plots of the observations with and without soft
		background adjustment within each class. The first column is for the original {
			simulated} observations, while the last three columns are the  scores after background
		adjustment using the methods GMM, SGMM and RB-SGMM respectively. The two rows correspond
		to Scenarios A and B.} 	\label{figure:bi.sim.contour} 	\begin{tabular}{c*{4}{c@{\hskip
					-2pt}}} 		\begin{sideways} \rule[0pt]{0.23in}{0pt} Scenario A \end{sideways} 		 &
		\includegraphics[trim = 80 25 82 25, clip, width =
		0.24\textwidth]{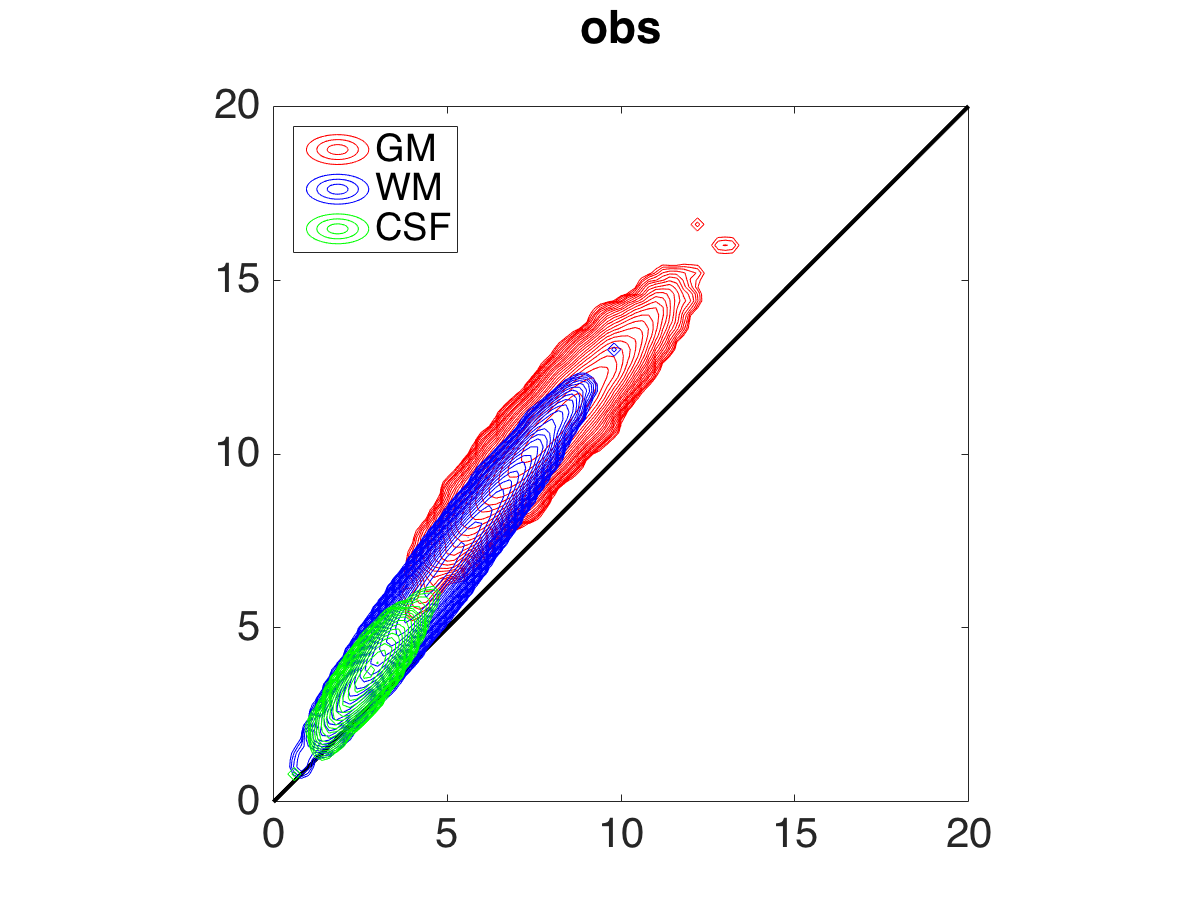} 		& \includegraphics[trim = 80
		25 82 25, clip, width = 0.24\textwidth]{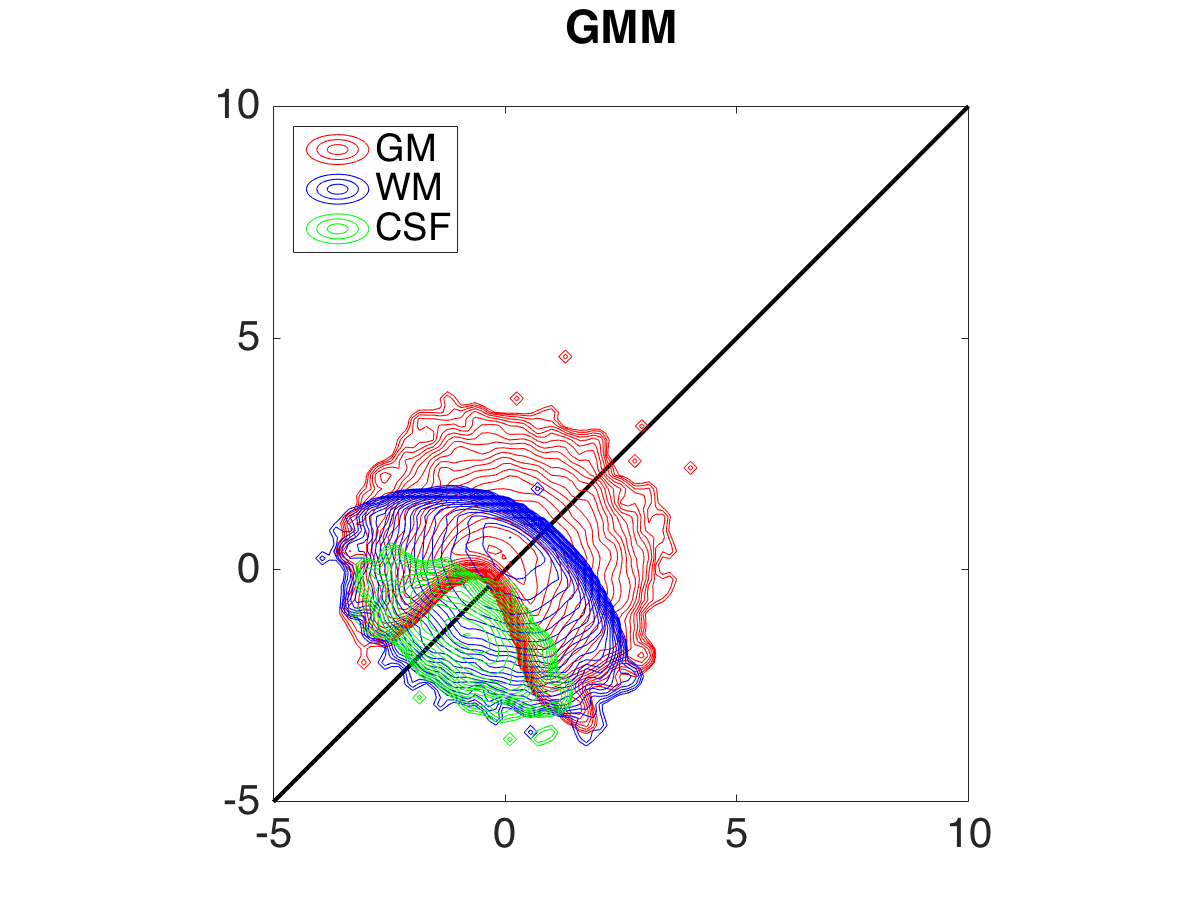} 		&
		\includegraphics[trim = 80 25 82 25, clip, width =
		0.24\textwidth]{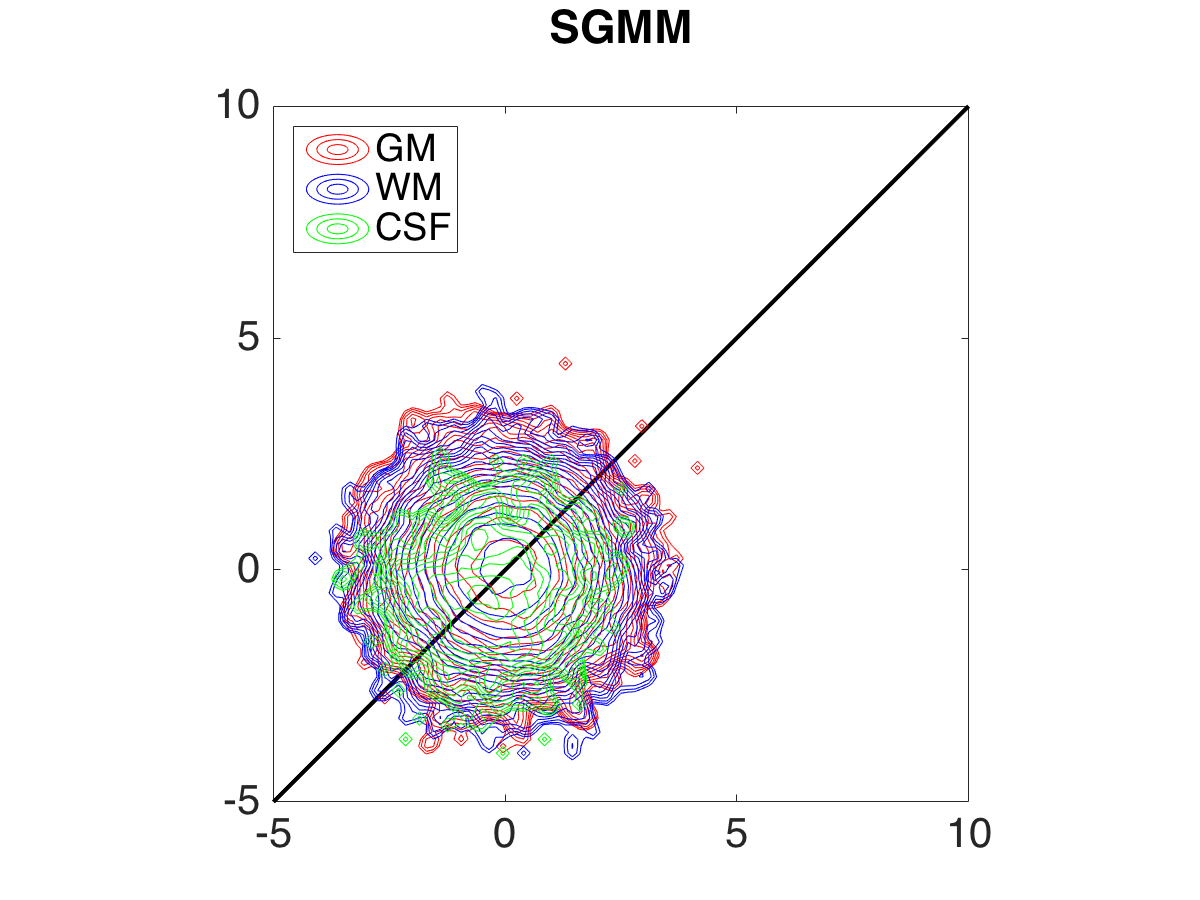} 		& \includegraphics[trim = 80
		25 82 25, clip, width = 0.24\textwidth]{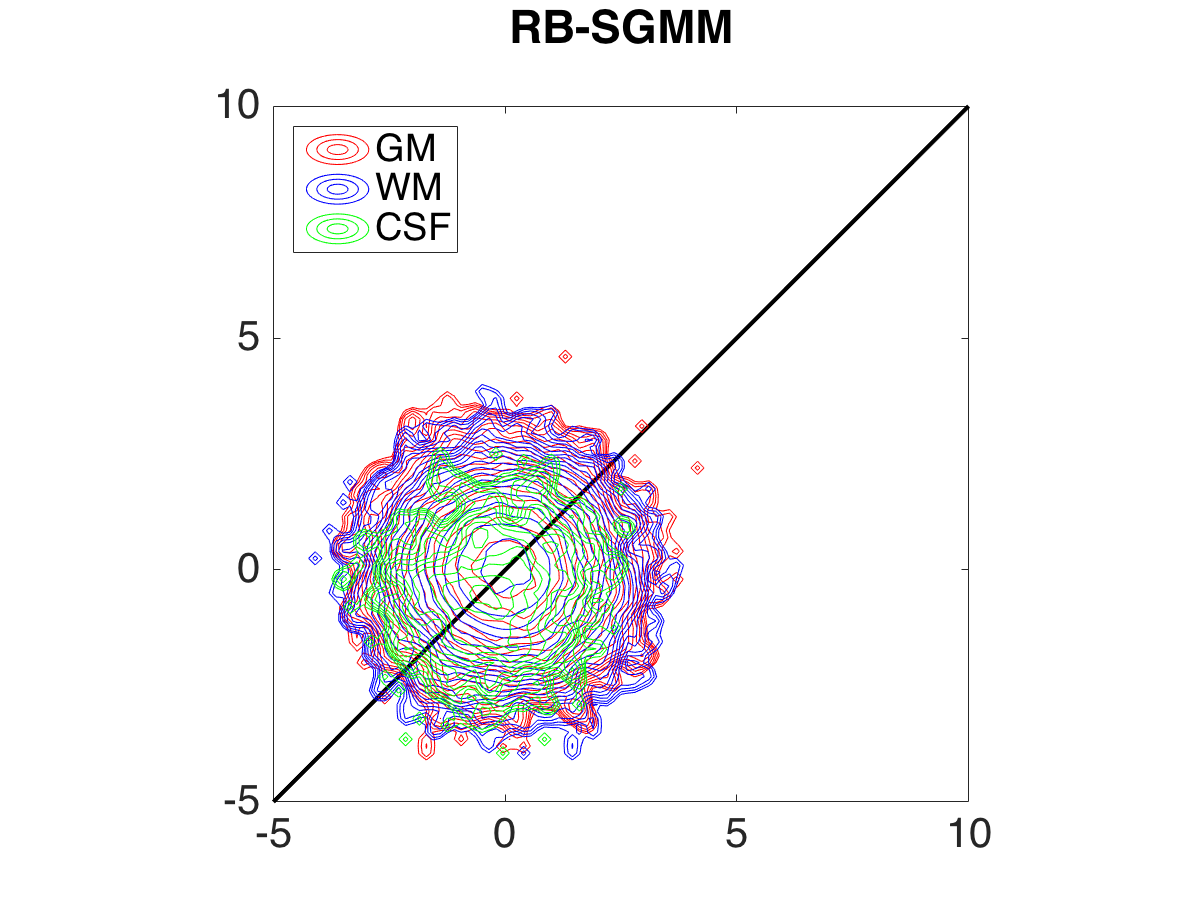} \\ 		 
		\begin{sideways} \rule[0pt]{0.23in}{0pt} Scenario B \end{sideways} 		&
		\includegraphics[trim = 80 25 82 25, clip, width =
		0.24\textwidth]{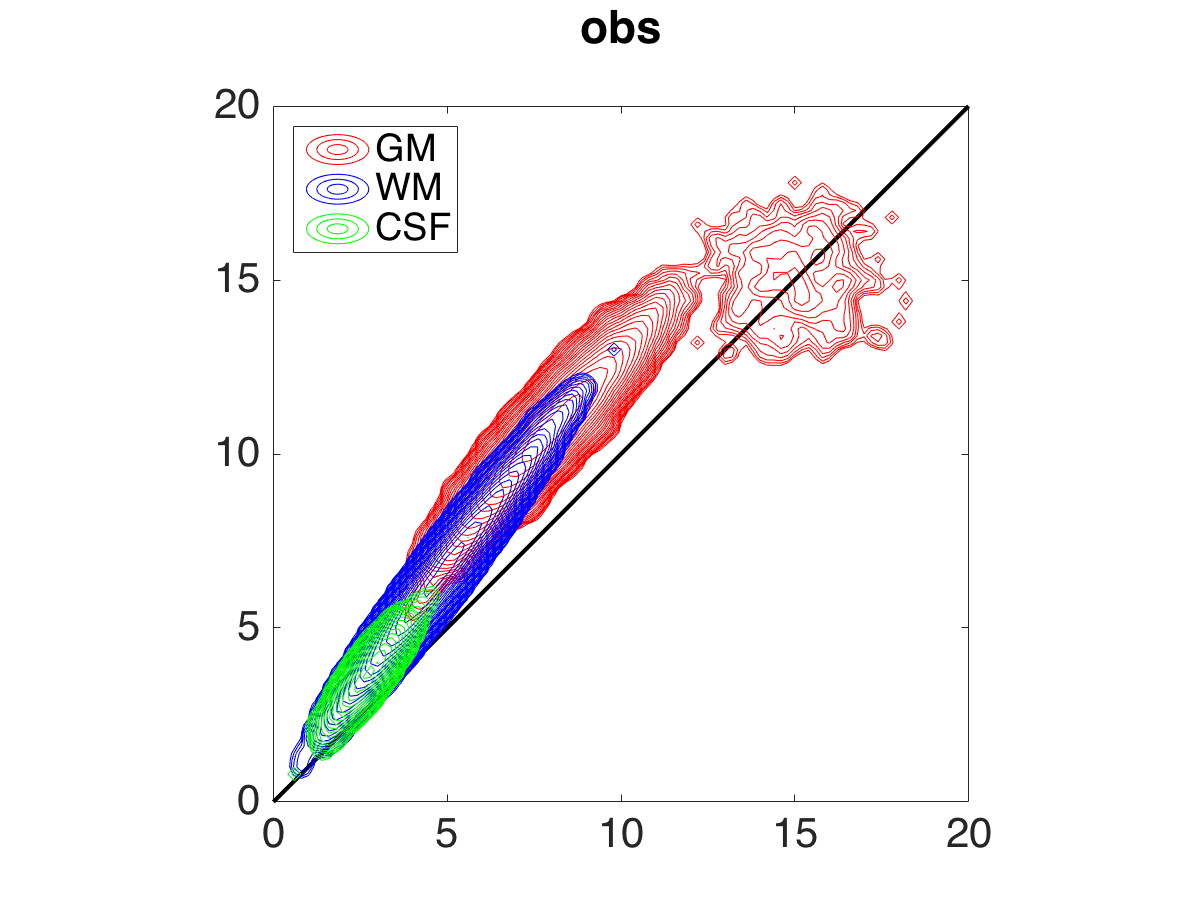} 		&  \includegraphics[trim = 80
		25 82 25, clip, width = 0.24\textwidth]{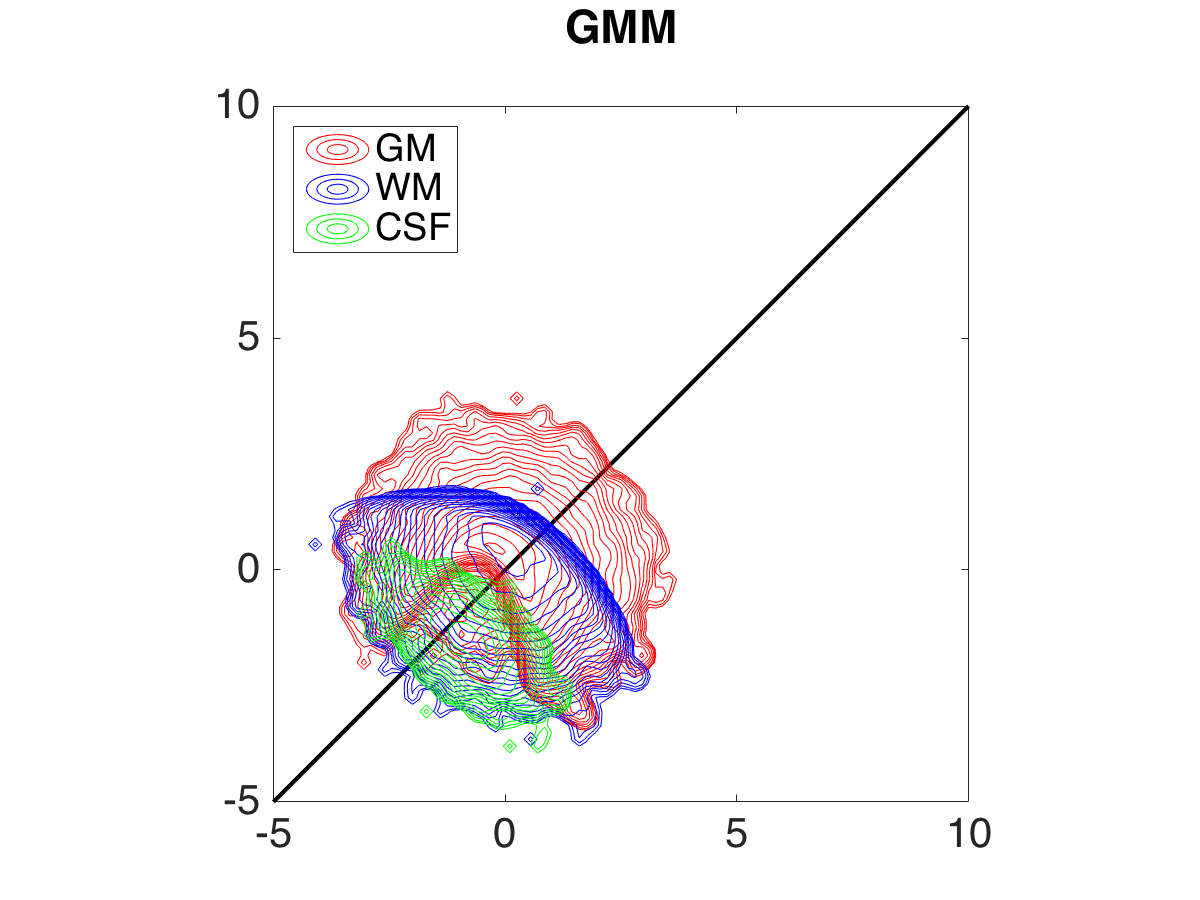} 		&
		\includegraphics[trim = 80 25 82 25, clip, width =
		0.24\textwidth]{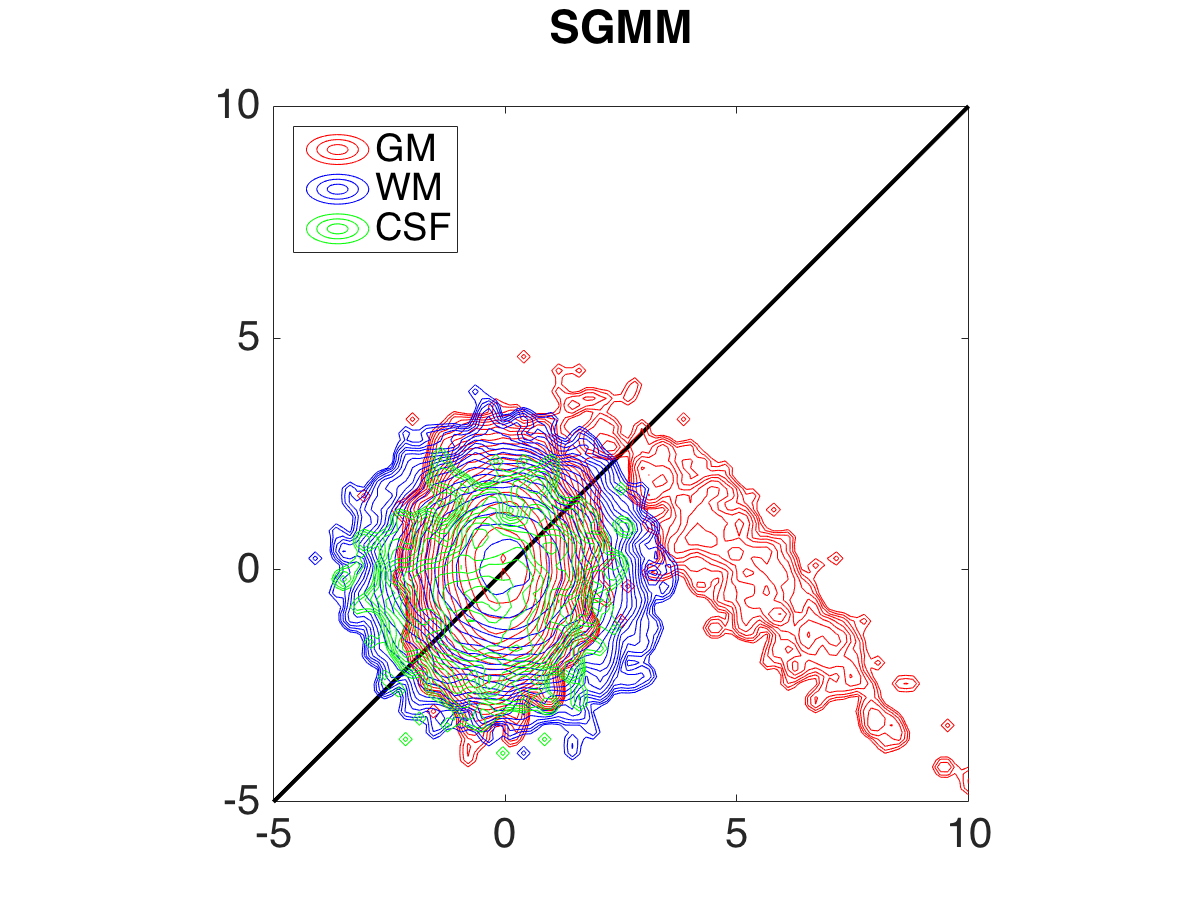} 		& \includegraphics[trim = 80
		25 82 25, clip, width = 0.24\textwidth]{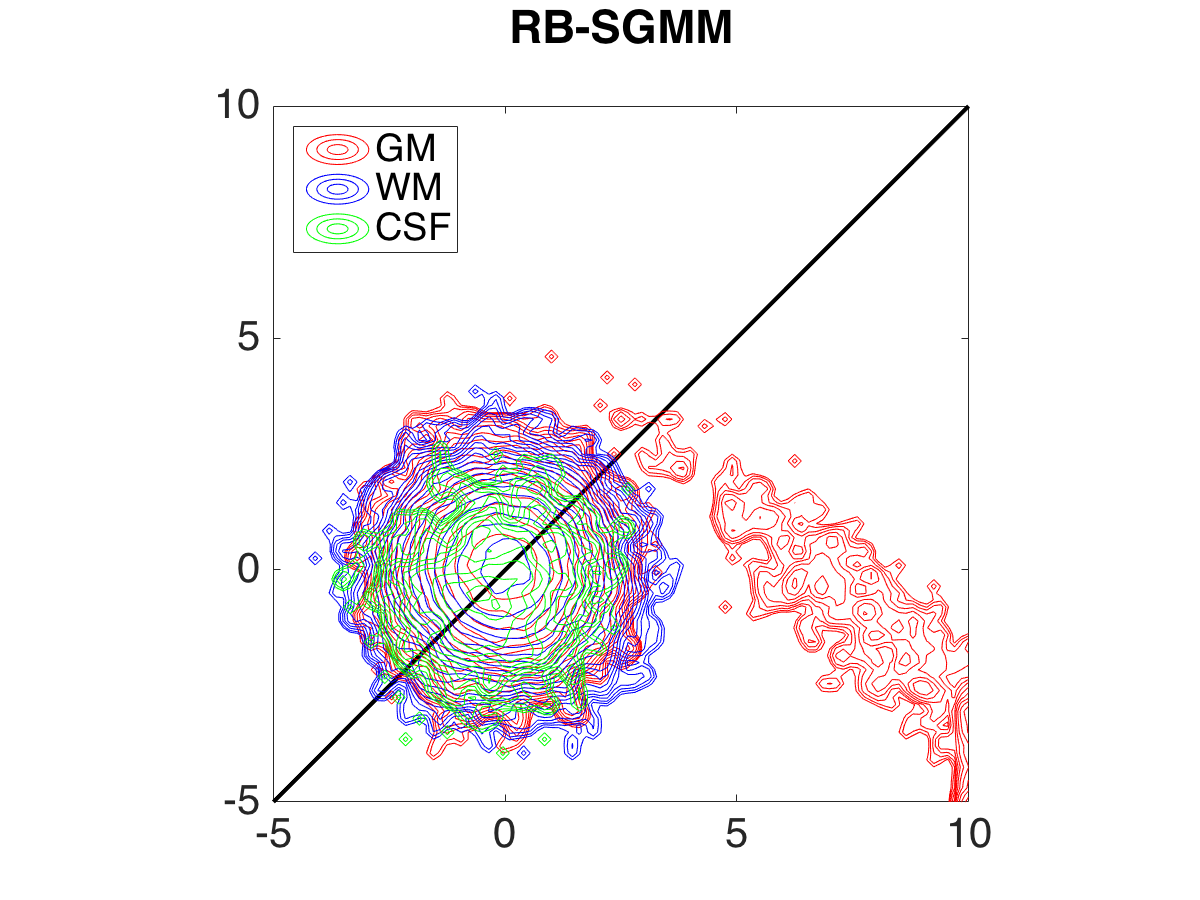} \\ 		 
		& Raw Observation & GMM & SGMM  & RB-SGMM 	
	\end{tabular} 
\end{figure}

To better appreciate the distribution of the standardized scores,
Figure~\ref{figure:bi.sim.contour} plots the density contour plots of the two scans from
a single simulation. While the standardized scores $T_S^{(1)}$ are obtained using soft
assignment, the three classes are separated in the plots by hard assignment to ease
visualization. In Scenario A, we can see that the original simulated observation has a
mixture structure with correlation, while the standardized scores are given a
distribution close to bivariate standard normal by the SGMM and RB-SGMM methods (first
row, last two columns). The GMM method fails to do so because of the lack of use of
spatial information and the inaccurate parameter estimates in Table~\ref{table:par.est.AB}. The RB-SGMM method shows its advantage over the SGMM method in
Scenario B, offering better standardization (particularly in the grey matter) and better
separation of the tumor pixels (second row, last two columns).

While Figure \ref{figure:bi.sim.contour} describes the distribution of the standardized scores across voxels, it is important to make sure that the tail probabilities dictated by the standard normal distribution is also valid at each voxel. To show this, we study the tail probabilities of $a^T T^{(1)}_S$ in Scenario A using the contrast $a = (-1, 1)^T/\sqrt{2}$, which is the one of interest in the data analysis. Similar to Section \ref{sec:oracle}, we evaluate the tail probabilities numerically using Monte Carlo simulation with $10^5$ replications at level $\alpha = 0.01$.

\begin{figure}
	\caption{Heatmaps of the size ratio $R(0.01)$ (left-side tail) for the bivariate simulations in Section~\ref{subsection:bivariate} with soft assignment: using the true parameters under various combinations of $(\pi_1, \pi_2)$ (a); using the estimated parameters by GMM (b) and SGMM (c).  }
	\label{fig:bi.sim.tail}
	\centering 
	\begin{tabular}{ccc}
		\includegraphics[trim = 30 15 52 28, clip, width = 0.31\textwidth, height = 0.19 \textheight]{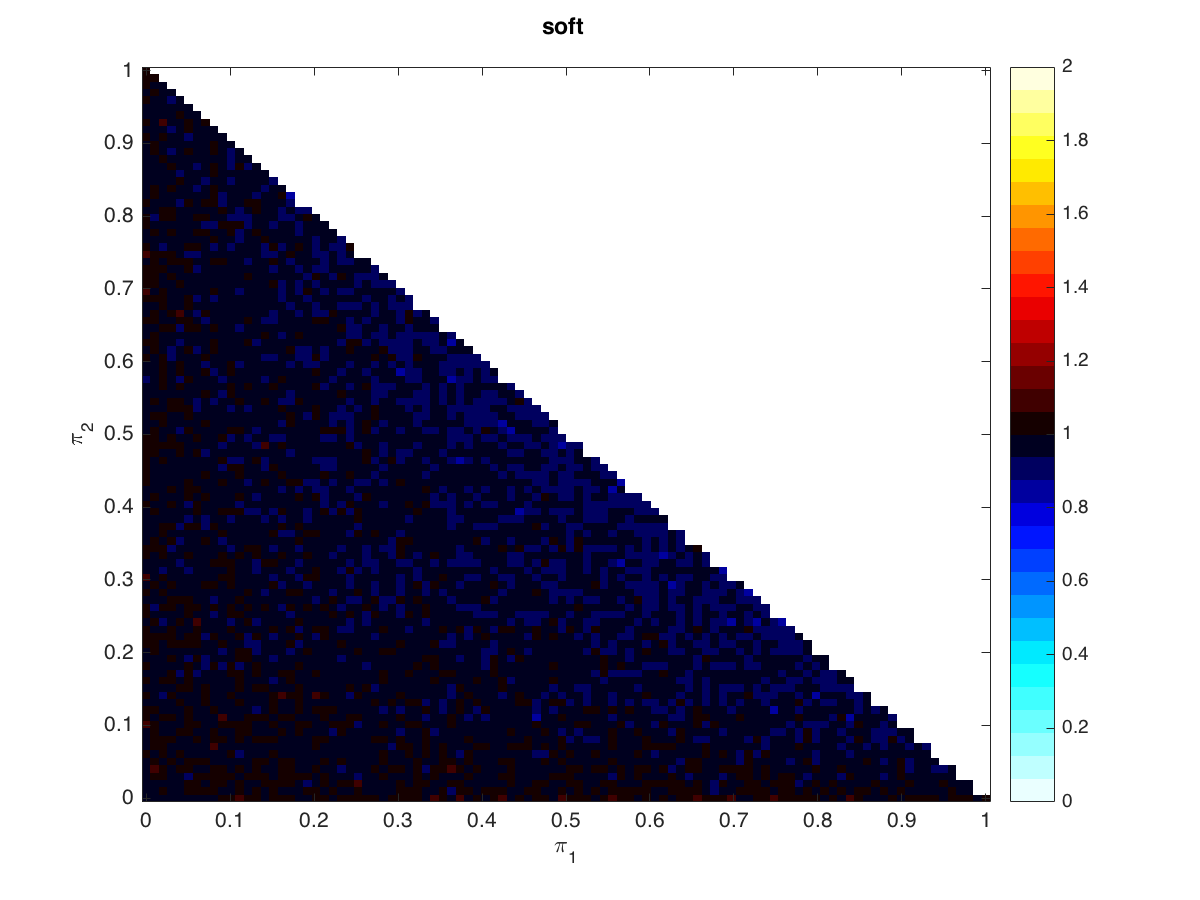}
		& \includegraphics[trim = 120 15 52 28, clip, width = 0.31\textwidth]{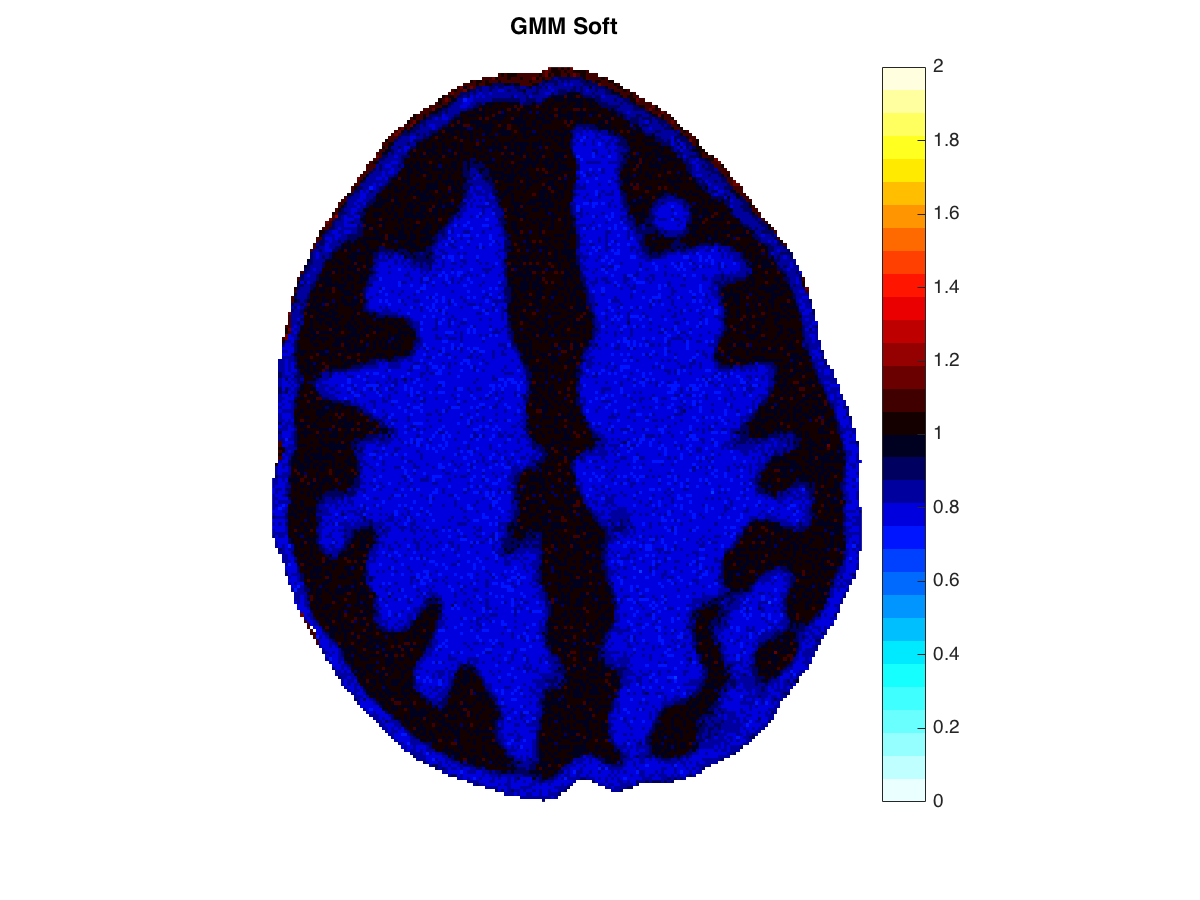}
		& \includegraphics[trim = 120 15 52 28, clip, width = 0.31\textwidth]{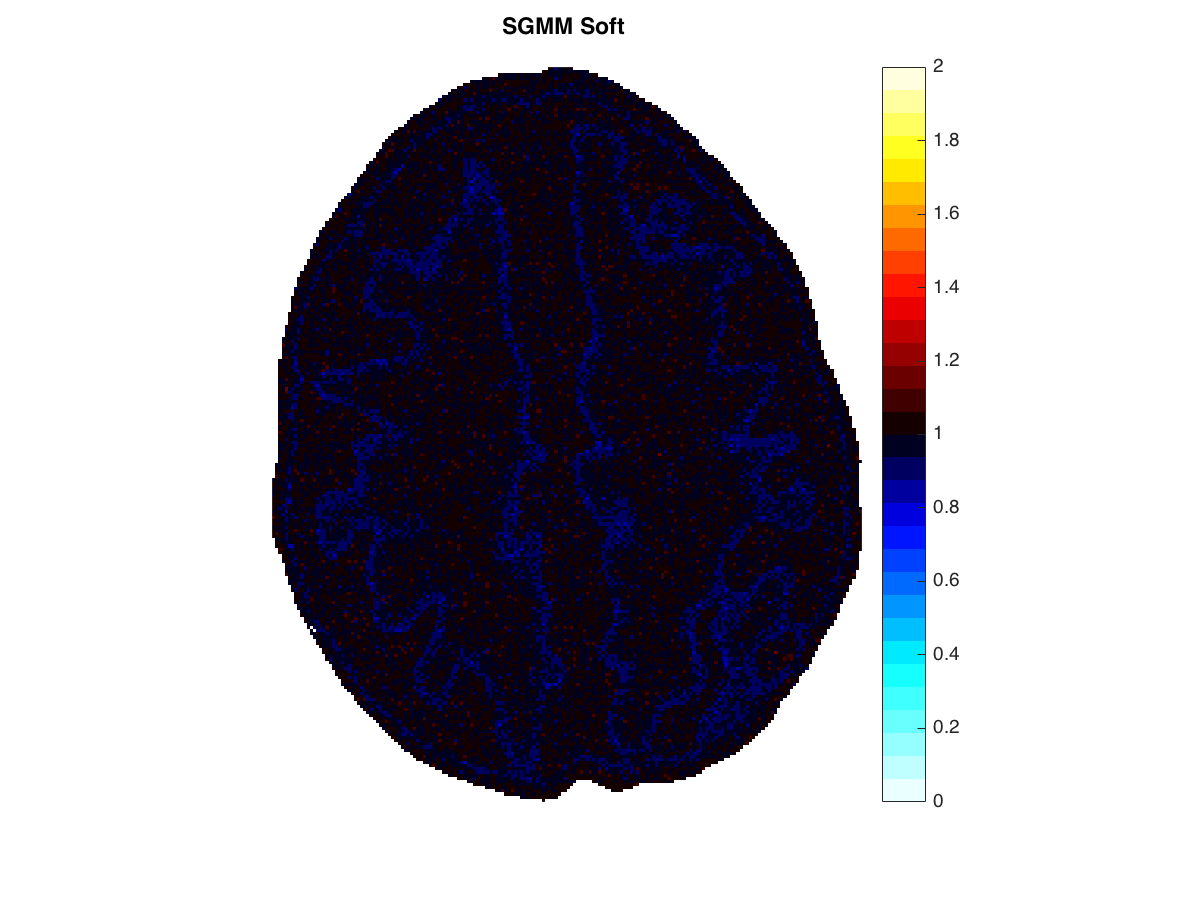} \\
		(a) Oracle & (b) GMM & (c) SGMM
	\end{tabular}
\end{figure}

Figure~\ref{fig:bi.sim.tail} plots the size ratio $R(0.01)$ at the left tail of the distribution of $a^T T^{(1)}_S$ using the true parameters (oracle) and the estimated parameters via GMM and SGMM. Panel (a) shows that the size ratio is very close to or smaller than 1 for almost all combinations of $(\pi_1, \pi_2)$ (note that $\pi_1 + \pi_2 = 1 - \pi_3 \le 1$). As expected from the univariate simulation results in Figure \ref{fig:univariate.right}, the GMM method (panel (b)) gives very conservative size ratios whose location depend highly on the anatomical region (in this case the white matter). In contrast, given the estimation accuracy of the SGMM method shown in Table \ref{table:par.est.AB}, it is not surprising that the voxelwise size ratios using SGMM (panel (c)) are very close to 1 for almost all voxels, being slightly conservative mainly in the thin transition region between the gray and white matter. This suggests that the standardized scores obtained via SGMM and soft assignment are valid for statistical inference according to the standard normal at each voxel, as desired. The results for hard assignment are very similar (not shown).

\section{PET data application} 
\label{section:data.application}
In this section, we provide more details about the PET data application discussed in the Introduction (Figure~\ref{fig:motivation}). We use the data produced by the lesion change detection study in~\citep{Qin+:} using the Hoffman 3-D brain phantom~\citep{Hoffman+:91}, which simulates pre- and post-treatment scans with a tumor lesion.
As described there, the brain phantom was filled with FDG radioactive fluid and PET scans were acquired on a GE Discovery ST PET-CT scanner. A malignant lesion was simulated within the central gray matter at a location superior and anterior within the brain, by placing a 1.5 cm diameter sphere of FDG. {The tumor-to-background-ratio (TBR) for Scan 1 was 2:1, which was changed to 1.5:1 for Scan 2. Due to the physical construction of the phantom, these two TBR levels were achieved by increasing the activity in the phantom background rather than changing the activity in the lesion (injecting more radio-tracer to the background, while keeping the activity concentration in the lesion constant), effectively producing a reduction in the lesion activity with respect to the background. Image registration was performed between the two scans. The first row of Figure~\ref{fig:motivation} shows one slice of the two scans and their difference (same slice as in~\cite{Qin+:}). 
	
	By design, there is a large background change but no change in the lesion. A direct difference between the two scans shows a global non-homogeneous background change while failing to detect changes in the lesion (Figure~\ref{fig:motivation}, Row 1 and Column 3). In contrast, the analysis based on the model-based standardized differences proposed in this paper is successful (Figure~\ref{fig:motivation}, Row 2 and Column 3). Specifically, the second row shows the standardized scores using the proposed robust EM
	algorithm RB-SGMM and background adjustment via the soft-assignment transformation $T_S^{(1)}$. The estimated background parameters are those given in~\eqref{eq:data.mean}. The standardized scores show a distribution close to standard normal with little anatomical structure except for the lesion. The standardized difference in the third column again has a distribution close to standard normal and exhibits the lesion change clearly at -6 standard deviations away from 0.
	
	To better appreciate the distributions, Figure~\ref{figure:bi.contour.phantom} plots the
	density contours of the original observations and standardized scores for the three
	estimation methods considered in this paper. As in Figure~\ref{figure:bi.sim.contour},
	while the standardized scores $T_S^{(1)}$ are obtained using soft assignment, the three
	classes are separated in the plots by hard assignment to ease visualization. Compared to
	GMM and SGMM, the robust RB-SGMM method offers the best standardization and separation of the pixels corresponding to the lesion.
	
	\begin{figure}
		\caption{Bivariate density contour plots of the phantom data with and without soft
			background adjustment within each class. The first column is for the original
			observation, while the last three columns are the scores after background adjustment
			using the methods of GMM, SGMM and RB-SGMM respectively. } 	
		\label{figure:bi.contour.phantom} 	
		\begin{tabular}{*{4}{c@{\hskip -2pt}}} 		
			\includegraphics[trim = 80 25 82 25, clip, width =
			0.24\textwidth]{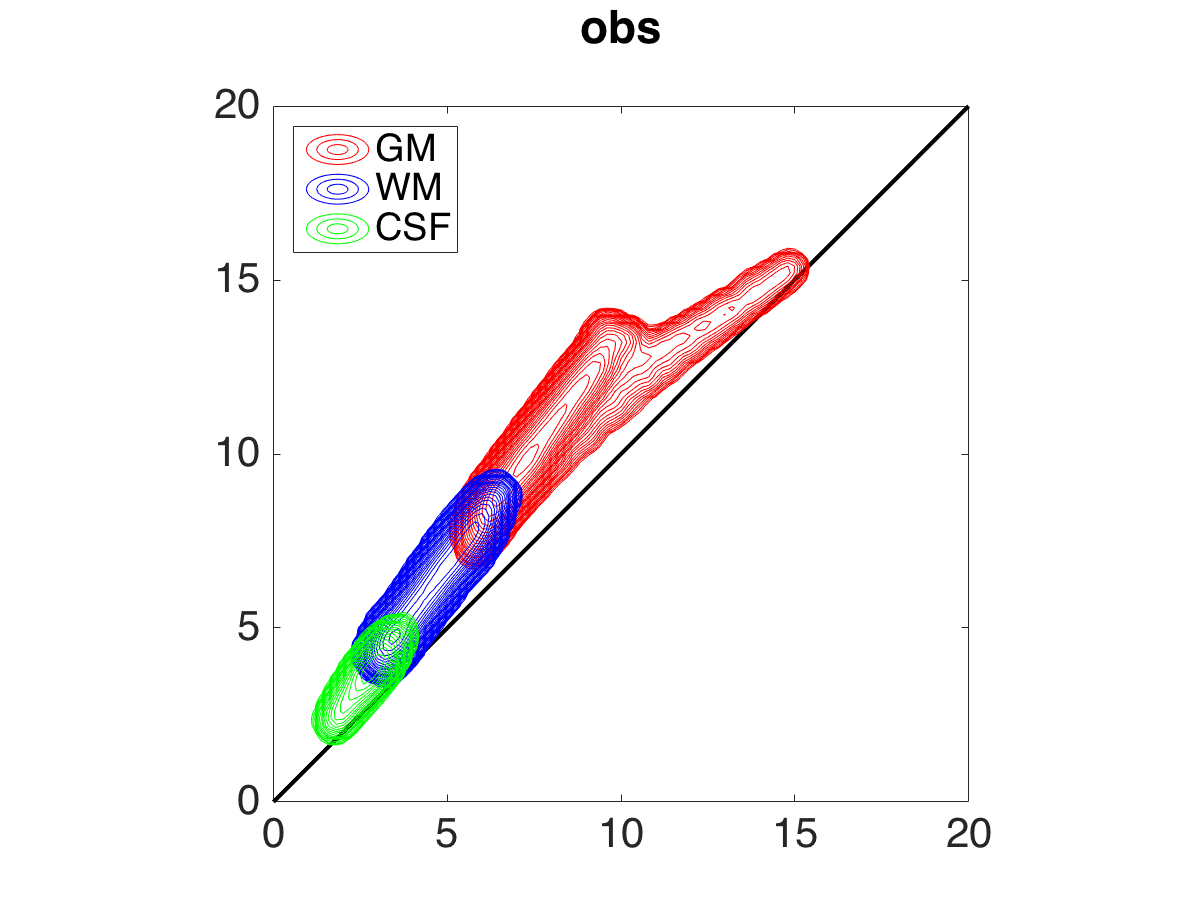} 		&  \includegraphics[trim = 80 25
			82 25, clip, width = 0.24\textwidth]{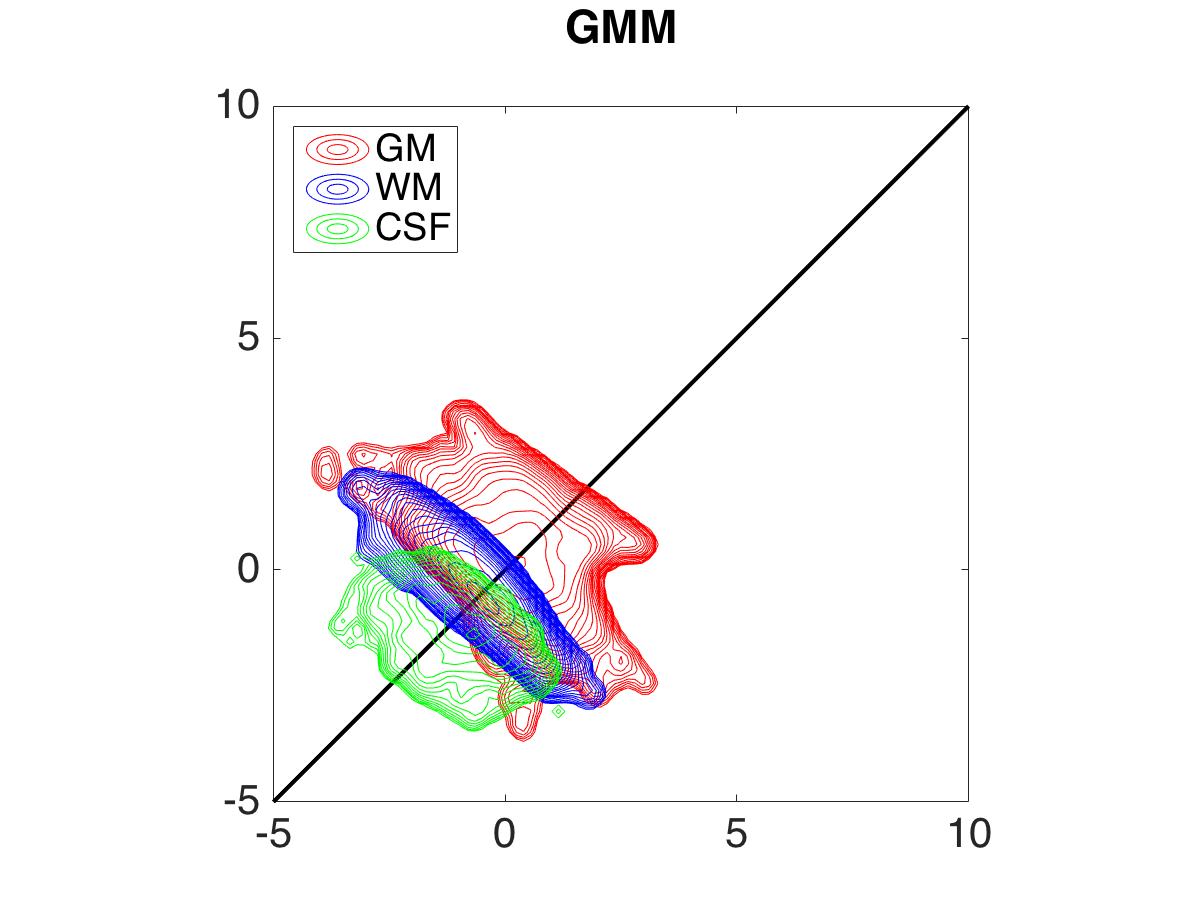} 		&
			\includegraphics[trim = 80 25 82 25, clip, width =
			0.24\textwidth]{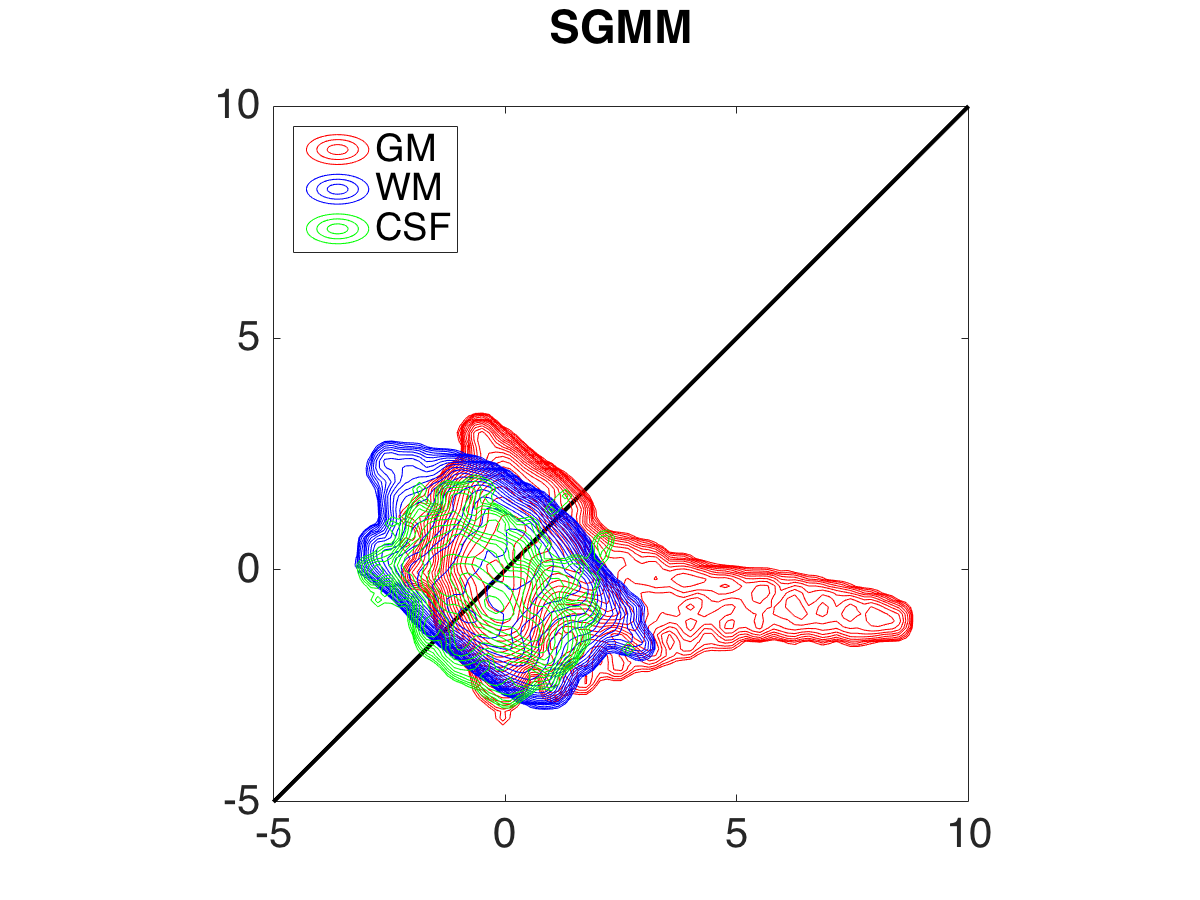} 		& \includegraphics[trim = 80 25
			82 25, clip, width = 0.24\textwidth]{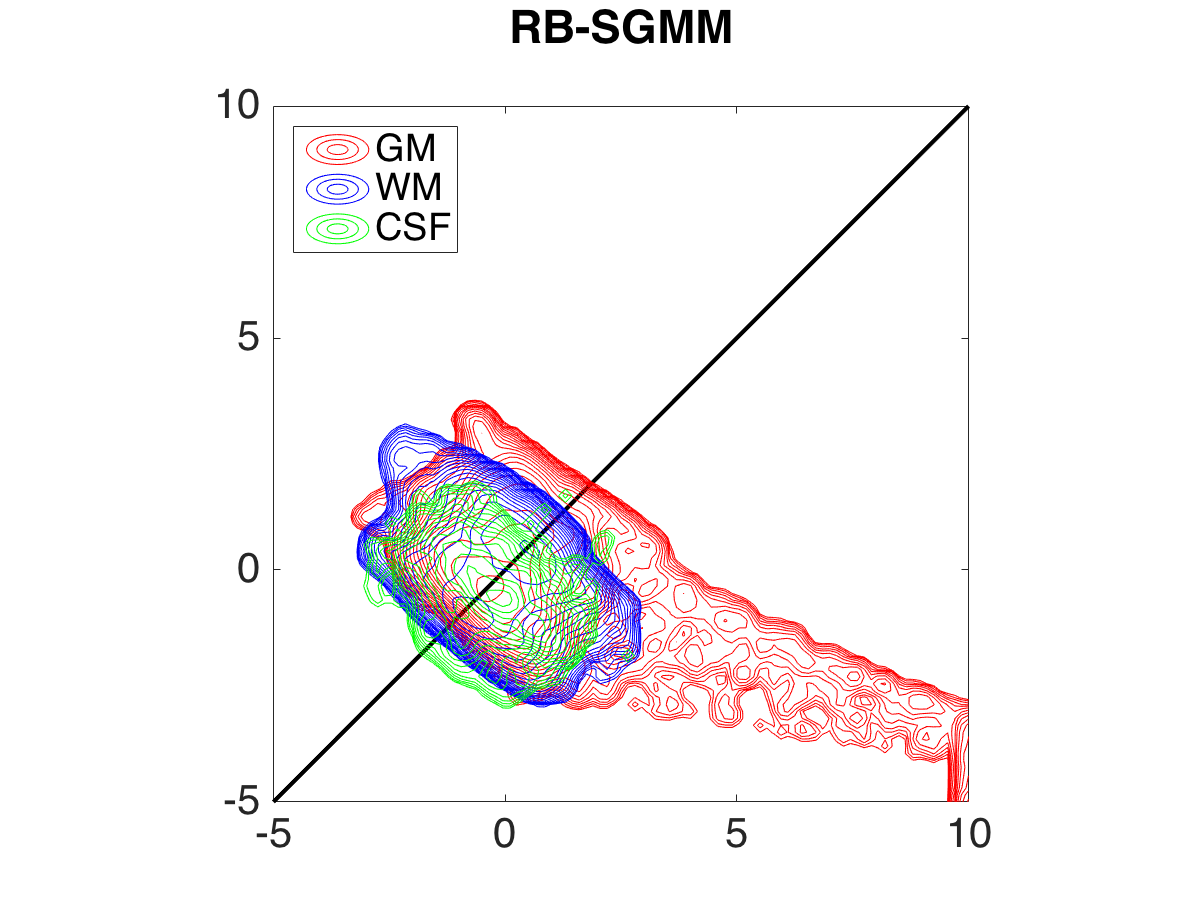} \\ 		Raw
			Observation & GMM & SGMM  & RB-SGMM 	
		\end{tabular} 
	\end{figure}
	\section{Discussion} \label{section:discussion}
	In this paper, we addressed the background adjustment problem in brain image analysis where there is interest in detecting outliers (i.e., tumor lesions) against the mixture model background. This problem motivated a robust EM algorithm to estimate the model parameters in a spatial Gaussian mixture model and standardization methods for background adjustment using both soft and hard assignment. We found that the proposed RB-SGMM method using soft assignment is justified for the purpose of hypothesis testing in the sense that tail probabilities at each voxel tend to be equal or smaller than those of the standard normal, thereby making testing accurate or conservative in a variety of
	scenarios.
	
	In terms of standardization, we found that the differences in performance between
	$T_S^{(1)}$, $T_S^{(2)}$ and $T_S^{(3)}$ are small, as long as soft assignment is used.
	We generally prefer $T_S^{(1)}$ because of its ease of analysis {(it depends on a lower dimensional parameter space)} and slightly better
	performance, and is the default method in the software package {\tt RB-SGMM-BA}. {Hard assignment is less reliable, performing worse than soft assignment for some mixture parameter combinations.}
	For univariate data, there exists the possibility of
	applying a quantile transformation as a way of transforming the observed mixture
	variables to standard normal. We did not consider the quantile transformation here
	because it does not have an obvious extension to bivariate or multivariate data, which
	has been the focus of this paper. Other simulations, not presented here, indicate that
	the quantile transformation for univariate data may be very sensitive to the estimates of
	the class-belonging probabilities $\pi_{ik}$ and yield undesired anti-conservative tail
	probabilities.
	
	Including a robust step in the SGMM algorithm to create RB-SGMM has been found to be critical. To create RB-SGMM, we modified the EM algorithm of~\cite{SPM:05} in such a way
	that we kept the number of classes in the mixture to represent the major tissue types,
	but made the M step robust to outliers. The proposed robust method replaces that of
	\citet{Qin+:}, which iteratively re-estimated the mixture parameters from observations
	whose standardized scores, after applying the contrast of interest, were in absolute
	value less than a constant $c$ (they used $c=2$). While also iterative, our proposed robust
	estimation is based on the theory of M-estimation and the EM
	algorithm, {yielding accurate results}. The use of a
	continuous weight function $\min(s, k_1(p))/s$ in the M-estimators is preferable to the
	discontinuous weight function that would correspond to the method of \citet{Qin+:} and is
	less sensitive to the tuning parameter $k_1(p)$ than to the constant $c$.
	A more sophisticated model could allow the tissue-belonging
	probabilities $\pi_{ik}$ to be modified not only according to regular anatomical
	variation via the $\gamma_k$ coefficients in \eqref{eq:b2pi} but also due to the
	anatomical deformations produced by the {lesions} themselves. A possible combination of SGMM
	and GMM including a non-spatial component for lesions may be an idea to consider in
	future work.
	
	Although tailored to brain image analysis, we emphasize that the developed standardization of GMM with robust estimation may be applicable to other settings (e.g., genomics), where there is interest in detecting signal against a mixture background.


\begin{supplement}\label{supp}
\sname{Supplement}
\stitle{Additional material}
\slink[doi]{ATTACHED TO THIS FILE}
\sdatatype{.pdf}
\sdescription{Supplementary materials contain: (A) proofs of all theorems and lemmas in the main paper; (B) a simulation study to compare the proposed robust EM with the multivariate $t$ mixtures method~\citep{peel2000robust} in a non-spatial setting; (C) additional simulation studies of the proposed RB-SGMM approach when lesions have smaller sizes and are non-circular.}
\end{supplement}

\bibliographystyle{apalike}
\bibliography{SGMM}

\begin{thebibliography}{}

\bibitem[Ashburner, 2012]{Ashburner2012}
Ashburner, J. (2012).
\newblock {SPM: A history}.
\newblock {\em NeuroImage}, 62(2):791--800.

\bibitem[Ashburner and Friston, 2005]{SPM:05}
Ashburner, J. and Friston, K.~J. (2005).
\newblock {Unified segmentation}.
\newblock {\em Neuroimage}, 26(3):839--851.

\bibitem[Bai et~al., 2013]{Bai2013}
Bai, B., Bading, J., and Conti, P.~S. (2013).
\newblock {Tumor quantification in clinical positron emission tomography}.
\newblock {\em Theranostics}, 3(10):787--801.

\bibitem[Besag, 1986]{Besag1986a}
Besag, J. (1986).
\newblock {On the statistical analysis of dirty pictures}.
\newblock {\em Journal of the Royal Statistical Society. Series B (Statistical
  Methodology)}, 48(3):259--302.

\bibitem[Borghammer et~al., 2009]{borghammer2009data}
Borghammer, P., Aanerud, J., and Gjedde, A. (2009).
\newblock {Data-driven intensity normalization of PET group comparison studies
  is superior to global mean normalization}.
\newblock {\em Neuroimage}, 46(4):981--988.

\bibitem[Campbell, 1984]{Campbell:84}
Campbell, N.~A. (1984).
\newblock {Mixture models and atypical values}.
\newblock {\em Mathematical Geology}, 16:465--477.

\bibitem[Chen et~al., 2001]{chen2001markov}
Chen, J.~L., Gunn, S.~R., Nixon, M.~S., and Gunn, R.~N. (2001).
\newblock {Markov random field models for segmentation of PET images}.
\newblock In {\em Biennial International Conference on Information Processing
  in Medical Imaging}, pages 468--474. Springer.

\bibitem[Dasgupta et~al., 2005]{dasgupta2005learning}
Dasgupta, A., Hopcroft, J., Kleinberg, J., and Sandler, M. (2005).
\newblock {On learning mixtures of heavy-tailed distributions}.
\newblock In {\em 46th Annual IEEE Symposium on Foundations of Computer Science
  (FOCS'05)}, pages 491--500. IEEE.

\bibitem[Dempster et~al., 1977]{Dempster+Laird+Rubin:77}
Dempster, A.~P., Laird, N.~M., and Rubin, D.~B. (1977).
\newblock {Maximum likelihood from incomplete data via the EM algorithm}.
\newblock {\em Journal of the Royal Statistical Society. Series B (Statistical
  Methodology)}, 39(1):1--38.

\bibitem[Devlin et~al., 1981]{Dev+Gna+Ket:81}
Devlin, S.~J., Gnanadesikan, R., and Kettenring, J.~R. (1981).
\newblock {Robust estimation of dispersion matrices and principal components}.
\newblock {\em Journal of the American Statistical Association},
  76(374):354--362.

\bibitem[Figueiredo and Jain, 2002]{figueiredo2002unsupervised}
Figueiredo, M. A.~T. and Jain, A.~K. (2002).
\newblock {Unsupervised learning of finite mixture models}.
\newblock {\em IEEE Transactions on Pattern Analysis and Machine Intelligence},
  24(3):381--396.

\bibitem[Guo et~al., 2014]{Guo+:14}
Guo, M., Yap, J.~T., den Abbeele, A.~D., Lin, N.~U., and Schwartzman, A.
  (2014).
\newblock {Voxelwise single-subject analysis of imaging metabolic response to
  therapy in neuro-oncology}.
\newblock {\em Stat}, 3(1):172--186.

\bibitem[Gupta and Chen, 2011]{gupta2011theory}
Gupta, M.~R. and Chen, Y. (2011).
\newblock {\em {Theory and use of the EM algorithm}}.
\newblock Now Publishers Inc.

\bibitem[Hanson, 2006]{Hanson2006}
Hanson, T.~E. (2006).
\newblock {Inference for mixtures of finite Polya tree models}.
\newblock {\em Journal of the American Statistical Association},
  101(476):1548--1565.

\bibitem[Hoffman et~al., 1991]{Hoffman+:91}
Hoffman, E.~J., Cutler, P.~D., Guerrero, T.~M., Digby, W.~M., and Mazziotta,
  J.~C. (1991).
\newblock {Assessment of accuracy of PET utilizing a 3-D phantom to simulate
  the activity distribution of [18F] fluorodeoxyglucose uptake in the human
  brain}.
\newblock {\em Journal of Cerebral Blood Flow {\&} Metabolism}, 11:A17----A25.

\bibitem[Huber, 1964]{Huber:64}
Huber, P.~J. (1964).
\newblock {Robust Estimation of a Location Parameter}.
\newblock {\em Annals of Mathematical Statistics}, 35(1):73--101.

\bibitem[Leahy and Qi, 2000]{leahy2000statistical}
Leahy, R.~M. and Qi, J. (2000).
\newblock {Statistical approaches in quantitative positron emission
  tomography}.
\newblock {\em Statistics and Computing}, 10(2):147--165.

\bibitem[Lee et~al., 2009]{Lee+:09}
Lee, Y.-Y., Choi, C.~H., Kim, C.~J., Kang, H., Kim, T.-J., Lee, J.-W., Lee,
  J.-H., Bae, D.-S., and Kim, B.-G. (2009).
\newblock {The prognostic significance of the SUVmax (maximum standardized
  uptake value for F-18 fluorodeoxyglucose) of the cervical tumor in PET
  imaging for early cervical cancer: preliminary results}.
\newblock {\em Gynecologic oncology}, 115(1):65--68.

\bibitem[Lin et~al., 2007]{lin2007finite}
Lin, T.~I., Lee, J.~C., and Yen, S.~Y. (2007).
\newblock {Finite mixture modelling using the skew normal distribution}.
\newblock {\em Statistica Sinica}, 17(3):909--927.

\bibitem[Lo et~al., 2001]{lo2001testing}
Lo, Y., Mendell, N.~R., and Rubin, D.~B. (2001).
\newblock {Testing the number of components in a normal mixture}.
\newblock {\em Biometrika}, 88(3):767--778.

\bibitem[Maronna, 1976]{Maronna:76}
Maronna, R.~A. (1976).
\newblock {Robust M-Estimators of Multivariate Location and Scatter}.
\newblock {\em The Annals of Statistics}, 4(1):51--67.

\bibitem[Maronna et~al., 2006]{Maronna+Martin+Yohai:06}
Maronna, R.~A., Martin, R.~D., and Yohai, V.~J. (2006).
\newblock {\em {Robust Statistics}}.
\newblock Wiley Series in Probability and Statistics. John Wiley {\&} Sons,
  Ltd, Chichester, UK.

\bibitem[McLachlan and Peel, 2000]{mclachlan2004finite}
McLachlan, G. and Peel, D. (2000).
\newblock {\em {Finite Mixture Models}}.
\newblock Wiley Series in Probability and Statistics. Wiley-Interscience, New
  York.

\bibitem[McLachlan and Basford, 1988]{McL+Bas:88}
McLachlan, G.~J. and Basford, K.~E. (1988).
\newblock {\em {Mixture models: Inference and applications to clustering}}.
\newblock Marcel Dekker, New York.

\bibitem[McLachlan and Krishnan, 2008]{McLachlan2007}
McLachlan, G.~J. and Krishnan, T. (2008).
\newblock {\em {The EM Algorithm and Extensions}}.
\newblock Wiley Series in Probability and Statistics. John Wiley {\&} Sons,
  Inc., Hoboken, NJ, USA, second edition.

\bibitem[O'Sullivan et~al., 2014]{o2014voxel}
O'Sullivan, F., Muzi, M., Mankoff, D.~A., Eary, J.~F., Spence, A.~M., and
  Krohn, K.~A. (2014).
\newblock {Voxel-level mapping of tracer kinetics in PET studies: A statistical
  approach emphasizing tissue life tables}.
\newblock {\em The Annals of Applied Statistics}, 8(2):1065--1094.

\bibitem[Peel and McLachlan, 2000]{peel2000robust}
Peel, D. and McLachlan, G.~J. (2000).
\newblock {Robust mixture modelling using the t distribution}.
\newblock {\em Statistics and Computing}, 10(4):339--348.

\bibitem[Qin et~al., 2017]{Qin+:}
Qin, L., Schwartzman, A., McCall, K., Kachouie, N.~N., and Yap, J.~T. (2017).
\newblock {Method for detecting voxelwise changes in
  fluorodeoxyglucose-positron emission tomography brain images via background
  adjustment in cancer clinical trials}.
\newblock {\em Journal of Medical Imaging}, 4(2):024006.

\bibitem[Qin and Priebe, 2013]{Qin2013}
Qin, Y. and Priebe, C.~E. (2013).
\newblock {Maximum L q -Likelihood Estimation via the Expectation-Maximization
  Algorithm: A Robust Estimation of Mixture Models}.
\newblock {\em Journal of the American Statistical Association},
  108(503):914--928.

\bibitem[Redner and Walker, 1984]{redner1984mixture}
Redner, R.~A. and Walker, H.~F. (1984).
\newblock {Mixture densities, maximum likelihood and the EM algorithm}.
\newblock {\em SIAM review}, 26(2):195--239.

\bibitem[Richardson and Green, 1997]{richardson1997bayesian}
Richardson, S. and Green, P.~J. (1997).
\newblock {On Bayesian analysis of mixtures with an unknown number of
  components (with discussion)}.
\newblock {\em Journal of the Royal Statistical Society: Series B (Statistical
  Methodology)}, 59(4):731--792.

\bibitem[Sanjay-Gopal and Hebert, 1998a]{Hebert1998}
Sanjay-Gopal, S. and Hebert, T. (1998a).
\newblock {Bayesian pixel classification using spatially variant finite
  mixtures and the generalized EM algorithm}.
\newblock {\em IEEE Transactions on Image Processing}, 7(7):1014--1028.

\bibitem[Sanjay-Gopal and Hebert, 1998b]{sanjay1998bayesian}
Sanjay-Gopal, S. and Hebert, T.~J. (1998b).
\newblock {Bayesian pixel classification using spatially variant finite
  mixtures and the generalized EM algorithm}.
\newblock {\em IEEE Transactions on Image Processing}, 7(7):1014--1028.

\bibitem[Soffientini et~al., 2016]{Soffientini2016}
Soffientini, C.~D., {De Bernardi}, E., Zito, F., Castellani, M., and Baselli,
  G. (2016).
\newblock {Background based Gaussian mixture model lesion segmentation in PET}.
\newblock {\em Medical Physics}, 43(5):2662--2675.

\bibitem[Soret et~al., 2007]{Soret2007a}
Soret, M., Bacharach, S.~L., and Buvat, I. (2007).
\newblock {Partial-volume effect in PET tumor imaging}.
\newblock {\em Journal of Nuclear Medicine}, 48(6):932--945.

\bibitem[Stephens, 2000]{stephens2000bayesian}
Stephens, M. (2000).
\newblock {Bayesian analysis of mixture models with an unknown number of
  components—an alternative to reversible jump methods}.
\newblock {\em The Annals of Statistics}, 28(1):40--74.

\bibitem[Takeda et~al., 2011]{Tak+:11}
Takeda, A., Yokosuka, N., Ohashi, T., Kunieda, E., Fujii, H., Aoki, Y., Sanuki,
  N., Koike, N., and Ozawa, Y. (2011).
\newblock {The maximum standardized uptake value (SUVmax) on FDG-PET is a
  strong predictor of local recurrence for localized non-small-cell lung cancer
  after stereotactic body radiotherapy (SBRT)}.
\newblock {\em Radiotherapy and Oncology}, 101(2):291--297.

\bibitem[{Thanh Minh Nguyen} and Wu, 2012]{Nguyen2012a}
{Thanh Minh Nguyen} and Wu, Q. M.~J. (2012).
\newblock {Gaussian-Mixture-Model-Based Spatial Neighborhood Relationships for
  Pixel Labeling Problem}.
\newblock {\em IEEE Transactions on Systems, Man, and Cybernetics, Part B
  (Cybernetics)}, 42(1):193--202.

\bibitem[Valk et~al., 2003]{Valk2003}
Valk, P.~E., Bailey, D.~L., Townsend, D.~W., and Maisey, M.~N. (2003).
\newblock {\em {Positron emission tomography: basic science and clinical
  practice}}.
\newblock Springer London.

\bibitem[Vehtari and Ojanen, 2012]{Vehtari2012a}
Vehtari, A. and Ojanen, J. (2012).
\newblock {A survey of Bayesian predictive methods for model assessment,
  selection and comparison}.
\newblock {\em Statistics Surveys}, 6(August):142--228.

\bibitem[Venturini et~al., 2008]{venturini2008gamma}
Venturini, S., Dominici, F., and Parmigiani, G. (2008).
\newblock {Gamma shape mixtures for heavy-tailed distributions}.
\newblock {\em The Annals of Applied Statistics}, 2(2):756--776.

\bibitem[Vlassis and Likas, 1999]{vlassis1999kurtosis}
Vlassis, N. and Likas, A. (1999).
\newblock {A kurtosis-based dynamic approach to Gaussian mixture modeling}.
\newblock {\em IEEE Transactions on Systems, Man, and Cybernetics-Part A:
  Systems and Humans}, 29(4):393--399.

\bibitem[Wahl et~al., 2009]{Wah+:09}
Wahl, R.~L., Jacene, H., Kasamon, Y., and Lodge, M.~A. (2009).
\newblock {From RECIST to PERCIST: evolving considerations for PET response
  criteria in solid tumors}.
\newblock {\em Journal of Nuclear Medicine}, 50(Suppl 1):122S--150S.

\bibitem[Young et~al., 1999]{Young+:99}
Young, H., Baum, R., Cremerius, U., Herholz, K., Hoekstra, O., Lammertsma,
  A.~A., Pruim, J., Price, P., and Others (1999).
\newblock {Measurement of clinical and subclinical tumour response using [18
  F]-fluorodeoxyglucose and positron emission tomography: review and 1999 EORTC
  recommendations}.
\newblock {\em European Journal of Cancer}, 35(13):1773--1782.

\bibitem[Zasadny and Wahl, 1993]{Zas+:93}
Zasadny, K.~R. and Wahl, R.~L. (1993).
\newblock {Standardized uptake values of normal tissues at PET with
  2-[fluorine-18]-fluoro-2-deoxy-D-glucose: variations with body weight and a
  method for correction.}
\newblock {\em Radiology}, 189(3):847--850.

\bibitem[Zhang et~al., 1994]{zhang1994}
Zhang, J., Modestino, J.~W., and Langan, D.~A. (1994).
\newblock {Maximum-likelihood parameter estimation for unsupervised stochastic
  model-based image segmentation}.
\newblock {\em IEEE Transactions on Image Processing}, 3(4):404--420.

\end{thebibliography}

\includepdf[pages=-]{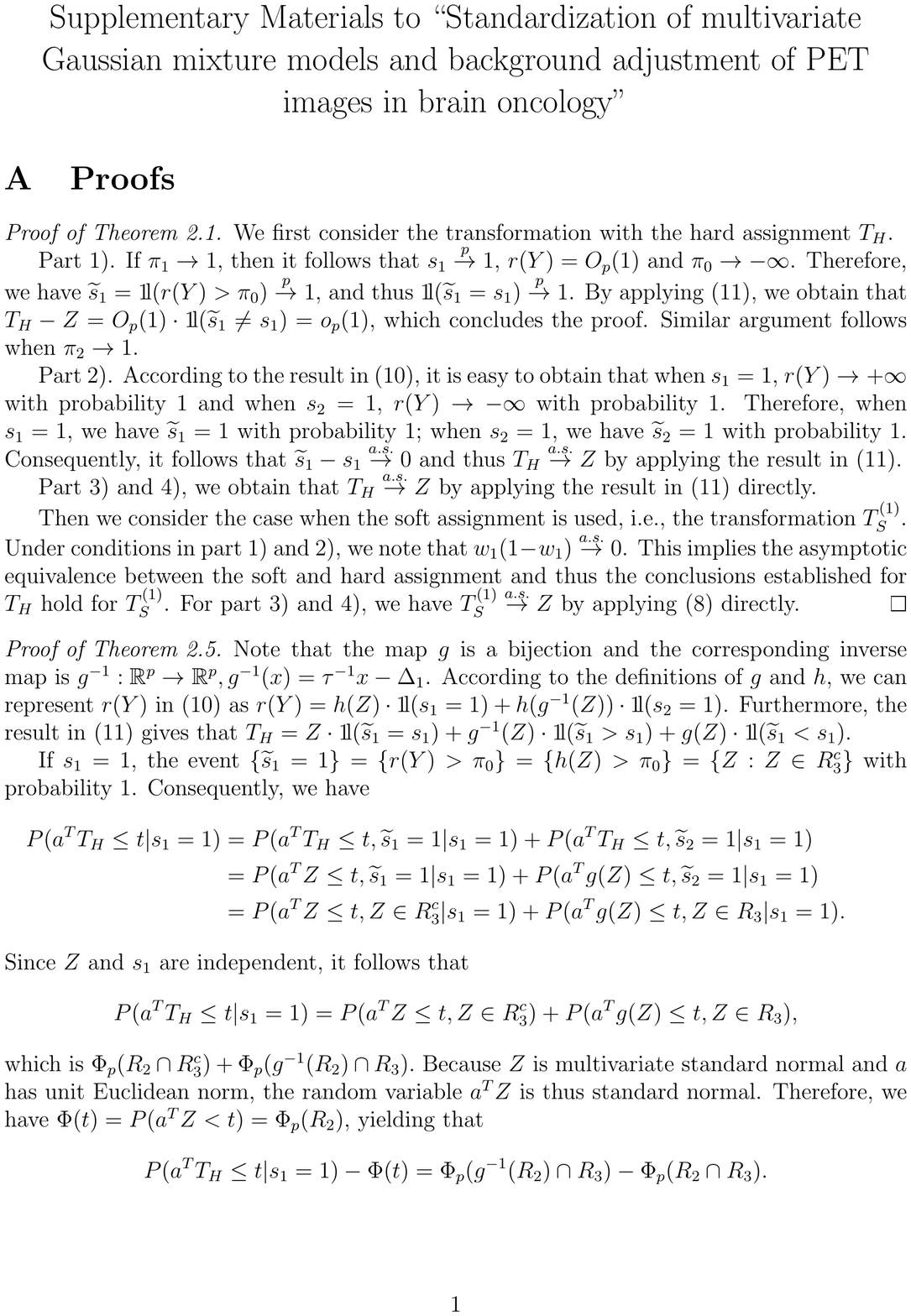}

\typeout{get arXiv to do 4 passes: Label(s) may have changed. Rerun}

\end{document}